\documentclass{elsart}
\usepackage{natbib} 
\usepackage{graphicx} 
\usepackage{amsbsy} 
\journal{New Astronomy}
\newcommand{\bS}{\mbox{\boldmath$S$}} 
 
\def\spose#1{\hbox to 0pt{#1\hss}}  
\def\lta{\mathrel{\spose{\lower 3pt\hbox{$\mathchar"218$}}  
        \raise 2.0pt\hbox{$\mathchar"13C$}}}       
\def\gta{\mathrel{\spose{\lower 3pt\hbox{$\mathchar"218$}}  
        \raise 2.0pt\hbox{$\mathchar"13E$}}} 
\def\gs{g~s$^{-1}$\ }  
 
\begin{document} 
\runauthor{J.P. Lasota} 
\begin{frontmatter} 
\title{The disc instability model of dwarf-novae and low-mass X-ray binary transients} 
\author{Jean-Pierre Lasota} 
\ead{lasota@iap.fr}

\address{Institut d'Astrophysique de Paris, 98bis Bd Arago, 75014 Paris, France}

\bigskip

\begin{abstract} 
The disc instability model which is supposed to describe outbursts of 
dwarf nova  and low-mass X-ray binary transient systems is presented and 
reviewed in detail. Various deficiencies of the model are pointed out 
and various remedies and generalizations are presented and discussed. 
\end{abstract}   
\begin{keyword} 
accretion discs; instabilities; dwarf novae; X-ray binaries; X-ray transients 
\end{keyword} 
\end{frontmatter}

{\small \sl In memory of Jan van Paradijs who inspired my research into 
the subject, and without whom this review would not have been written}
 
\section{Introduction} 
 
Dwarf novae are erupting cataclysmic variable stars (CVs) (Warner 1995a). 
In these binary systems outbursts take place in the accretion disc which 
is formed around the central white dwarf by matter transfered from the 
low-mass, Roche-lobe filling companion star. Low-mass X-ray binary 
transients (LMXBTs) are similar binary systems in which the white dwarf 
is replaced by a neutron star or a black hole (van Paradijs \& 
McClintock 1995). Such X-ray transient systems are also called ``Soft 
X-ray transients" or ``X-ray Novae". Both designations are misleading 
and we will use here only the term ``LMXBT". (The designation ``dwarf 
nova" is also misleading but it has stuck and cannot be changed). It is 
believed that outbursts in both dwarf novae and LMXBTs are driven by the 
same mechanism: disc instability. This mechanism was proposed for dwarf 
novae more than 25 years ago (Osaki 1974). The instability was 
identified 4 years later (Hoshi 1979) and the model itself began to take 
shape in the early 80's (Meyer \& Meyer-Hofmeister 1981; Smak 1982,1984b; 
Cannizzo, Ghosh \& Wheeler 1982; Faulkner, Lin \& Papaloizou 1983; 
Mineshige \& Osaki 1983). There are excellent articles reviewing the 
model -- Smak (1984c), Cannizzo (1993a) and Osaki (1996) to mention only a 
few -- so writing a new review on this venerable and well reviewed 
subject requires a good reason. 
 
Such a reason was unintentionally suggested by Trimble \& Aschwanden 
(2000) in their article reviewing astrophysics in 1999. After remarking 
that the dwarf-nova outburst mechanism is not known, the authors say 
with some sarcasm: ``[The] Hameury et al. (1999) conclusion that there 
are several kinds of such mechanisms is not likely to get anybody into 
serious trouble". This is true. But it is also rather new. Some fifteen 
years ago a similar conclusion did get some people into serious trouble. 
(For the history of the subject see the review by Cannizzo, 1993a). Until 
recently it was ``generally accepted" that the thermal-viscous disc 
instability model (DIM) is {\sl the} model of dwarf-nova outbursts. The 
main reason for this was that the competing model posited outbursts that 
were due to a mass-transfer instability, i.e. an instability in the 
secondary star. The DIM had a big advantage: it had a physical mechanism 
which was responsible for the instability. The mass-transfer instability 
model was lacked this essential ingredient because no plausible 
mechanism could be found. (The `plausible' mechanism proposed by 
Hameury, King \& Lasota 1986 for the dwarf-nova cousins, the ``Soft 
X-ray Transients", was shown not to work in practice by Gontikakis \& 
Hameury, 1993). Observations also rather favoured the DIM (with some 
exceptions, see e.g. Smak 1991). Thus the DIM was widely considered to 
be the winner and {\sl the} model. 
 
These days things are different. It is still considered that the DIM is 
the dwarf nova model, but instead of the pure, original version, in 
which a disc receiving matter at a constant rate was undergoing 
outbursts due to a thermal-viscous instability only, one finds a rather 
impure mixture in which the mass-transfer rate varies on all possible 
time-scales and the outburst properties depend on the irradiation of 
both the disc and the companion star. In some cases the thermal-viscous 
instability needs help from a `tidal' instability and big holes are cut 
in the disc's inner regions. Heating by the mass-transfer stream impact 
and by dissipation of the tidal torque was also found to play a role in 
the outburst. All that is left of the original DIM is the thermal 
instability itself but even this, one would think fundamental, 
ingredient of the model is now looked at with some suspicion.  
 
A short summary of some of the new ingredients can be found in Smak 
(2000). Although some of the modifications to the model date back a long 
time, most were successfully incorporated into the DIM only in the last 
few years. That is the real reason for this review.  
 
It does not intend to show that the DIM is the model of dwarf-nova 
outbursts and of transient behaviour of the low-mass X-ray binaries. Nor 
will it show that it is {\sl not} the model. Its aim is to show and 
discuss how the DIM works and why it needs modifications. Some of these 
modifications are required simply because we know from observations that 
physical processes ignored in the DIM operate in real binaries. Others 
are needed because the model, in its standard form, fails to reproduce 
some of the fundamental properties of the observed outbursts. Luckily, 
when some of the neglected processes are included in the model, they 
help to repair the failures. The price to pay, however, is that the 
number of free parameters increases and instead of a single model one 
now has several, none of them flawless. In recent years substantial 
progress has been made in understanding the fundamental physical process 
which is supposed to drive accretion in discs: turbulent viscosity. 
Paradoxically this has made things even more difficult for the DIM, 
since it challenges one of its main tenets: that both heating and 
angular momentum are due to the same, local mechanism. So if the DIM is 
to survive, it might require more than just a modification. 
 
Finally, a personal remark: in a recent scientific meeting a speaker (a 
very good and clever observer) showed a transparency with side by side 
the light curve of the best-observed dwarf nova SS Cyg (see Fig. 
\ref{comp_bc}) and several light-curves calculated by a well known 
specialist of the subject. ``As you can see", the speaker said, ``the 
model works pretty well". Only two people started to laugh: the author 
of the calculations, present at the meeting, and the author of this 
review. We were laughing because we would have made such a comparison to 
show that the model {\sl does not} work (see Sect. \ref{standard}). I 
hope this review will prevent people from saying that the DIM works when 
it does not, or at least that it will make more than two people laugh 
when such assertions are made. 
 
\begin{itemize} 
 
\item{} The review begins (Section 2) with a presentation and discussion 
of the DIM's equations. One could have chosen to begin by describing the 
observations the model is supposed to reproduce or imitate, but this 
review is about a model and it seemed preferable to start with its 
description. An excellent description of dwarf novae can be found in the 
`CV--Bible' by Warner (1995a); LMXTB properties are reviewed in Chen, 
Shrader \& Livio (1997) and Tanaka \& Shibazaki (1996). Too often, 
however, it is forgoten that a model like the DIM is a physical model, 
i.e. it consists of partial differential equations {\sl and} boundary 
conditions. It is not a vague scenario which says that discs are 
sometimes unstable.  
 
\item{} Section 3 is devoted to thermal equilibria and stability 
conditions and various generalizations of the DIM. In the first two 
sections after the introduction dwarf novae and LMXTBs are discussed 
together.  
 
\item{} The next Section, 4, deals mainly with dwarf novae. However, if 
a problem is common to the two classes of systems it is often discussed 
in this section. Here, at last, one can find concise presentations of 
various classes of dwarf novae. Also here we begin to confront the model 
with observations. This section contains a description of a dwarf nova 
outburst which might appear rather lengthy but it is necessary in order 
to understand later sections of the review. Section 4 contains a 
discussion of the DIM outburst, the heating and cooling fronts, the front 
`reflections' (reflares), the inside-out and outside-in outbursts and 
related problems (the `UV-delay'). This section ends with the discussion 
of the outburst recurrence times.  
 
\item{} Section 5 contains models of various classes of dwarf nova 
stars. All of these models are a generalization of the DIM, the previous 
section having shown that the standard DIM is unable to reproduce even 
the simplest properties of dwarf-nova outbursts. This section ends with 
a discussion of the problems caused by dwarf nova quiescence.  
 
\item{} Models of Low Mass X-ray Binary Transient systems are presented 
in Section 6. The review ends with a section discussing the main 
questions to be answered and the future of the DIM and its successors. 
 
Although this review contains more than 200 references it really uses the 
work of only three groups of researchers. The reason is given by Smak 
(1998): ``Of all models calculated in the past, only those of Smak (1984b, 
[...1998]), Ichikawa \& Osaki (1992), and Hameury et al. (1998,[1999]) used 
correct outer boundary conditions, describing the deposition of the 
stream material in the outermost parts of the disc, and the effects of 
the angular momentum".  
 
\end{itemize}

\section{Equations} 
 
The DIM is often introduced by showing the \bS-shaped curve which 
represents local disc equilibria.  The \bS-shape is due to the 
presence of a thermal-viscous instability: the middle part of the \bS 
represent unstable equilibria. With this \bS-curve one then can show 
schematically how the disc's state may perform a ``limit cycle" 
oscillating between hot and cold states. Such a schematic picture, is 
useful in understanding the basic reason for the disc instability, but 
it can be misleading, because although the instability is local the 
outburst itself is a global process and quite often the resulting local 
disc behaviour does not correspond to the simple, schematic \bS-curve 
diagram. We therefore start the presentation of the model by the 
equations which are assumed to represent an accretion disc subject to a 
thermal and viscous instability. We will come to the \bS-curve later, in 
Sect. \ref{scurve}. 
 
In the disc instability model we assume that the disc is geometrically 
thin and in particular that its angular momentum is always Keplerian. 
This allows a separation of its vertical and radial structures (see e.g. 
Pringle 1981). Several, more or less equivalent, versions of the disc's 
structure equations have been used in the literature. Here we will 
follow the version of Hameury et al. (1998, thereafter HMDLH), which in 
turns closely follows that of Smak (1984b). 
 
\subsection{Radial equations} 
 
The mass and angular momentum conservation equations in a 
geometrically thin accretion disc can be written as: 
\begin{equation} 
{\partial \Sigma \over \partial t} = - {1 \over r} {\partial \over \partial 
r} (r \Sigma v_{\rm r}) + {1 \over 2 \pi r} {\partial \dot{M}_{\rm ext} \over 
\partial r} 
\label{eq:consm} 
\end{equation} 
and 
\begin{eqnarray} 
{\partial \Sigma j_K \over \partial t} = - {1 \over r} {\partial \over \partial 
r} (r \Sigma j_K v_{\rm r}) & + & {1 \over r} {\partial \over \partial r} 
\left(- {3 \over 2} r^2 \Sigma \nu \Omega_{\rm K} \right) + \nonumber \\ 
& & { j_{\rm 2} \over 2 \pi r} {\partial \dot{M}_{\rm ext} \over \partial r} - 
{1 \over 2 \pi r} {\mathcal T}_{\rm tid}(r), 
\label{eq:consj} 
\end{eqnarray} 
where the surface density $\Sigma=2 \int_{0}^{+\infty} \rho dz$, with 
$\rho$ the mass density, $\dot{M}_{\rm ext}(r,t)$ is the rate at which 
mass is incorporated into a ring at radius $r$, $v_{\rm r}$ the radial 
velocity in the disc, $j_K = (GM_1r)^{1/2}$ is the specific Keplerian 
angular momentum of material at radius $r$ in the disc, 
$\Omega_K=(GM_1/r^3)^{1/2}$ is the Keplerian angular velocity ($M_1$ 
being the mass of the accreting object), $\nu$ is the kinematic 
viscosity coefficient, and $j_{\rm 2}$ the specific angular momentum of 
the material transferred from the secondary. ${\mathcal T}_{\rm tid}$ is the torque 
due to tidal forces, whose form will be specified later (Eq. 
\ref{eq:defT}). 
 
The energy conservation (thermal equation) can be written as: 
\begin{equation} 
{\partial T_{\rm c} \over \partial t} = { 2 (Q^+ -Q^- +(1/2)Q_i + J) \over C_P \Sigma} 
 - {\Re T_{\rm c} 
\over \mu C_P} {1 \over r} {\partial (r v_{\rm r}) \over \partial r} - 
v_{\rm r} {\partial T_{\rm c} \over \partial r}, 
\label{eq:heat} 
\end{equation} 
where $Q^+$ and $Q^-$ are respectively the heating and cooling rates per unit surface. 
Their usual form is  
\begin{equation} 
Q^+=\frac{9}{8} \nu \Sigma \Omega_{\rm K}^2  
\label{qplus} 
\end{equation} 
and $Q^- =\sigma T_{\rm eff}^4$, where $T_{\rm eff}$ is the effective temperature.  
$C_P$ is the specific heat at constant pressure. 
 
The term $J$ accounts for the radial energy flux carried by viscous 
processes and/or by radiation. There is no generally accepted form of 
this term; for particular versions see e.g. HMDLH and Cannizzo (1993b). 
(The form used in HMDLH assumes ``viscous" radial energy-transport.) 
Since it contains a second derivative with respect to radius, this term 
can be neglected when considering equilibria of geometrically thin 
discs  because then the energy equation is just $Q^+=Q^-$ and radial 
gradients are small compared to the vertical ones.  
During outbursts, when steep temperature gradients are present in 
the disc, it can play an important role in front propagation (Menou, 
Hameury \& Stehle 1999). 
 
Eq. (\ref{eq:heat}) also contains terms describing heating of the outer 
disc by the impact of the stream of matter transferred from the 
secondary (the ``hot spot") and by the tidal torque ${\mathcal T}_{\rm tid}$. These 
terms, denoted by $Q_i$, are not usually taken into account. Except for 
Ichikawa \& Osaki (1992), only Buat-M\'enard, Hameury \& Lasota 
(2001a,b) calculated models which include these two additional heat 
sources. As we will see later, including these terms helps repair 
several of the DIM's deficiencies. Heating by impact and by tidal 
torques can be treated only in a rather simple way. The tidal torque is 
taken from Papaloizou \& Pringle (1977): 
\begin{equation} 
{\mathcal T}_{\rm tid} = c \omega r \nu \Sigma \left( {r \over a} \right)^{5}, 
\label{eq:defT} 
\end{equation} 
where $\omega$ is the angular velocity of the binary orbital motion, $c$ 
is a numerical coefficient taken so as to give a stationary (or time 
averaged) disc radius equal to a chosen value, and $a$ is the binary 
orbital separation. The viscous dissipation induced by this torque is 
written as 
\begin{equation} 
Q_{\rm tid}(r)=\left(\Omega_K(r) - \Omega_{\rm orb}\right){\mathcal T}_{\rm tid}(r) 
\label{qtid} 
\end{equation} 
 
The heating by the hot spot is difficult to evaluate in a way that would 
be both physically consistent and numerically interesting. Buat-M\'enard 
et al. (2001a) assumed that the stream impact heats an annulus fraction 
$\Delta r_{\rm hs}$ of the disc with an efficiency $\eta_{\rm i}$. The 
heating rate $Q_{\rm i}$ is then taken as: 
\begin{equation} 
Q_{\rm i}(r) = \eta_{\rm i} \frac{G M_1 \dot{M_{\rm tr}}}{2 r_{\rm 
out}} \frac{1}{2 \pi r_{\rm out} \Delta r_{\rm hs}} 
\exp \left(-\frac{r_{\rm out}-r}{\Delta r_{\rm hs}}\right) 
\end{equation} 
where $\dot{M_{\rm tr}}$ the mass transfer rate from the secondary and $r_{\rm 
out}$ the accretion disc outer radius. This assumes that the difference 
between the stream and the Keplerian kinetic energy is released in a 
layer whose width is $\Delta r_{\rm hs}$ with an exponential 
attenuation. 
 
\subsection{Boundary conditions} 
 
The partial differential equations which describe the radial disc 
structure must be completed by boundary conditions. Some of these 
boundary conditions affect the solutions in a fundamental way (see Sect. 
\ref{standard}). This is the case of the outer boundary conditions for the 
mass and angular momentum equations. 
 
\subsubsection{Outer boundary} 
 
A proper treatment of the precise way in which matter is incorporated 
into the disc is a very difficult problem in itself, and it can be 
included in the DIM only in the simplest fashion. The assumption  
apparently reasonable but not always confirmed by observations 
is that mass addition at the outer edge of the disc occurs in a 
very narrow region, so that the disc edge is very sharply defined. One 
can then write $\dot{M}_{\rm ext}(r) = \dot{M}_{\rm tr} \delta 
(r_0(t)-r)$ and $\Sigma = \Sigma_0 Y(r_0(t)-r)$, where $\dot{M}_{\rm 
tr}$ is the mass transfer rate from the secondary star, $Y$ is the 
Heavyside function, $\delta$ is the Dirac function and $\Sigma_0$, the 
surface column density, is a smoothly varying function. Cancelling 
the $\delta$ terms in the equation for mass and angular momentum 
conservation yields two boundary conditions, which can be written in the 
form (HMDLH): 
\begin{equation} 
\dot{M}_{\rm tr} = 2 \pi r \Sigma_0 (\dot{r}_0 - v_{\rm r,0}) 
\label{eq:bc1} 
\end{equation} 
and 
\begin{equation} 
\dot{M}_{\rm tr} \left[ 1 - \left( {r_{\rm k} \over r_0}\right)^{1/2} 
\right] = 3 \pi \nu \Sigma_0, 
\label{eq:bc2} 
\end{equation} 
where the index 0 denotes quantities measured at the outer edge, and 
$r_{\rm k}$ is the circularization radius, i.e. the radius at which the 
Keplerian angular momentum is that of the matter lost by the secondary 
star ($j_{\rm 2}$ in Eq. \ref{eq:consj}), and $\dot{r}_0$ is the time 
derivative of the outer disc radius. It is worth noting that in this 
formulation the presence of a torque ${\mathcal T}_{\rm tid}$ is necessary; no 
steady solutions exist when ${\mathcal T}_{\rm tid}=0$. 
 
Conditions given by Eqs. (\ref{eq:bc1}) and (\ref{eq:bc2}) take into 
account the fact that the outer edge of the disc can vary with time;  
its position is controlled by the tidal torque ${\mathcal T}_{\rm tid}$.  
 
The expression for the tidal torque requires  careful reanalysis. 
Observations of accretion disc radii (Harrop-Allin \& Warner 1996) in 
quiescent dwarf novae give values smaller than those calculated in the 
model and as suggested by Smak (2000) the culprit could be Eq. 
(\ref{eq:defT}),  in particular the parameter $c$ which determines the 
disc's size (Buat-M\'enard et al. 2001a). We will come back  
to this issue in more detail later 
when discussing the nature of  `inside-out' and `outside-in' 
outbursts (Sect. \ref{io}). 
 
When modelling dwarf-nova outbursts it is often assumed that the outer 
edge of the disc is fixed at a given radius, in which case 
Eq.~(\ref{eq:bc1}) is used with $\dot{r}_0 = 0$, and Eq.~(\ref{eq:bc2}) 
is replaced by $r = r_0$; the tidal torque ${\mathcal T}_{\rm tid}$ is also 
neglected in Eq.~(\ref{eq:consj}). This is equivalent to assuming that 
${\mathcal T}_{\rm tid}$ is negligible at $r < r_0$ and becomes infinite at $r = 
r_0$ (Cannizzo calls this a ``brick wall''). In view of the steep 
functional dependence of ${\mathcal T}_{\rm tid}(r)$, this might seem a reasonable 
approximation; however, this is not the case as can be seen in Fig. 
\ref{comp_bc} and discussed in HMDLH, and most of the results obtained 
with the ``brick wall'' conditions are of no interest when applied to 
real systems (see e.g. Smak 1998 for a discussion of this problem). An 
intermediate formulation was proposed by Mineshige \& Osaki (1985) who 
assumed that the viscous stresses vanish at a fixed outer radius. This 
enables matter carrying angular momentum to leave the disc at its outer 
edge, at a rate comparable to the mass transfer rate.

\subsubsection{The inner boundary} 
 
The inner boundary condition is usually taken to be the no-stress 
condition ($\nu \Sigma=0$), which in practice amounts to $\Sigma = 0$ at 
the inner edge of the disc. 
 
There a several reasons, discussed later in the article, why a 
dwarf~-~nova or LMXBT accretion disc may not extend down to the surface 
of the accreting body, or to the last stable orbit, but instead be 
truncated at some larger radius. In the case of white dwarfs and neutron 
stars the disc around them might be truncated by the magnetic field of 
the accreting body. In such a case the inner disc radius is given 
by the magnetospheric radius (see e.g. Frank, King \& Raine 1992): 
\begin{equation} 
r_{\rm mag} = 9.8 \times 10^8 
\left(\frac{\dot{M}}{10^{15}{\rm g\ s^{-1}}}\right)^{-2/7}  
\left(\frac{M_1}{\rm M_{\odot}}\right)^{-1/7}  
\left(\frac{\mu}{10^{30}\ {\rm G~cm}^3}\right)^{4/7} \; {\rm cm}, 
\label{rmag} 
\end{equation} 
where $\mu$ is the white dwarf's magnetic moment.  
 
But the disc could also be truncated by evaporation of its inner 
regions. No complete theory of this process exists and one uses {\sl ad 
hoc} formulae which are supposed to be a reasonable approximation of 
what the models (e.g. Meyer \& Meyer-Hofmeister 1994; Esin, McClintock 
\& Narayan 1997; Kato \& Nakamura 1999; Liu et al. 1999; Manmoto et al. 
2000; Shaviv, Wickramasinghe \& Wherse 1999) suggest. For example, Menou 
et al. (2000) and Dubus, Hameury \& Lasota (2001, hereafter DHL) use the 
following prescription: 
\begin{equation} 
\dot M_{\rm disc}(r_{\rm in})=\dot M_{\rm evap}(r_{\rm in}),  
\ \ \ \ \ 
\dot M_{\rm evap}(r)=\frac{0.08 \dot M_{\rm Edd}} 
{\left( {r}/{r_s} \right)^{1/4} +  
{\mathcal E}\left({r}/{800 r_s} \right)^2}. 
\label{evap}  
\end{equation} 
where $\dot M_{\rm Edd}=L_{\rm Edd}/0.1c^2=1.39 \times 10^{18} M/{\rm 
M}_{\odot}$ \gs is the Eddington rate corresponding the Eddington 
luminosity assuming a 10\% accretion efficiency; c is the speed of light 
and $r_S= 2GM_1/c^2$ is the Schwarzschild radius. ${\mathcal E}$ is an 
`evaporation efficiency' factor (see Menou et al. 2000).

\subsubsection{The thermal equation} 
 
The thermal equation is a second order partial differential equation in 
$r$, so that two boundary conditions are required. However, except 
across a transition front between a hot and a cool region, the dominant 
terms in Eq.~(\ref{eq:heat}) are $Q^+$ and $Q^-$. The highest order 
terms in the thermal equation are therefore negligible in almost all of 
the disc. These boundary conditions are thus of no physical importance. 
One can take, as for example in HMDLH, $\partial T_{\rm c} / \partial r 
= 0$ at both edges of the disc.

\subsection{Vertical structure equations} 
\label{svert} 
 
In a geometrically thin disc one can decouple the vertical and radial 
structures (e.g. Pringle 1981; Frank et al. 1992). There are several 
ways of describing the accretion disc vertical structure. In the DIM, 
equations describing the disc's vertical structure are usually written 
in a form similar to that used for stellar structure: 
\begin{eqnarray} 
\lefteqn{{dP \over dz} = -\rho g_{\rm z} = -\rho \Omega_{\rm K}^2 z, } 
\label{eq:stra}\\ 
\lefteqn{{d \varsigma \over dz} = 2 \rho,}  
\label{eq:strd} \\ 
\lefteqn{{d\ln T \over dz} = \nabla {d \ln P \over dz},}  
\label{eq:strb}\\ 
\lefteqn{{dF_{\rm z} \over dz } = {3 \over 2} \alpha 
 \Omega_{\rm K} P + \frac{dF_t}{dz},}  
\label{eq:strc} 
\end{eqnarray} 
where $g_{\rm z} = \Omega_{\rm K}^2 z$ is the vertical component of gravity, 
$\varsigma$ is the surface density between $-z$ and $+z$, 
and $\nabla=d\ln T/d \ln P$ the temperature gradient of the structure.  
$F_t$ is a time-dependent contribution that includes terms 
resulting from heating/cooling and contraction/expansion (see below).  
 
In the case of radiative energy transport the temperature gradient can  
be written as  
\begin{equation} 
\nabla_{\rm rad} = {\kappa_{\rm R} P F_{\rm z} \over 4 P_{\rm rad} c g_{\rm z}}, 
\label{radtr} 
\end{equation} 
where $P_{\rm rad}$ is the radiation pressure and $\kappa_{\rm R}$ is 
the Rosseland mean-opacity coefficient. This formulation assumes that 
the disc is optically thick. Since it would be practically impossible to 
include an accurate treatment of the radiative transfer (such as that of 
Shaviv \& Wehrse 1991) in time-dependent calculations, a grey-atmosphere 
type approximation must be used when the disc is optically thin (see 
e.g. Smak 1984b; HMDLH). 
 
The DIM attributes the instability to partial hydrogen ionization, which 
implies that convection may become the dominant mode of energy transport, 
as these two processes often come together. Since we don't really know 
how convection operates in accretion discs this is one of the weak 
points of the model. The most efficient way of treating a convective 
disc ($\nabla=\nabla_{\rm conv}$) is to use the mixing-length 
approximation which is used in stellar models. The role of convection in 
accretion disc structure is very nicely described by Cannizzo in his 
1993 review to which the reader is referred for further discussion. The 
form of the equilibrium solutions is affected by the presence of 
convection (see Fig. \ref{fig:cnc}) but as explained in Sect. 
\ref{scurve} this, fortunately, is not very important in practice. 
 
Much more important is the term $F_t$ which is supposed to describe the 
contribution of all the non-equilibrium processes to the vertical disc 
structure. A correct inclusion of these effects would require a 2D 
treatment, which for the moment is out of the question. Instead one 
assumes that $F_t$ is proportional to $P$, or an {\it ansatz} roughly 
equivalent to this (e.g. Mineshige \& Osaki 1983). In such a case 
Eq.~(\ref{eq:strc}) can be replaced by: 
\begin{equation} 
{dF_{\rm z} \over dz } = {3 \over 2} \alpha_{\rm eff} \Omega_{\rm K} 
P, 
\label{eq:strcc} 
\end{equation} 
where $\alpha_{\rm eff}$ (the subscript has nothing to do with its 
namesake in the effective temperature !) may be considered as some 
effective viscosity parameter. This coefficient is not known {\it a 
priori} and is different from the true viscosity parameter (Eq. 
\ref{qplus2})); the difference $\alpha - \alpha_{\rm eff}$ is a measure 
of the departure from thermal equilibrium. The vertical structure, and 
hence the disc effective temperature at any given point, can then be 
calculated assuming that the disc is in thermal equilibrium, but with 
some unspecified $\alpha_{\rm eff}$ different from $\alpha$. The same 
method is used for the heating due to tidal torques and accretion stream 
impact (Buat-M\'enard et al. 2001a).  
 
Such an approximation is valid for perfect gas and homologous cooling 
and contraction (Smak 1984b), but in general is unjustified and can be 
the source of serious errors. For example, HMDLH estimate that the 
cooling fluxes in time-dependent accretion discs are only determined to 
within 50\%. This basic fact should not be forgotten when comparing the 
model's results with observation. Too good a correspondence should be 
treated with more suspicion than enthusiasm. 
 
The $\alpha_{\rm eff}$-approach is just an {\sl ansatz} replacing 
the required 2D treatment. 
Recently, Truss et al. (2000) used a 2D SPH code to simulate dwarf nova 
outbursts. To judge the value of this interesting approach  one should 
 wait for a more realistic physics to be added to the scheme.

\subsubsection{Boundary conditions} 
 
Vertical structure equations must be completed by boundary conditions. 
In the optically thick case Eqs. (\ref{eq:stra}--\ref{eq:strc}) are 
integrated between the disc midplane and the photosphere ($\tau_{\rm 
s}=2/3$), where $\tau(z)=\int_0^z \rho \kappa_{\rm R} dz$ is the optical 
depth. The boundary conditions are $z = 0$, $F_{\rm z} = 0$, $T = T_{\rm 
c}$, $\varsigma = 0$ at the disc midplane, and  
\begin{equation} 
T^4(\tau_{\rm s}) = T^4_{\rm eff} + T^4_{\rm irr} 
\label{bcondv} 
\end{equation} 
where $\sigma T^4_{\rm eff}=Q^+$, at the photosphere (Dubus et al. 1999;
hereafter DLHC).  
The term $T^4_{\rm irr}$ allows one to take into account the effects of  
irradiation (see Sect. \ref{irrsc}). 
 
\subsection{Viscous heating and angular momentum transport} 
 
The disc instability model uses the ``$\alpha$"-disc framework of 
Shakura \& Sunyaev (1973). The original $\alpha$ prescription for 
turbulent viscosity was based on a purely dimensional analysis. The 
origin of this viscosity was not specified. Nowadays it is widely 
believed that turbulence in accretion discs is due to magneto-rotational 
Balbus-Hawley instability (see Balbus \& Hawley 1998 for a review and 
Hawley 2000 for 3D global simulations). Recently, Balbus \& Papaloizou 
(1999) showed that $\alpha$-models can be obtained from the mean-flow 
dynamics of MHD turbulence. In particular, the turbulent viscous heating 
per unit surface can be derived to be: 
\begin{equation} 
Q^+= - \Sigma \tau_{r\varphi}\frac{d\Omega}{d \ln r} 
\label{vish} 
\end{equation} 
which, for Keplerian disc, is equivalent to Eq. (\ref{qplus}).

Assuming $\tau_{r\varphi}= \alpha P$, where P is the (usually total:  
gas plus radiation) 
pressure and $\alpha (\leq 1)$ is a parameter (the ``viscosity 
parameter"), this can be also written as: 
\begin{equation} 
Q^+= {3 \over 2} \alpha \Omega_{\rm K} P 
\label{qplus2} 
\end{equation} 
which is (for Keplerian discs) equivalent to the kinematic viscosity 
coefficient written as 
\begin{equation} 
\nu=\frac{2}{3} \alpha \frac{c_s^2}{\Omega_K}. 
\label{alfn} 
\end{equation} 
 
Although it is reassuring that the $\alpha$ prescription corresponds to 
at least one mechanism which produces turbulent viscosity one should be 
aware that in actual simulations ``$\alpha$" exhibits large (vertical) 
gradients and time fluctuations (Hawley 2000; Hawley \& Krolik 2000) so 
it is not clear that it can be used in the way it is in the disc 
instability model. However, until now simulations have not produced a 
{\sl geometrically thin disc} (see, however, Armitage, Reynolds \& 
Chiang 2000). Since it is just such a disc that is supposed to be an 
$\alpha$-disc, the problem of the physical meaning of the 
$\alpha$-prescription is still open. 
 
The solution of this problem is important for the DIM because this model 
is intimately related to the main tenet of the $\alpha$-prescription, that 
both heating and angular momentum transport are due to the same, local 
mechanism. I will come back to this in Sect. \ref{quidn}.

\section{Thermal equilibria: the \bS-curve} 
\label{scurve} 
 
\subsection{The ``standard" case} 
\label{stansc} 
 
Let us first consider thermal equilibria of an accretion disc in which 
heating is due only to local turbulence, leaving the discussion of the 
effects of irradiation and tidal dissipation and/or stream impact to 
Sections \ref{irrsc} and \ref{tssc}. We put therefore $T_{\rm 
irr}=Q_i=0$. 
 
The thermal equilibrium in the disc is defined by the equation $Q^-=Q^+$ 
(see Eq. \ref{eq:heat}), i.e. by 
\begin{equation} 
\sigma T_{\rm eff}^4=\frac{9}{8} \nu \Sigma \Omega_{\rm K}^2. 
\label{termeq} 
\end{equation} 
 
\begin{figure} 
\resizebox{\hsize}{!}{\includegraphics{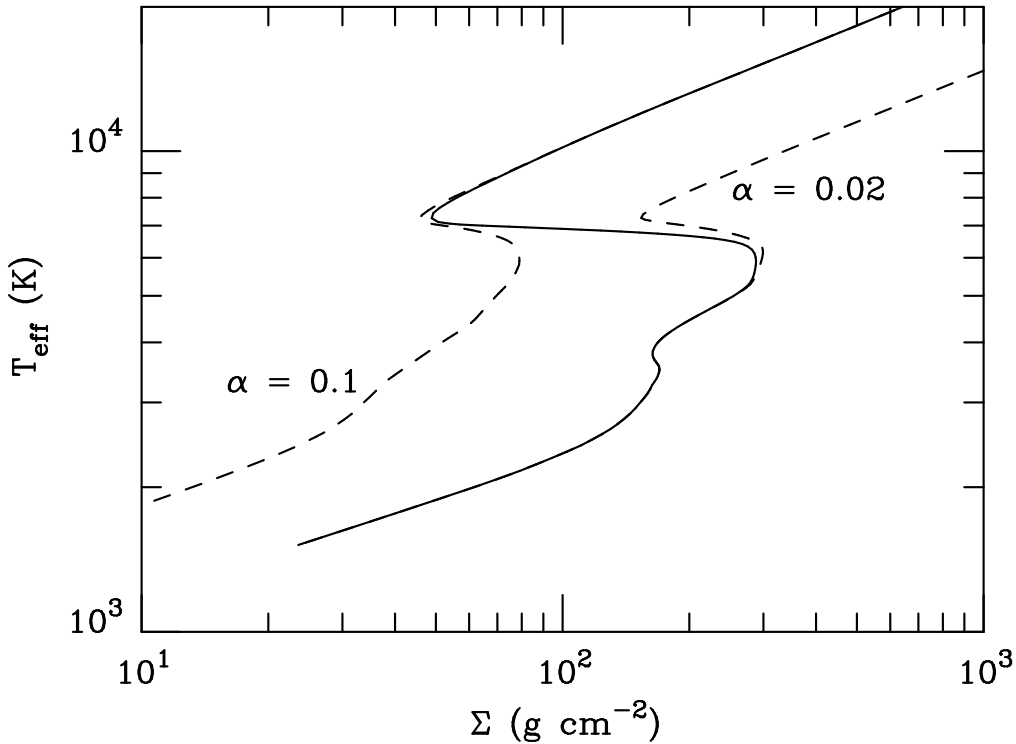}} 
\caption{The ``effective" $\Sigma - T_{\rm eff}$ curves when $\alpha$ is 
assumed to have different (constant) values on the upper and lower 
branches. The dashed curves correspond to constant $\alpha$, $\alpha = 
\alpha_{\rm hot}$ (left) and $\alpha = \alpha_{\rm cold}$ (right). The 
solid line is obtained for $\alpha$ given by Eq. (\ref{eq:alpha}). The 
radius of interest is 10$^{10}$ cm, and the primary mass 1.2 M$_\odot$} 
\label{fig:2alsc} 
\end{figure}

\begin{figure}  
\resizebox{\hsize}{!}{\includegraphics{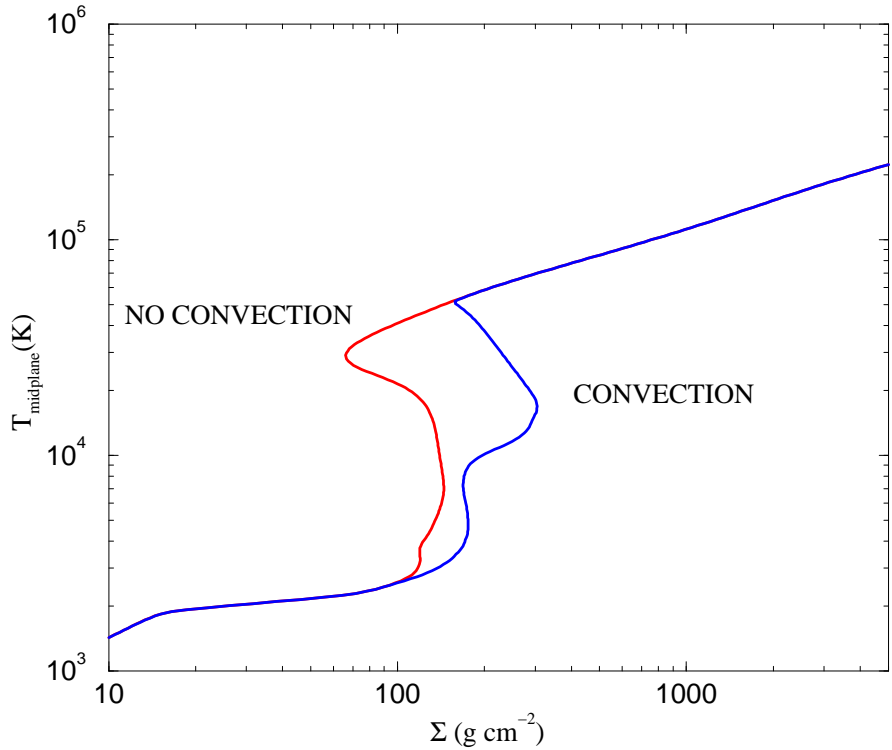}} 
\caption{The $\Sigma - T_{\rm c}$ curves calculated with and without 
convective energy transport, with $\alpha_{\rm cold}=0.035$, 
$M_1=0.6$M$_\odot$. The radius is $10^{10}$ cm}.  
\label{fig:cnc} 
\end{figure}

In general, $\nu$ is a function of density and temperature. The energy 
transfer equation (e.g. Eq. \ref{radtr}) provides a relation between the 
effective and the disc midplane temperatures so that thermal equilibria 
can be represented as a $T_{\rm eff}\left(\Sigma\right)$~ --~relation. In 
the case of a dwarf-nova disc this relation forms an {\bS} on the 
($\Sigma, T_{\rm eff}$) plane as in Figs. (\ref{fig:2alsc}) and 
(\ref{fig:cnc}). 

Each point on the ($\Sigma, T_{\rm eff}$) \bS-curve represents an 
accretion disc's thermal equilibrium at a given radius. The middle 
branch of the \bS-curve corresponds to thermally unstable equilibria, so 
that if anywhere in the disc the effective temperature corresponds to 
this middle branch the disc cannot be in a stable equilibrium. A stable 
disc equilibrium can be represented only by a point on the lower cold or 
the upper hot branch of the \bS-curve. This means that the surface 
density in the cold state must be lower than the maximal value on the 
cold branch: $\Sigma_{\rm max}$, whereas the surface density in the hot 
state must be larger than the minimum value on this branch: $\Sigma_{\rm 
min}$. Both these critical densities are functions of the viscosity 
parameter, the mass of the accreting object and the distance from the 
center. Numerical fits to the critical densities are given in the 
Appendix and plotted as a function of radius in Figs. \ref{smax}, ref{sigr}, 
\ref{tistr}, \ref{profhc} and 
\ref{radnoill}. The outburst properties depend strongly on the fact that 
both critical densities increase with distance. To the critical 
densities correspond critical values of the effective and midplane 
temperatures. Their values are almost independent of the parameters of 
the system because they represent critical temperatures at which the 
disc is, e.g., fully ionized or neutral, i.e. they are universal. When the 
\bS-curve is assumed to represent both thermal {\sl and} viscous 
equilibria one can also relate $\Sigma_{\rm max}$ and $\Sigma_{\rm min}$ 
to critical values of the accretion rate (see below). 
 
The disc is thermally stable if radiative cooling varies faster with  
temperature than viscous heating (see e.g. Frank et al. 1992). In  
other words it is stable if 
\begin{equation}  
{d\ln \sigma T_{\rm eff}^4 \over d\ln T_{\rm c}} > {d\ln F_{\rm vis} \over  
		d\ln T_{\rm c}}  
\label{stabil}  
\end{equation}  
Using Eq. (\ref{diff3}) this can be transformed into  
\begin{equation}  
{d\ln T_{\rm eff}^4 \over d\ln T_{\rm c}} =  
	{d\ln \kappa_{\rm R} \over d\ln T_{\rm c}}  
\label{cool}  
\end{equation}  
 
In a gas pressure dominated disc $F_{\rm vis} \propto T_{\rm c}$ (see e.g. 
Frank et al. 1992). The thermal instability (the upper bend in the 
\bS-curve) is due to a rapid change of opacities with temperature when 
hydrogen begins to recombine. At high temperatures ${d\ln \kappa/d\ln 
T_{\rm c}}\approx - 4$. At temperatures close to 7000 K, the temperature 
exponent becomes large and positive ${d\ln \kappa_{\rm R}/ d\ln T_{\rm c}} 
\approx 7 - 10$, so in the end cooling is decreasing with temperature. 
The disc is then unstable and its equilibria correspond to the middle 
branch of the \bS-curve. On the lower branch the disc is cold and 
and stable. 
 
The position of the critical points of the \bS-curve depends on the 
cooling processes which are taken into account in the model. Fig. 
\ref{fig:cnc} shows two \bS-curves calculated with and without 
convection. One can see that including convection increases the value of 
the critical midplane temperature, thus giving a larger amplitude of the 
temperature jump which occurs when a thermal instability appears in the 
disc. 
 
This is, however, not very important in practice because the actual 
\bS-curve used in the DIM is a combination of two \bS-curves with two 
different $\alpha$'s. It was realized quite early in the development of 
the model that in order to reproduce observed amplitudes and durations 
of the various phases of the dwarf-nova outburst cycle the parameter 
$\alpha$ must have different values in outburst and in quiescence (Smak 
1984b). In other words, $\alpha$ must be different on the upper and lower 
branches of the \bS-curve. It is then required that $\Sigma_{\rm 
min}=\Sigma_{\rm min}\left(\alpha_{\rm hot}\right)$ and $\Sigma_{\rm 
max}=\Sigma_{\rm max}\left(\alpha_{\rm cold}\right)$, where $\alpha_{\rm 
hot}$ and $\alpha_{\rm cold}$ are the values of the viscosity parameter 
at the upper and lower branches respectively. The transition between the 
two values must be sharp, which is assured by making $\alpha$ a function of 
temperature (Ichikawa \& Osaki 1992, use a formula which gives a not very 
sharp transition). HMDLH, for example, use the formula 
\begin{eqnarray} 
\log (\alpha)=\log(\alpha_{\rm cold})& +& \left[ \log(\alpha_{\rm hot})- 
\log( \alpha_{\rm cold} ) \right]~\nonumber~\\~&&~\times \left[1+  
\left( \frac{2.5 \times 10^4 \; \rm K}{T_{\rm c}} \right)^8 
\right]^{-1} . 
\label{eq:alpha} 
\end{eqnarray} 
A different prescription must be used in the case of an irradiated disc 
because the local state of the disc is determined by a non-local 
quantity: the irradiation from the disc's central regions (Hameury, 
Lasota \& Dubus 1999; DHL). 
 
In Figs. \ref{fig:2alsc}, \ref{fig:cnc} and \ref{fig:std} one can see a 
second lower critical value of $\Sigma$ which results from  
convection and changes in molecular opacity (see e.g. Cannizzo \& 
Wheeler 1984). This lower bend on the \bS-curve appears at relatively 
low temperatures, i.e. at large radii and for low $\alpha$'s. This lower 
bend plays an important role in the outburst cycle because it is where 
the system leaves the cold branch of the \bS-curve during the rise to 
outburst. This should be kept in mind when one estimates characteristic 
parameters of an outburst such as recurrence times. In the models, 
oscillations between the lower and the higher parts of the cold branch 
are sometimes obtained. It is not clear if they correspond to anything 
observed in real systems.  
 
There have been attempts to make $\alpha$ a function of the temperature not 
only in the narrow region between the cold and hot branches but also 
along the \bS-curve. In such models the viscosity parameter is also a 
function of the radius. Such an {\sl ansatz} allows interesting 
experiments with the DIM but all physical arguments presented in its 
favour are unfounded or just based on inconsistent numerical schemes (we 
will mercifully pass over in silence the relevant references). This does 
not mean that in `reality' $\alpha$ is not a function of radius and time 
(see e.g. Hawley \& Krolik 2000). It means only that there is nothing 
in the model itself that would require such a dependence. In this review 
I will therefore consider only models in which $\alpha_{\rm hot}$ and 
$\alpha_{\rm cold}$ are both constant. The only exception will be the 
form of $\alpha$ assumed by Ichikawa \& Osaki (1992) in order to 
suppress so called `inside-out' outbursts. 
 
The \bS-curve represents {\sl thermal} but not necessarily {\sl viscous} 
equilibria. To illustrate this let us consider, for example, a hot 
optically thick disc. From Eqs. (\ref{eq:strb})(\ref{eq:strc}) one can 
then obtain the following relation: (e.g. DLHC): 
\begin{equation} 
T^4_c  \equiv T^4(\tau_{\rm tot}) =  {3 \over 8} \tau_{\rm tot} T_{\rm eff}^4  
\label{diff2} 
\end{equation} 
where $T_c$ is the midplane temperature and $\tau_{\rm tot} = 
\int_0^{+\infty} \kappa_{\rm R} \rho dz$ is the total optical depth, 
with $\kappa_{\rm R}$ being the Rosseland mean opacity. Writing the mean 
opacity in Kramers form $\kappa_{\rm R}=\kappa_0 \rho T^{-3.5}$, one can 
easily find that $T_{\rm eff}\sim \Sigma^{5/14}$ which corresponds to 
the slope of the upper, hot branch of the \bS-curve (see e.g. Frank et 
al. 1992). We have nowhere assumed that the disc is in a viscous 
equilibrium, i.e. we did not assume that $\dot M=const$ in the disc. If 
one makes this assumption one obtains the well known formula 
\begin{equation} 
\sigma T_{\rm eff}^4=\frac{3 GM\dot M}{8\pi r^3}f, 
\label{D} 
\end{equation} 
which relates the effective temperature to the accretion rate ($f$ is an 
inner boundary-condition factor, see e.g. Frank et al. 1992). Equation 
(\ref{D}) results from $\dot M= 3\pi \nu \Sigma f^{-1}$, which in turn is 
obtained from the angular-momentum conservation equation when one 
assumes that the disc is stationary. Assuming only thermal equilibrium 
gives instead, the relation 
\begin{equation}  
\dot M= 3\pi \nu \Sigma \left[ 2 {\partial \ln \nu \Sigma  
\over \partial \ln r} + 1 \right]. 
\label{noneq} 
\end{equation} 
and Equation  (\ref{D}) has to be replaced by 
\begin{equation} 
\sigma T_{\rm eff}^4=\frac{3}{8\pi}\left[ 2 {\partial \ln \nu \Sigma  
\over \partial \ln r} + 1 \right]^{-1}\frac{GM\dot M}{r^3}. 
\label{neqqpl} 
\end{equation}

During a dwarf nova outburst a point representing a local (at a given 
radius) accretion disc's state moves in  the $\Sigma - T_{\rm eff}$ 
plane as shown on Fig. (\ref{fig:std}). A point out of the \bS-curve is 
out of thermal equilibrium. In the region to the right of the \bS-curve 
heating dominates cooling, so that the temperature increases and the 
system-point moves up towards the hot branch. On the left the is the 
case opposite and the point moves down towards the cool branch. These 
upward and downward motions take place in thermal time since they 
correspond to the heating and cooling of a disc's ring (in Fig. 
(\ref{fig:std}) one can see non-local effects, which will be discussed 
later). During decay from outburst and during the quiescent phase of the 
outburst cycle, the system-point moves along, respectively, the upper 
and lower branches. This evolution happens in viscous time, which in a 
geometrically thin disc is much longer that thermal time. During this 
viscous evolution the disc is in thermal equilibrium but, obviously, not 
in viscous equilibrium. In such a case Eq. (\ref{neqqpl}) (Idan et al. 1999).  
 
\subsection{Effects of irradiation} 
\label{irrsc} 
 
When a disc is irradiated the same equations are solved as in the 
preceding section but now with the full boundary condition 
\begin{equation} 
T^4(\tau_{\rm s}) = T^4_{\rm eff} + T^4_{\rm irr} 
\label{bcondv} 
\end{equation} 
where $\sigma T^4_{\rm eff}=Q^+$. 
Irradiation therefore changes the relation between the midplane and the 
surface temperature. In the case of large optical depths this relation 
can be written as (DLHC; Lasota 1999) 
\begin{equation}  
T^4_{\rm c} \equiv T^4(\tau_{\rm tot}) =  
		 {3 \over 8} \tau_{\rm tot} T_{\rm eff}^4 + T^4_{\rm irr}  
\label{diff3}  
\end{equation} 
where $\tau_{\rm tot}$ is the total optical depth of the disc. 
 
We will now see how irradiation modifies the \bS-curve. For the moment  
the irradiation temperature will be treated as a free parameter. In 
applications this temperature is often related to the accretion rate at 
the inner disc and depends on distance and other parameters. This 
implies that thermal equilibria depend on non-local quantities, and 
makes the concept of a \bS-curve less simple and useful. The \bS-curve, 
however, helps to understand how irradiation modifies the 
disc's stability properties. 
\begin{figure} 
\resizebox{\hsize}{!}{\includegraphics{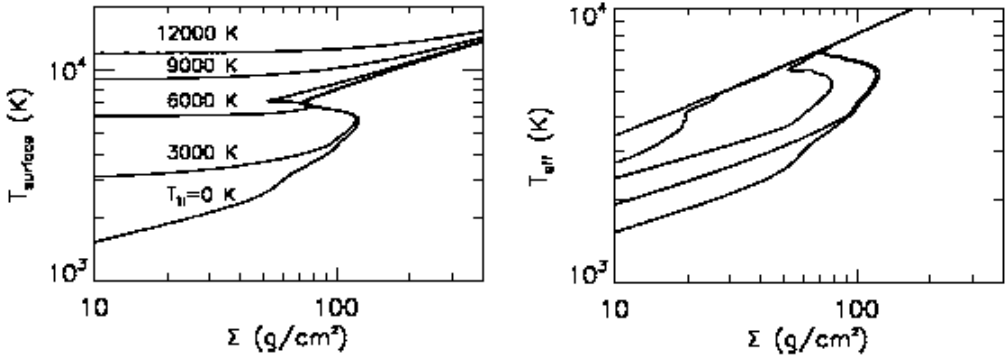}}
\caption{The $\Sigma$ -- $T_{\rm surface}$, $\Sigma$ --  
$T_{\rm eff}$ ``\bS-curves"  for $r =3\cdot 10^{10}$ cm,  
$M=10M_{\odot}$, $\alpha$ = 0.1, and $T_{\rm irr}=  
[0,3,6,9,12]\times 10^3$ K. (From DLHC)} 
\label{sir} 
\end{figure} 
 
The effective and midplane temperatures at which a non-irradiated hot 
disc becomes thermally unstable are given by Eqs. (\ref{te}), (\ref{tc}) 
leads to $\tau_{\rm tot} \sim 10^2$ (see Eq. \ref{diff3}) so the disc is 
then optically thick. 
\begin{figure} 
\centering 
{\includegraphics[scale=1,totalheight=7cm]{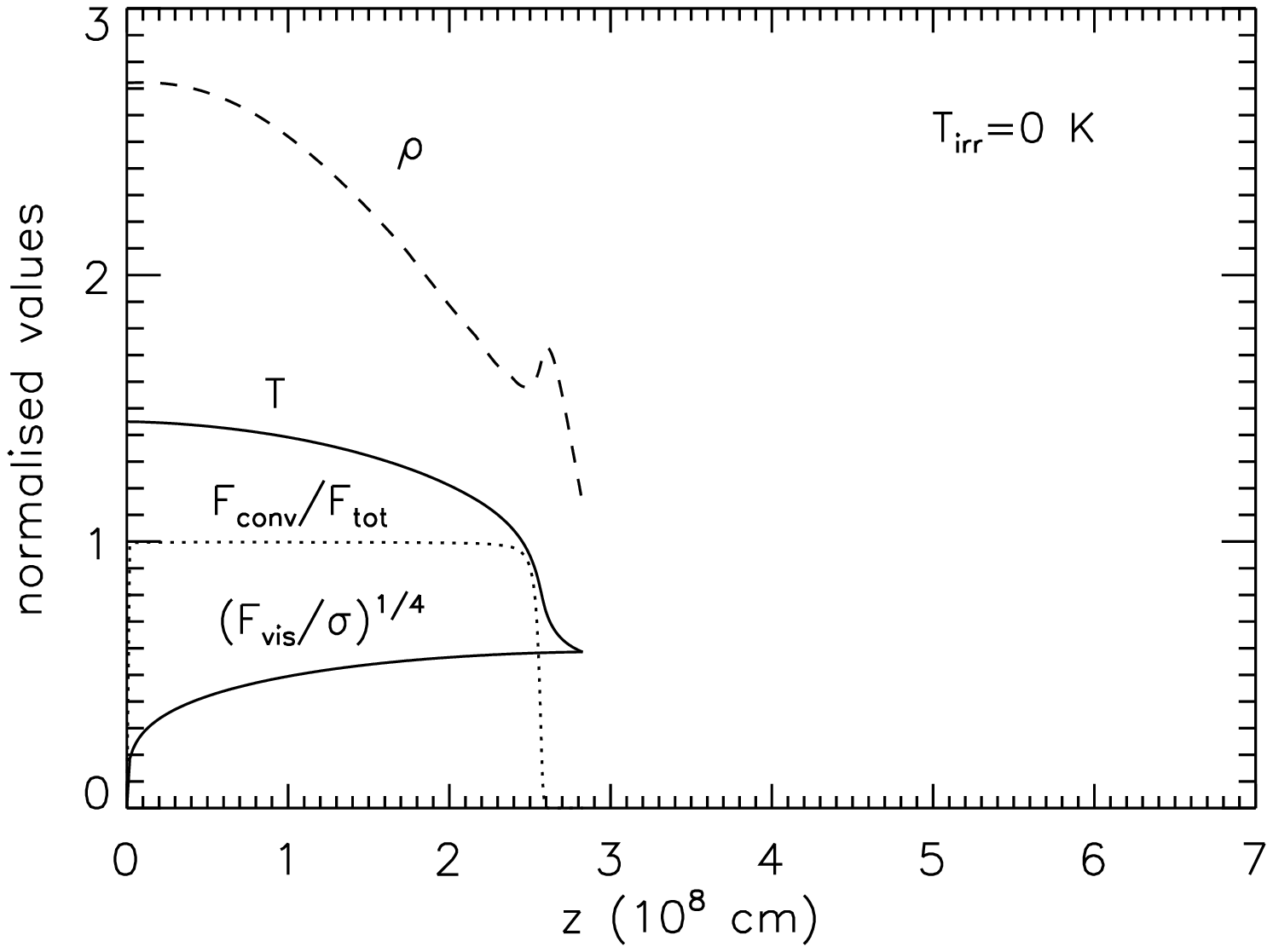}} 
{\includegraphics[scale=1,totalheight=7cm]{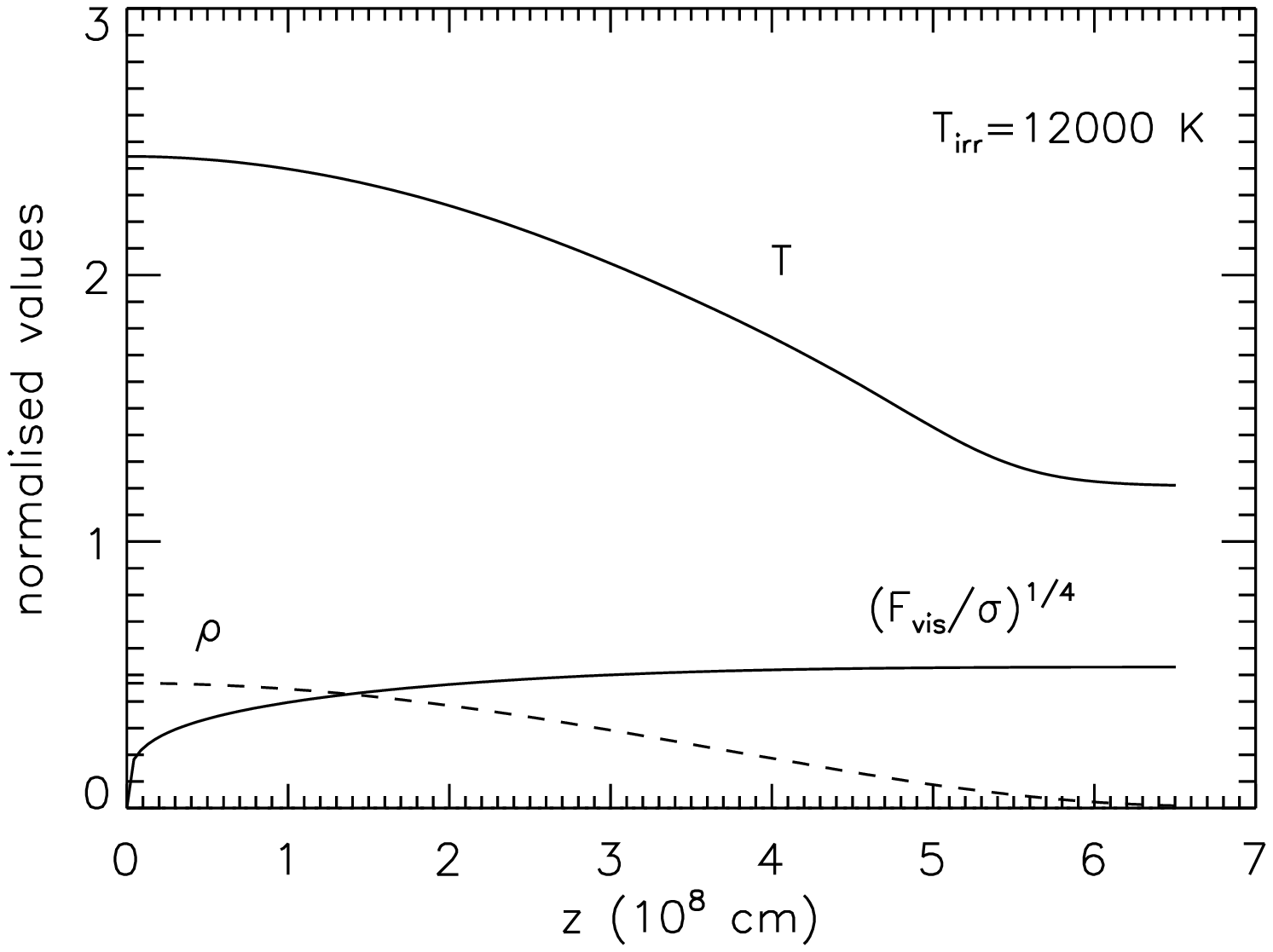}} 
\caption{Upper panel: vertical structure of an accretion disc around a 
$M=10 M_{\odot}$ compact object at $r =3\times 10^{10}$ cm. 
$\alpha\approx 0.1$, $\dot M \approx 10^{16}$ g s$^{-1}$ (i.e. $T_{\rm 
eff}\approx 5700$ K) and $T_{\rm irr}=0$. Both the disc temperature $T$ 
and the temperature corresponding to the viscous flux $(F_{\rm 
vis}/\sigma)^{1/4}$ are plotted in units of $10^4$ K. At the 
photosphere, the latter gives the effective temperature $T_{\rm eff}$. 
Since $T_{\rm irr}=0$, the surface temperature $T(\tau_{\rm s})= T_{\rm 
eff}$. The dashed line is the density in units of $10^{-7}$ g cm$^{-3}$ 
and the dotted line is the ratio of the convective to the total fluxes 
(between 0 and 1). For these parameters, the section of the disc lays on 
the lower stable branch. Lower panel: the same but for $T_{\rm 
irr}=12000$ K. Here, the convective flux is negligible with $F_{\rm 
conv}/F_{\rm tot} \approx 0$. The disc height has increased along with 
the irradiation temperature (DLHC).}  
\label{vert}  
\end{figure} 
 
Eq. (\ref{cool}) changes now to 
\begin{equation}  
{d\ln T_{\rm eff}^4 \over d\ln T_{\rm c}} =  
4\left[1 - \left({T_{\rm irr} \over  
	T_{\rm c}}\right)^{4}\right]^{-1} -  
	{d\ln \kappa_{\rm R} \over d\ln T_{\rm c}}  
\label{ircool}  
\end{equation}  
where Eq. (\ref{diff3}) was used. 
As can be seen from Eq. (\ref{ircool}), even a 
moderate $T_{\rm irr}/ T_{\rm c}$ ratio modifies the transition from 
stable to unstable configurations, thereby pushing it to lower 
temperatures. This effect is clearly seen in Fig. \ref{sir}. Irradiation 
has a second effect on the disc's stability: it suppresses or alters 
convection, as can be seen in Fig. \ref{vert}.

The \bS-shape is suppressed when the surface temperature is higher than 
the ionization temperature. Then, since the temperature must increase 
towards the midplane, the disc will always be ionized and no thermal 
instability is possible. 
 
Overall, irradiation lowers $\Sigma_{\rm min}$ as pointed out by van 
Paradijs (1996) and also lowers the critical mass-transfer rate at 
which hot discs become unstable. 
\begin{figure} 
\resizebox{\hsize}{!}{\includegraphics[angle=-90]{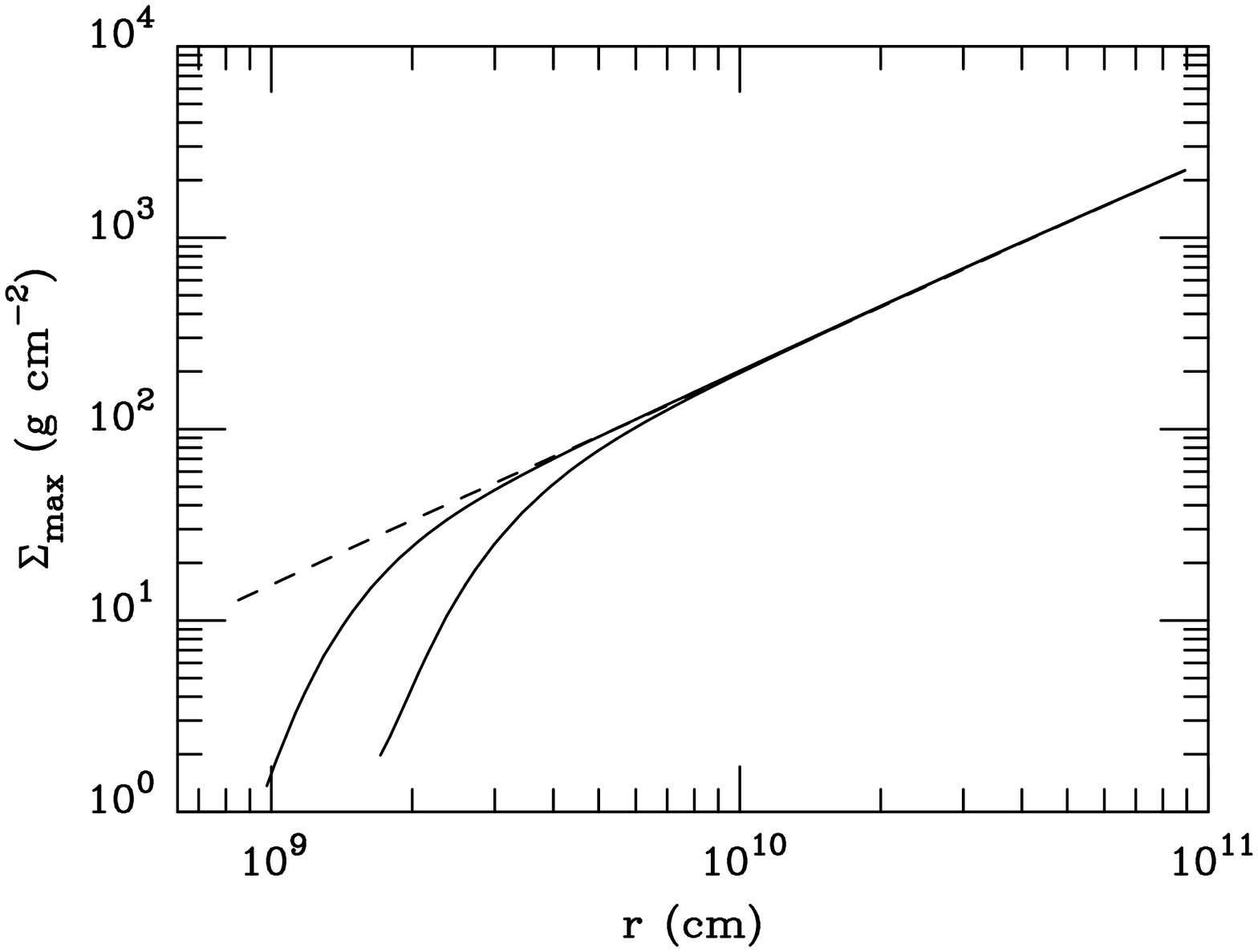}} 
\caption{Maximum value of $\Sigma$ on the lower stable branch of the $\Sigma 
- T_{\rm eff}$ curve. The primary is a 0.6 M$_\odot$, $8.5 \times 10^8$ cm 
white dwarf, with effective temperature 25,000 K (lower solid curve), or 
15,000 K (upper solid curve). The unilluminated case (dashed curve) is also 
shown for comparison (Hameury et al. 1999).} 
\label{smax} 
\end{figure} 
This effect is important in low mass X-ray binaries where the outer disc 
is subject to strong X-ray irradiation (van Paradijs \& McClintock 1995) 
by a central source. In cataclysmic variables this effect is not 
important but as pointed out by Smak (1989) and more recently by King 
(1997) a hot white dwarf may affect the inner disc structure. This will be 
discussed in the Section \ref{global}. 
 
\subsubsection{Irradiation by a hot white dwarf} 
 
The irradiation of the inner disc by a hot white dwarf and its influence 
on the dwarf-nova outburst cycle was discussed in detail by Hameury 
et al. (1999). 
\begin{figure} 
\resizebox{\hsize}{!}{\includegraphics{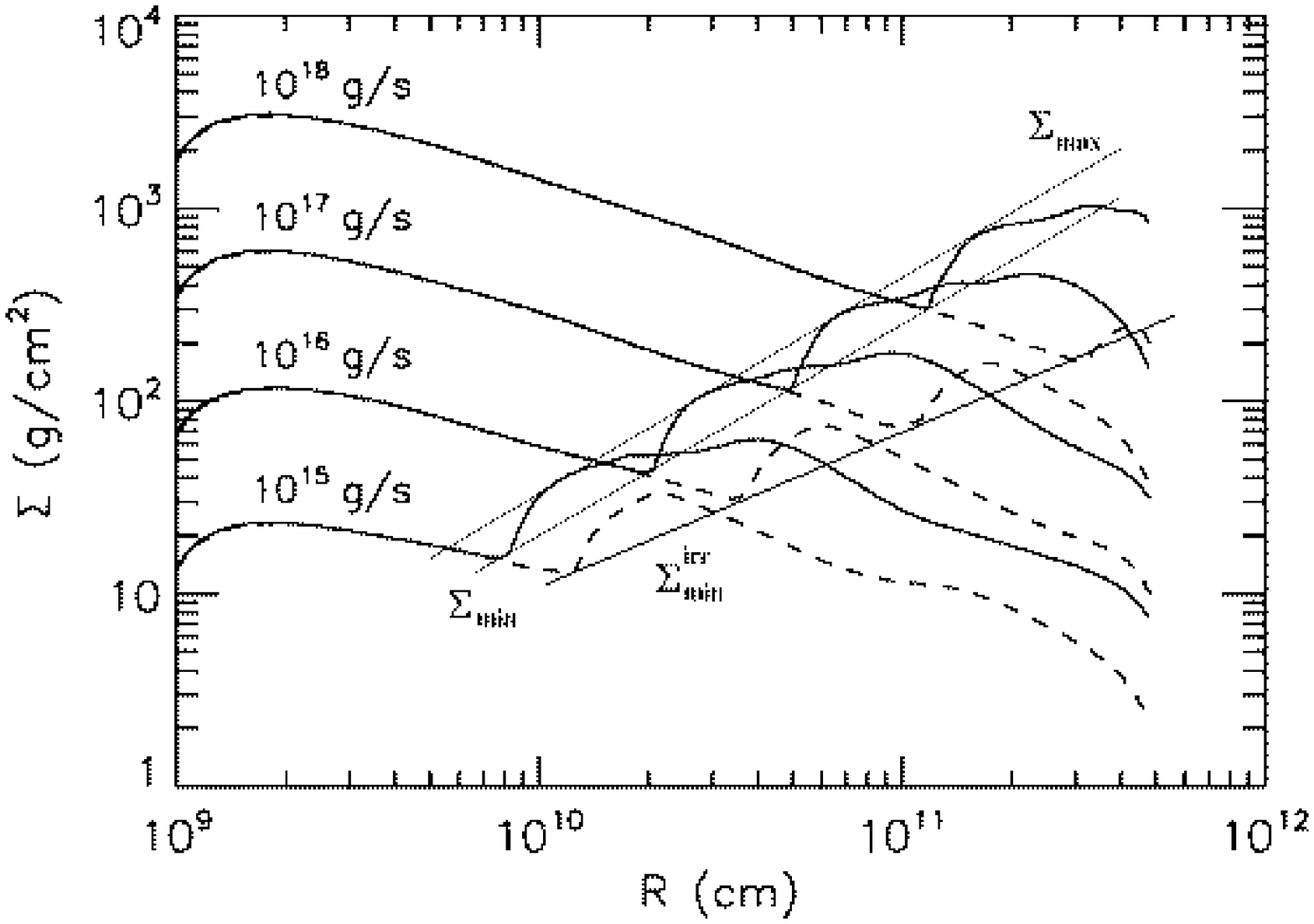}} 
\caption{Stationary accretion disc surface density profiles for 4 values of  
accretion rate (as labelled) and $M=10 M_{\odot}$, $\alpha=0.1$. The continuous  
line corresponds to the un-irradiated disc, the dotted lines to the  
irradiated configuration. (Following DLHC).} 
\label{sigr} 
\end{figure} 
They considered the irradiation temperature given by (e.g. Smak 1989) 
\begin{equation} 
T_{\rm irr}^4 = (1-\beta) T_*^4 {1 \over \pi} [\arcsin \rho -\rho 
(1-\rho^2)^{1/2} ] 
\label{eq:till} 
\end{equation} 
where $\rho = R_*/r$, $R_*$ and $T_*$ are the white dwarf radius and 
temperature. $1 - \beta$ is the fraction of the incident flux which is 
absorbed in optically thick regions, thermalized and reemitted as 
photospheric radiation. 
 
In this case irradiation has two effects. First, it stabilizes the disc 
near the white dwarf, modifying the \bS-curve in a way similar to that 
shown in Fig. \ref{sir}. Second, as shown in Figure \ref{smax}, 
irradiation has a strong {\sl destabilizing} effect farther away from 
the white dwarf. Figure \ref{smax} shows $\Sigma_{\rm max}$, as a 
function of radius (see Hameury et al. 1999 for details). One sees that 
$\Sigma_{\rm max}$ can be reduced by more than one order of magnitude as 
compared with the un-irradiated case. By lowering the critical density 
irradiation makes this part of the disc more subject to instability. The 
reason is very simple: if the inner part of the disc is hot and stable 
but the outer part is cold, somewhere in between, there must be  an 
unstable region. This, of course, affects the outburst cycle. 
 
King (1997) suggested that irradiation by a hot white dwarf would force 
the inner part of a quiescent dwarf-nova disc to remain hot and 
optically thin. Hameury et al. (1999) shows that this is not what the 
DIM predicts, at least in the case of U Gem-type dwarf novae. They argue 
that it seems more likely that the inner part of the disc is evaporated 
(Meyer \& Meyer-Hofmeister 1994) and the dwarf nova disc is truncated. 
The role of the white dwarf irradiation in this process is not clear. 
The observation of the low state of TT Ari by G\"ansicke et al. (1999), 
just shows that in this system the inner disc is truncated, but contrary 
to authors assertion is not a confirmation of King's scenario. In any 
case the structure of low state accretion disc in TT Ari, {\sl not} a 
dwarf nova, should be different from a dwarf-nova quiescent disc.

\subsection{A global view of \bS-curves} 
\label{global} 
 
The \bS-curve is a local concept but it can also be used in a global 
way. Figure~\ref{sigr} shows radial surface-density profiles of 
stationary discs around a 10 M$_{\odot}$ accreting body (a `black- 
hole'). Profiles for 4 values of the accretion rate are shown. Each 
$\Sigma(r)$ curve has a (local) minimum corresponding to $\Sigma_{\rm 
min}$ and a maximum corresponding to $\Sigma_{\rm max}$. The figure, also 
shows the effects of irradiation by a point source located at the disc's 
inner regions (see DLHC for details). As mentioned above, irradiation 
lowers the value of $\Sigma_{\rm min}$. These new, lower values are 
marked by an asterisk. For a given accretion rate if $\Sigma > 
\Sigma_{\rm min}$ the disc is on the hot stable branch. On a local 
\bS-curve such inequality would only mean that a particular ring is hot 
and stable; here it means that the whole disc is hot and stable. One 
can see that in the unstable disc regions the surface density increases 
with radius. In the cold stable parts things are more complicated, as 
could be expected from the presence of two critical `$\Sigma_{\rm max}$' 
in the \bS-curve. 
\begin{figure} 
\centering 
\resizebox{12truecm}{!}{\includegraphics[angle=-90]{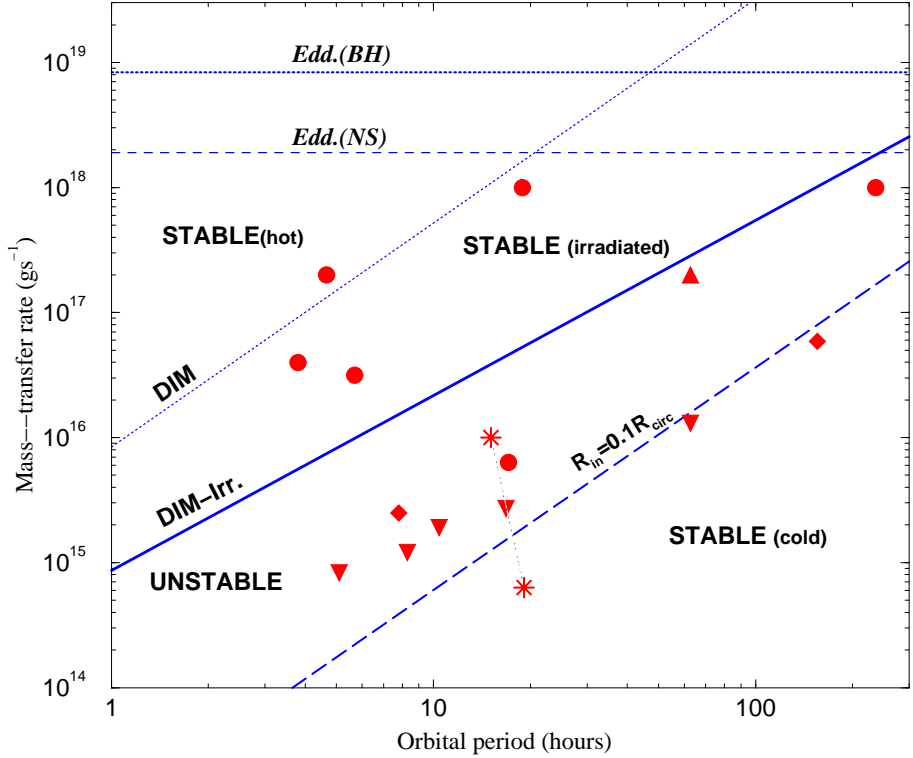}} 
\caption{Stability limits and parameters of Low Mass X-ray Binaries 
(following van Paradijs 1996; Menou et al. 1999b; Lasota 2000b). Filled 
circles represent steady (i.e. non-transient) LMXBs containing neutron 
stars. The two asterisks correspond to two neutron-star LMXBTs. Diamonds 
represent black-hole LMXBTs with known recurrence times and 
down-pointing triangles those where only the lower limits for the 
recurrence time are known. The up-pointing triangle corresponds to 
GRO~J1655-40 with the recurrence time between the 1994 and 1996 
outbursts (see text). The two horizontal lines marked `{\sl Edd.}' 
correspond to the Eddington limit for, respectively, a $6 {\rm 
M_{\odot}}$ (`{\sl BH}') and 1.4 ${\rm M_{\odot}}$ (`{\sl NS}') 
accreting body . The dotted line marked `DIM' represents the stability 
limit for non-irradiated (Eqs. \ref{mdoth}), the solid line `DIM-irr.' 
the corresponding limit for irradiated discs (Eq. \ref{mdotir} with 
${\mathcal C} =5\times 10^{-3}$); all systems above this line should be 
stable. Discs truncated at $R_{\rm in}=0.1 R_{\rm circ}$ should be cold 
and stable below the dashed line correspondingly marked.} 
\label{stab} 
\end{figure}

From Figure \ref{sigr} one can immediately estimate what is the minimum 
mass-transfer rate for which an accretion disc with a given outer radius 
is hot and stable. For example a disc with $r_{\rm out}\approx 2 \times 
10^{10}$ cm will be stable for mass-transfer rates $\gta 10^{16}$ g 
s$^{-1}$ but a larger non-irradiated disc would be unstable. Irradiation 
extends the stability range of $M_{\rm tr}= 10^{16}$ g s$^{-1}$ up to 
$4\times 10^{10}$ cm. The corresponding stability criterion can be obtained 
from Eq. (\ref{mdmin}) with $r=r_{\rm out}$. Using $r_{\rm out}=0.9 R_L$ 
where \begin{eqnarray} 
R_{\rm L1}&\approx& 0.462 \left(\frac{M_2}{M_2 + M_1}\right)^{1/3} a \nonumber \\ 
a&=&3.53 \times 10^{10}  (M_1 + M_2)^{1/3} P^{2/3}_{hr} {\rm cm}, 
\label{kepler} 
\end{eqnarray} 
($R_{\rm L1}$ is the {\sl mean} Roche-lobe radius), one obtains 
\begin{equation} 
\dot M_{\rm hot}^{\rm n-irr}= 2.67 \times 10^{16} \alpha_{\rm h}^{0.01} 
\left(\frac{M_2}{M_1}\right)^{0.89} P_{\rm hr}^{1.79} 
\label{mdoth} 
\end{equation} 
     for the minimum accretion rate in a hot and stable non-irradiated disc. 
 
For a LMXB disc irradiated by a point source the criterion is 
\begin{equation} 
\dot M_{\rm hot}^{\rm irr}= 3.37 \times 10^{15}  
\left(\frac{M_2}{M_{\odot}}\right)^{0.7}  
\left(\frac{M_1}{M_{\odot}}\right)^{-0.4} 
\left(\frac{\mathcal C}{5 \times 10^{-3}}\right)^{-0.5} 
P_{\rm hr}^{1.4} 
\label{mdotir} 
\end{equation} 
where ${\mathcal C}$ is defined in (Shakura \& Sunyaev 1973): 
\begin{equation} 
\sigma T^4_{\rm irr} = {\mathcal C} \frac{L_{X}}{4 \pi R^2}  
\label{C1} 
\end{equation} 
(the definition of this 
quantity here differs by a factor of 10 from the one used in DLHC). 
The difference between Eq. (\ref{mdotir}) and similar equations 
published in DLHC and Lasota (2000b) is due to different 
approximations used for the outer disc radius and to typing errors. One 
should keep in mind, however, that the main uncertainty is contained in 
the factor ${\mathcal C}$. 
\begin{figure} 
\centering 
\resizebox{12truecm}{!}{\includegraphics{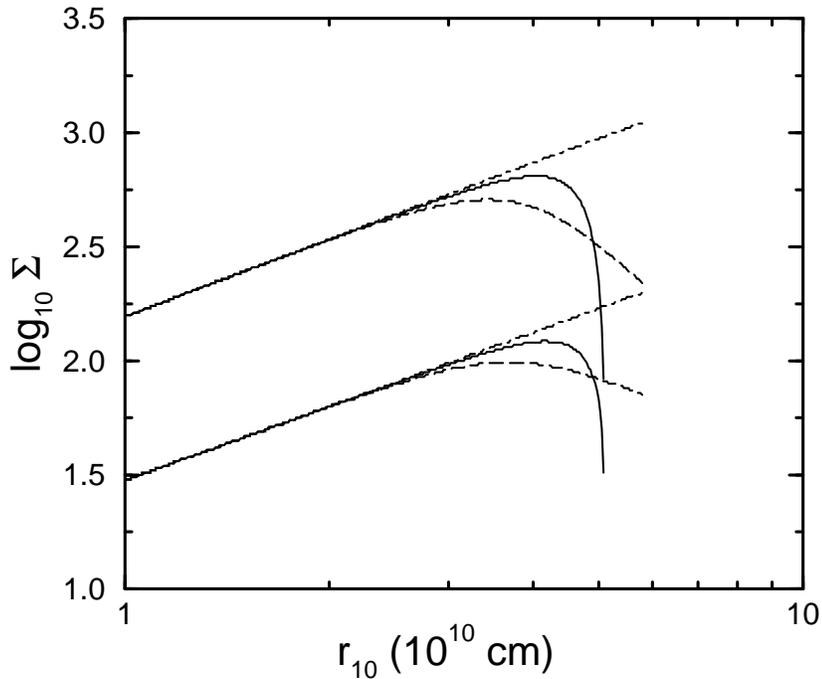}} 
\caption{$\Sigma_{\rm min}$ (lower curves) and $\Sigma_{\rm max}$ (upper 
curves) for a white dwarf mass 1.2 M$_{\odot}$, outer disc radius 
$5.4 \times 10^{10}$ cm, $\alpha_{\rm cold}=0.04$ and $\alpha_{\rm hot}=0.2$.  
The dot-dashed 
lines show the results of the standard DIM, the solid lines have been 
obtained when the stream impact is taken into account, and the dashed 
lines when tidal effects are included. (Buat-M\'enard et al. 2001a.)} 
\label{tistr} 
\end{figure} 
 
In Figure \ref{stab} most of the steady X-ray binaries are above the 
line $\dot M_{\rm tr}= \dot M_{\rm hot}^{\rm irr}$ but would be unstable 
according to the criterion $\dot M_{\rm tr} > \dot M_{\rm hot}^{\rm n- 
irr}$ (van Paradijs 1996). This means, probably, that Eq. (\ref{mdotir}) 
gives a reasonable criterion for stability of irradiated discs in LMXBs. 
The two odd sources which find themselves below the critical line but 
show no signs of being transient are AC 211 ($P_{\rm orb}=17.1$ hr)and 
Cyg X-2 ($P_{\rm orb}=236$ hr). It seems that the mass-transfer rates 
used for them in the figure are underestimated. Cyg X-2 could still be 
in a slightly super-Eddington regime (see King \& Ritter 1999). AC 211 
should rather have a mass-transfer rate $\sim 3\times 10^{17} $ g 
s$^{-1}$ (Podsiadlowski, private communication) and according to Homer 
\& Charles (1998) it could even be accreting at a very super-Eddington 
rate, both variants thus solving the apparent discrepancy between the 
properties of these two sources and their position in Fig. \ref{stab}. 
 
Conditions $\dot M_{\rm tr} > \dot M_{\rm hot}$ are only necessary for 
stability. A disc could be stable for such mass-transfer rates if it 
was in a cold state everywhere, i.e. if $\dot M_{\rm tr} < \dot 
M_{\rm cold}$ for all $r$. One can guess from Fig. \ref{sigr} that for 
discs extending down to small radii this would imply very low, usually 
uninteresting mass-transfer rates. One can also see, however, that for 
an {\sl inner} disc radius $r_{\rm in} \gta 10^{10}$ cm (i.e. in this 
case for $r_{\rm in} \gta 10^4r_S$) an accretion disc will be cold and 
stable for $M_{\rm tr}\lta 10^{15}$ g s$^{-1}$ . These are values of 
inner disc radii that were suggested for quiescent LMXBTs (e.g. Narayan, 
McClintock \& Yi 1996, Narayan, Barret \& McClintock 1997; Lasota, 
Narayan \& Yi 1996; Hameury et al. 1997; Lasota 2000a; Esin, McClintock 
\& Narayan 1997; Menou, Narayan \& Lasota 1999). Such discs could be 
stable or marginally stable. Figure \ref{stab} shows the line
representing a possible inner disc radius, assuming that it represents a 
fraction 0.1 of the circularization radius $r_{\rm k}$ (see 
Menou et al. 1999b for details). Clearly black-hole transient systems 
are close to the stability limit. Menou et al. (1999b) suggested that 
they could represent the high mass-transfer tip of a larger population 
of faint and stable binary systems. 
 
Similar ideas were proposed earlier for dwarf-nova quiescent discs 
(Lasota, Hameury \& Hur\'e 1995; Hameury, Lasota \& Hur\'e 1997; Meyer 
\& Meyer- Hofmeister 1994; Warner, Livio \& Tout 1996 and more recently 
Menou 2000b). 
 
\subsection{Effects of stream-impact heating and tidal dissipation} 
\label{tssc} 
 
Heating by stream impact and/or by tidal dissipation will of course 
change the properties of thermal equilibria. Figure \ref{tistr} shows how 
these effects change the values of the critical surface-densities. 
Additional heating lowers both $\Sigma_{\rm min}$ and $\Sigma_{\rm 
max}$. It also diminishes the interval between the two. Lower values of 
$\Sigma_{\rm max}$ make it easier to trigger outbursts in the outer disc 
regions. The situation is `symmetrical' in radius with respect to the 
effects of the heating of the inner disc by the hot white dwarf 
(Fig. \ref{smax}). Lower $\Sigma_{\rm min}-\Sigma_{\rm max}$ should 
shorten the duration of quiescence.

\section{Dwarf novae} 
\label{dn} 
\begin{figure} 
\centering 
\resizebox{\hsize}{!}{\includegraphics[angle=-90]{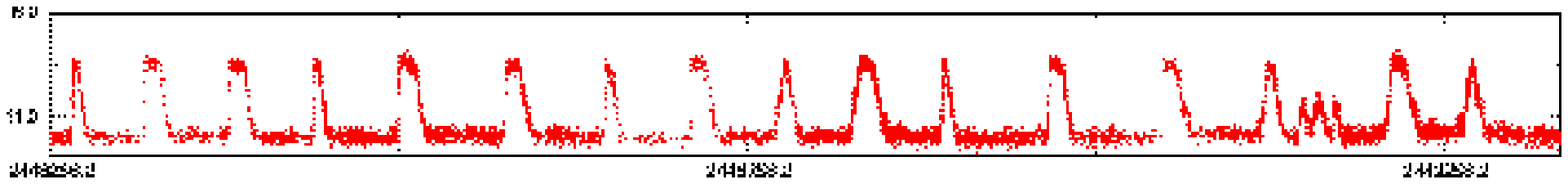}} 
\resizebox{\hsize}{!}{\includegraphics{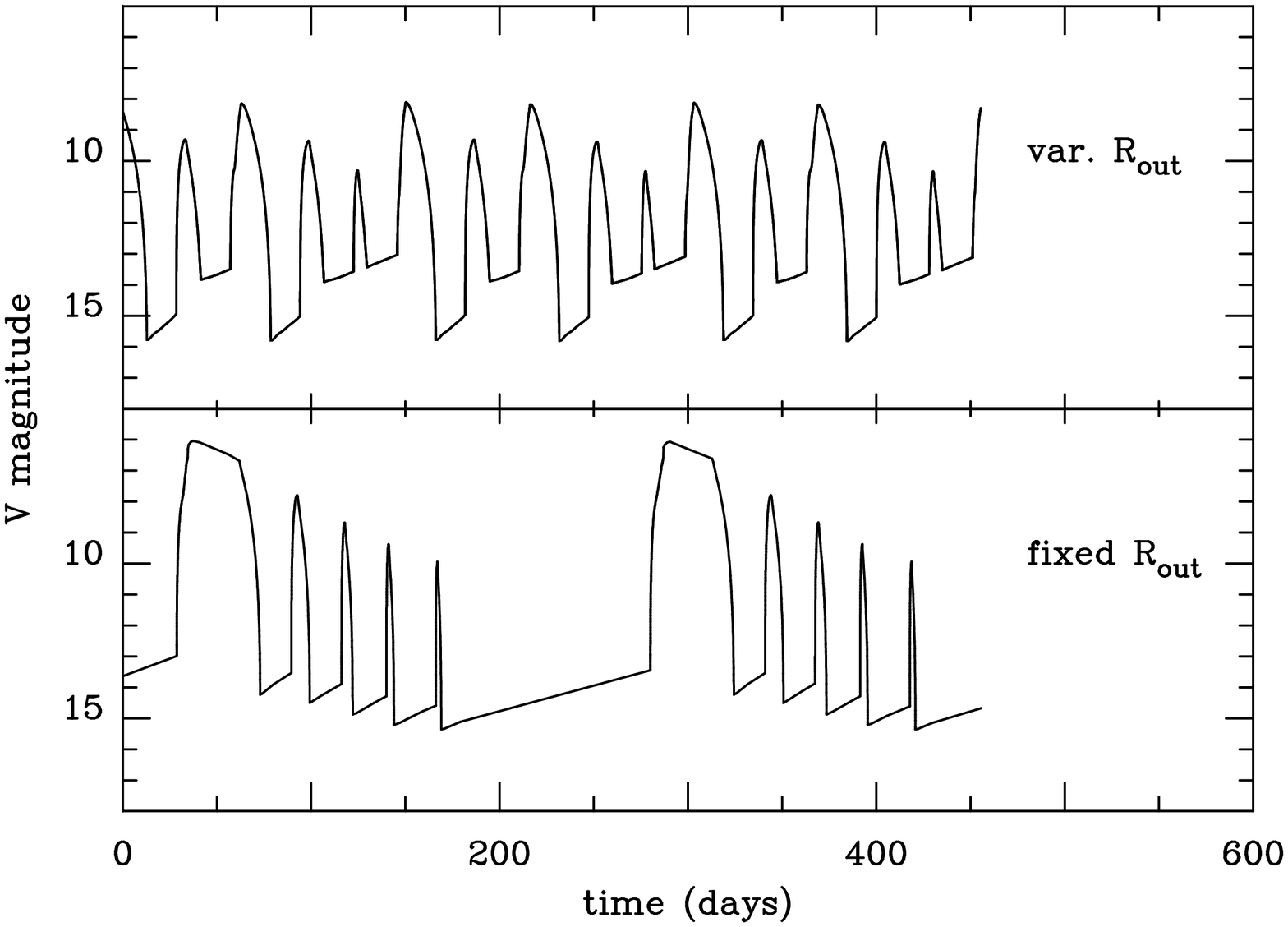}} 
\caption{Uppermost panel: the light-curve of the U Gem-type dwarf-nova SS 
Cyg. The figure shows the daily mean $m_V$ as a function of the Julian 
Day. Data from AAVSO. Lower panels: DIM light-curves calculated for an 
accretion disc around a 1.2M$_{\odot}$ white dwarf with mass-transfer 
$\dot M_{\rm trans}=6.35 \times 10^{16}$ g s$^{-1}$, $\alpha_{\rm 
cold}=0.02$ and $\alpha_{\rm hot}=0.1$. The inner radius $R_{\rm 
in}=5\times 10^8$ cm. In the lower panel the outer disc radius $R_{\rm 
out}$ is fixed at $4\times 10^{10}$ cm; in the upper panel the outer 
disc radius varies around an average value of $4\times 10^{10}$ cm 
(HMDLH).} 
\label{comp_bc} 
\end{figure} 
We will first deal with dwarf nova outbursts. The standard version of 
the DIM was designed to explain and reproduce these phenomena, or at 
least a subclass of them, since dwarf novae show at least three types of 
outburst cycles. According to the type the outburst dwarf novae can be 
of U Gem, SU UMa and Z Cam type. The types {\sl are} named after the 
prototypes, which still belong to the class they defined one of the few 
reassuring things one can say about dwarf novae. The outbursts 
themselves are divided into `normal' outbursts and `superoutbursts'. All 
three dwarf-nova types show normal outbursts and {\sl only} SU UMa stars show 
superoutbursts. There are important exceptions, however. Some SU UMa stars 
show {\sl only} superoutbursts and are sometimes classified as a 
separate class of WZ Sge type dwarf novae. One definition of a 
superoutburst would allow one to say that at least one U Gem-type dwarf 
nova showed a superoutburst: the prototype itself. Z Cams are particular 
in that during decline from outburst they get caught in a `standstill', 
$\sim$ 0.7 mag below maximum, which can last from ten days to years. 
 
Normal outbursts have amplitudes of 2-5 mag and last 2-20 days. Their 
recurrence times are typically from $\sim 10$ days to years. As can be 
seen in Fig. \ref{comp_bc} there are basically two types of normal 
outbursts: narrow and symmetrical and wide and asymmetrical. The bimodality 
of width distribution is a general property of normal outbursts (van 
Paradijs 1983). The uppermost panel of Fig.~\ref{comp_bc} shows that 
both types of outbursts have very similar amplitudes.  
 
Superoutbursts have amplitudes brighter by $\sim$ 0.7 magnitudes, they 
last 5 times longer and their recurrence time is longer than that of 
normal outbursts. When both types, normal and superoutbursts, are present 
in a given system, superoutbursts are separated by a sequence of normal 
outbursts. When normal outbursts are absent, recurrence times are 
unusually long: 33 years in WZ Sge. Van Paradijs (1983) suggested that 
superoutbursts are a universal property of dwarf novae, assigning them 
to the class of wide outbursts. We know now, however, that this 
is not the case because the three types of outbursts -- narrow, wide and 
superoutbursts -- are observed to occur in the same systems (TU Men and U 
Gem, see Smak 2000 and references therein). 
 
SU UMa star superoutbursts are characterized by the appearance of a 
``superhump": a feature in the light-curve which is modulated at a 
period slightly longer than the orbital period. All SU UMa stars have 
periods shorter than 3 hours. Such systems have low secondary to primary 
mass-ratios and, therefore, large Roche-lobes around the primary (white 
dwarf). This suggests that the superhump is tidally induced. A popular 
model attributes the superhump to the 3:1 resonance operating in a 
local--viscosity driven disc (Whitehurst 1988, 1994; see Lubow 1994 for 
a review). The validity of this model (but not the relevance of the 3:1 
resonance) has been recently, and rather convincingly, put in doubt by 
Kornet \& R\'o\.zyczka (2000). This would also cast doubt on the 
so-called tidal-thermal disc instability model (see e.g. Osaki 1996). We 
will discuss these and related problems in Sects. \ref{tt} and 
\ref{quidn}.  
 
\subsection{The ``standard" model} 
\label{standard} 
 
First we will present the ``standard" version of the DIM, i.e. the 
model in which heating is due only to local viscous dissipation and the 
mass-transfer rate is kept constant. In other words, in Eq. 
(\ref{eq:heat}) $Q_i=0$ and $T_{\rm irr}$=0 in the boundary condition 
Eq. (\ref{bcondv}); 
$\rm d\dot{M}_{\rm tr} / {\rm d} t=0$. This version of the model 
was supposed to be able to reproduce outburst properties of U Gem type 
dwarf novae. In Figure \ref{comp_bc} are shown two light-curves 
calculated for the presumed parameters of SS Cyg, the system whose 
observed light-curve is also shown in Fig. \ref{comp_bc}. Let us first 
deal with the lower-panel light-curve of Fig. \ref{comp_bc}. It is shown 
here for `historical' reasons and also as a warning. This light-curve 
does not look like the SS Cyg light-curve but is reminiscent of 
light-curves observed in other systems. The sequence of several narrow 
outbursts followed by a wide, `flat-top' outburst is reminiscent of SU 
UMa-type dwarf-novae (which will be discussed in Sect. \ref{tt}). 
However, the form of this light-curve results {\sl purely from an 
incorrect boundary condition}, which here assumes that the outer disc 
radius is fixed during the outburst cycle. This is incorrect, because the 
position of the disc's outer radius is controlled by the tidal torque 
and in any case this radius is observed to vary (e.g. Paczy\'nski 1965; 
Smak 1971, 1984a; O'Donoghue 1986; Wood et al. 1989b; Wolf et al. 1993; 
Harrop-Allin \& Warner 1996). Since it represents an unphysical system, 
the lower panel of Fig. \ref{comp_bc} should be forgotten, as should  
many if not all  the numerous results which assume the same boundary  
condition. 
 
The upper panel shows a light-curve calculated using the correct 
boundary conditions. Despite this (or maybe because of that) the 
result is not very satisfactory, as noticed by Smak (1999a) who used 
this figure from HMDLH to illustrate the deficiencies of the standard 
DIM. Why it took so long to arrive at this obvious conclusion is, one 
might hope, an interesting problem in the sociology of science. The 
standard model fails here at least on two counts (Smak 1999a, 2000): 
\begin{itemize} 
\item  The calculated light-curve shows a significant increase of the disc's 
luminosity in quiescence, whereas in the observed light-curves this 
luminosity is constant (Fig. \ref{comp_bc}) or decreasing.  
\item In the calculated light-curve the narrow outbursts have a much 
lower amplitude than the wide ones, whereas in the observed light-curves 
these two types of outbursts differ mainly in their duration and only 
very little in their amplitude (Fig. \ref{comp_bc}). 
\end{itemize} 
Clearly something must be changed in the model. Before trying to 
understand what must (and what may) be changed, let us look 
in detail at how the model works. Later we will see that one can 
find remedies for the second problem. The 
first is still with us and may require a drastic modification of 
the model. 
 
\subsubsection{The rise to outburst} 
 
\begin{figure} 
\resizebox{\hsize}{!}{\includegraphics{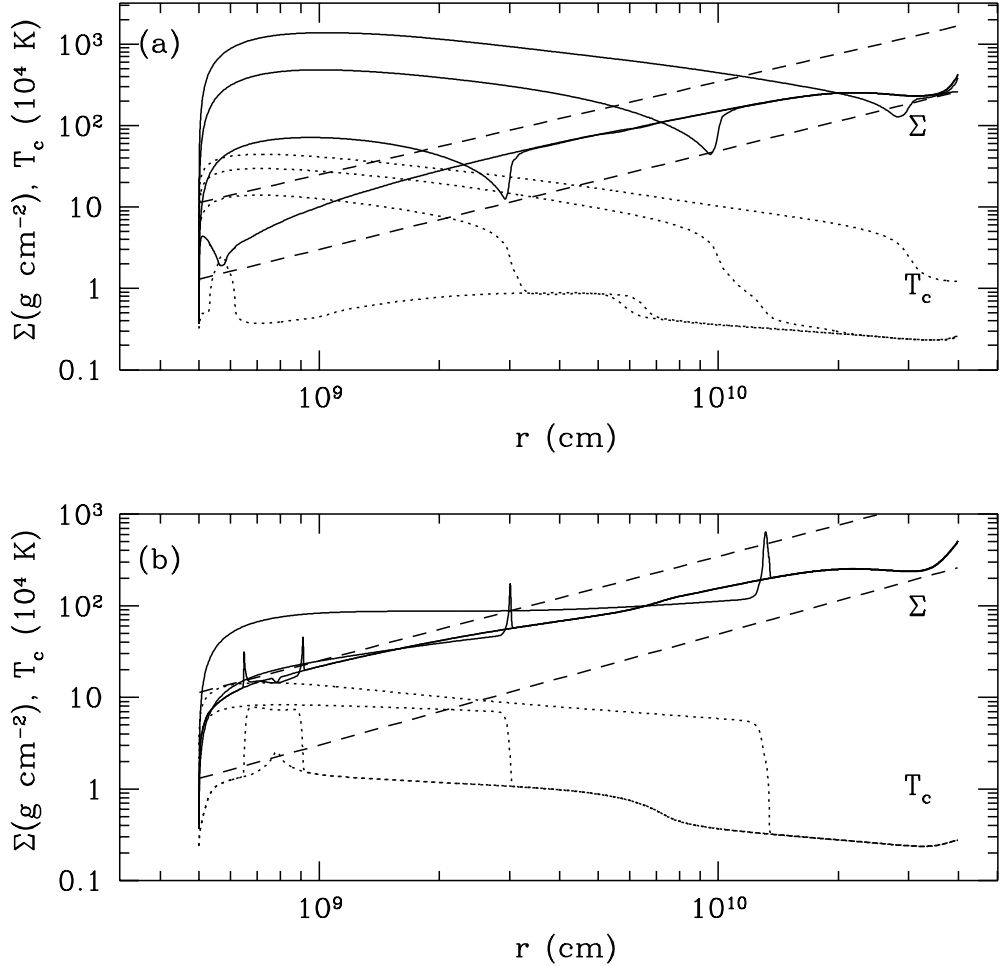}} 
\caption{Typical profiles of the surface density $\Sigma$ (solid lines) 
and the midplane temperature $T_{\rm c}$ (dotted lines) observed during the 
evolution of the thin disc. 
(a) Inward propagation of a cooling front and the associated density 
rarefaction wave. 
(b) Outward propagation of an inside-out heating front and the associated 
density spike. 
The dashed lines represent $\Sigma_{\rm min}$ (upper curve) and $\Sigma_{max}$ 
(lower curve). (HMDLH).} 
\label{profhc} 
\end{figure}

Let us start with the state of the disc represented by the 
upper curve in Figure \ref{sig}, the one that almost touches the 
critical $\Sigma_{\rm max}$ line. This is the end of the quiescent 
interval of the outburst cycle. The quiescent disc had accumulated 
matter near the inner edge of the disc. This is where the surface-density 
profile will cross the critical line, triggering an outburst. In the 
local picture this corresponds to leaving the lower branch of the 
\bS-curve. The next `moment' (in a thermal time-scale) is 
represented in the lower panel (b) of Figure \ref{profhc}. This is when a 
large spike forms in the midplane temperature profile and when the 
surface-density profile is already above the critical line. The disc is 
undergoing a thermal runaway at $r\approx 8\times 10^9$ cm. The midplane 
temperature rises to $\sim 70000$ K. This raises the viscosity which 
leads to an increase of the surface-density. It is here that Eq. 
(\ref{eq:alpha}) enters into the game. If the increase in viscosity were 
due only to the rise in the temperature through the speed of sound ($\nu 
\propto c_s^2$, Eq. (\ref{alfn})) the resulting outburst would have 
nothing to do with the observed ones. 
 
We now have a spike in the density profile and a flat-topped temperature 
profile with steep gradients. Because of these gradients the density 
spike and the temperature gradients (heating fronts) will start to 
propagate. The detailed structure of a heating front will be discussed 
in Sect. \ref{fronts}. What is shown in Fig. \ref{profhc} is a so-called 
`inside-out' outburst, because the outburst starts in the disc's inner 
regions and then propagates outwards. This concerns the main outburst; 
but since inside-out outbursts never start exactly at the inner edge 
there is always an associated outside-in outburst, as clearly seen in 
the temperature and density profiles of Fig. \ref{profhc} (see also Fig. 
\ref{ref1}, where it is seen in the $\dot M_{\rm in}$-curve). These 
short-lived outbursts cannot be observed in reality (in Fig. \ref{ref1} 
this `outburst' is invisible in the V - curve). One also should keep in 
mind that outside-in outbursts may start quite far away from the outer 
disc's edge (see Sect. \ref{io}). 
 
In our case (Fig. \ref{profhc}), the heating front reached the outer 
disc radius. Such an outburst corresponds to the large outbursts in Fig. 
\ref{comp_bc}. Smaller outbursts are produced when the front does not 
reach the outer disc regions. In an inside-out outburst the 
surface-density spike has to propagate against the surface-density 
gradient because in a quiescent disc $\Sigma\sim r^{1.14}$, roughly 
parallel to the critical surface-density. Most of the disc's mass is 
therefore contained in the outer regions. A heating front will be able 
to propagate if the post-front surface-density is larger than 
$\Sigma_{\rm min}$ -- in other words, if it can bring successive rings 
of matter to the upper branch of the \bS-curve. If not, a {\sl cooling 
front} will appear just behind the $\Sigma$ spike, the heating front 
will die-out and the cooling front will start to propagate inwards (the 
heating-front will be `reflected', see Menou et al. 1999a). 
 
The difficulty inside-out fronts encounter when propagating is due to 
angular-momentum conservation. In order to move outwards the 
$\Sigma$-spike has to take with it some angular momentum because the 
disc's angular momentum increases with radius (the specific angular 
momentum is Keplerian, which is a good approximation even in the front 
region, according to Ludwig \& Meyer 1998). For this reason inside-out 
front propagation induces a strong outflow. In order for matter to be 
accreted, a lot of it must be sent outwards. That is why during an 
inside-out dwarf-nova outburst only $\sim 10\%$ of the disc's mass is 
accreted onto the white dwarf. Another reason is that the propagation of 
the cooling front during the outburst's decay also induces a strong 
outflow (see below). 
 
\subsubsection{The limit-cycle} 
 
\begin{figure} 
\resizebox{\hsize}{!}{\includegraphics{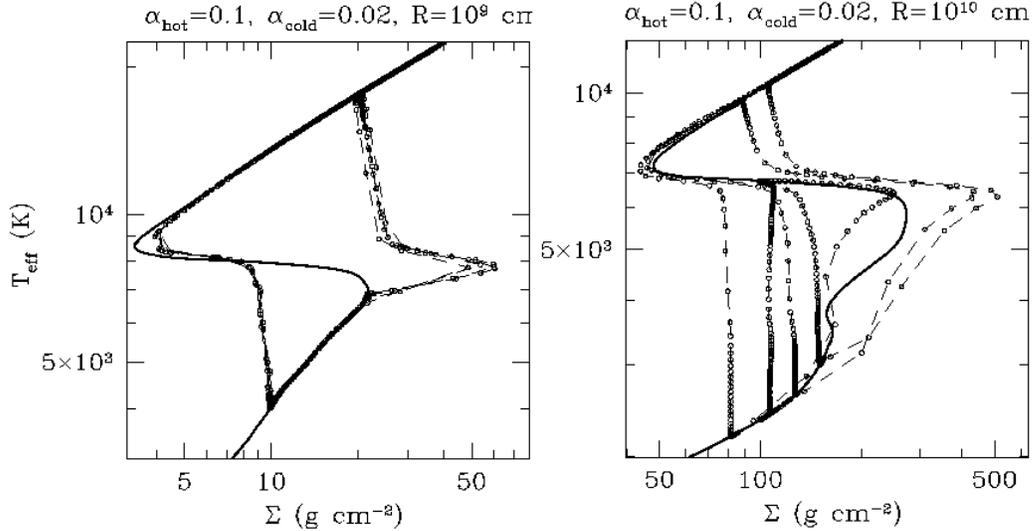}} 
\caption{Cycles in the $\Sigma-T_{\rm eff}$ plane that a disc annulus 
undergoes during several inside-out outbursts. Two annuli are shown: at 
$r=10^{9}$ and $10^{10}$ cm. The outer disc radius (fixed in this 
calculation) is at $r_{\rm out}=4 \times 10^{10}$ cm, the mass-transfer 
rate is $\dot M_{\rm transf}= 6.66 \times 10^{16}$ g s$^{-1}$, 
$\alpha_{\rm cold}=0.02$, $\alpha_{\rm hot}=0.1$. The white dwarf mass is 1.2 
M$_{\odot}$. The \bS-curve and the trajectories of system-points are 
shown. Note that after getting onto the upper branch a system-point 
goes first up and then down this part of the thermal equilibrium curve. 
The density of points corresponds to the time spent in a given state so 
that during quiescence (lower branch) and during outburst, when the 
point is moving along the \bS-curve, the density of points is such that 
they look like a `continuous line'. From Menou et al. (1999a) } 
\label{fig:std} 
\end{figure}

All this can also be seen in Fig. \ref{fig:std}, where we observe 
the `limit cycle' which the local state of the disc 
undergoes during outbursts. The left panel 
of this figure shows the situation well inside the disc; the right panel 
represents the disc's outer regions. In both cases the path of the system in 
the $\Sigma-T_{\rm eff}$ plane looks different from what is often shown 
in schematic drawings found in the literature. There, the transition to the 
upper branch of the \bS-curve is happening at constant $\Sigma$ (the 
downward transition to the lower branch is shown in the same fashion). 
The argument in favour of such a representation of the cycle undergone 
locally by the disc is that the thermal time-scale for temperature 
variations is much shorter than the viscous time-scale at which  
surface density changes. This is true, however, only in an equilibrium 
disc. Here, the transition to the hot state (upper branch) is due to the 
passage of a front, i.e. to the passage of strong temperature and  
density gradients. In such a case, Fig. \ref{fig:std} clearly shows that 
the two timescales are comparable. 
\begin{figure} 
\centering 
\resizebox{10truecm}{!} 
{\includegraphics[angle=-90]{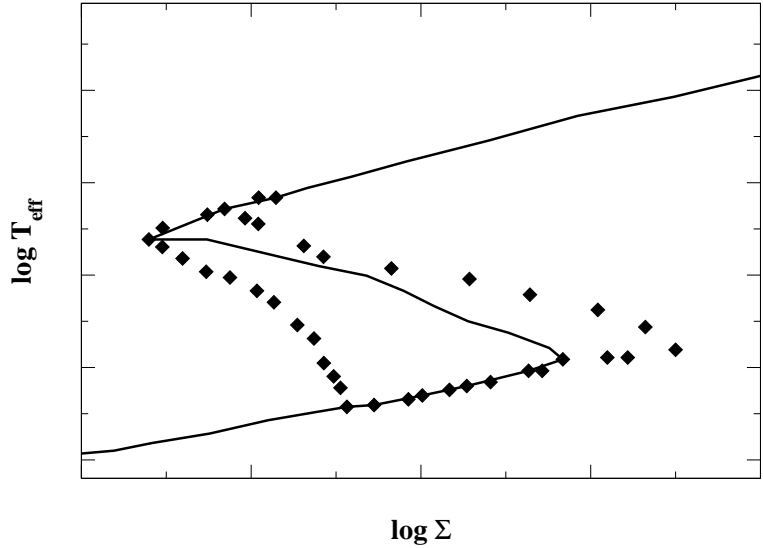}} 
\caption{The limit cycle close to the outer edge of the disc during an 
outside-in outburst. Unlike the preceding figure, the system point 
almost immediately starts going down the hot branch. One can also see 
that the maximum temperature reached on the upper branch diminishes with 
increasing distance from the center. After Smak (1998)} 
\label{sma} 
\end{figure} 
 
The passage of the front is clearly seen: after leaving the the 
\bS-curve the system moves towards high $\Sigma$'s. It also moves 
towards high temperatures, but the effect of the $\Sigma$ spike is 
clearly marked, especially on the left panel. Closer to the disc's outer 
radius things are more complex. One can still see the passage of the 
spike in surface density, but now the `attraction' of the $\Sigma_{\rm 
min}$ critical point is stronger than it was at smaller radii. The 
temperature increases all the time, but only during the last stage of 
the transition towards the upper branch, in the wake of the front, does 
happen at roughly constant $\Sigma$. With the distance to the outer disc 
radius decreasing, the value that $\Sigma$ reaches when the system 
arrives at the upper branch decreases: it becomes closer to $\Sigma_{\rm 
min}$. If it is lower than $\Sigma_{\rm min}$ the heating front will not 
make it to the outer disc's edge. In the right panel in Fig. 
\ref{fig:std} one front was not able to bring the system to the upper 
branch. Two other fronts died out before reaching the disc edge. That is 
why this panel shows different trajectories at each cycle. Closer to the 
center all the cycles are the same. Fig. \ref{sma} shows the limit cycle 
for an outside-in outburst close to the outer disc radius (Smak 1998). 
The rise-to-outburst path is here parallel to the unstable middle 
branch, which indicates that the process is occurring close to thermal 
equilibrium. The negative slope reflects an outflow (see Eq. 
\ref{noneq}). The maximum effective temperature which is reached in this 
case is very close to that corresponding to $\Sigma_{\rm min}$. Since 
things are happening close to the outer disc edge non-local effects are 
important: the variations of the tidal torque with viscosity and disc 
expansion strongly affect local disc behaviour. 
 
\subsubsection{Front propagation: inside-out vs outside-in} 
 
A sequence of fronts which do not reach the disc outer radius 
accumulates matter in the outer regions and in the end enables a front 
to travel through the whole extent of the disc. This produces a sequence 
of several narrow outbursts followed by one wide outburst. However, 
since such narrow outbursts have, inescapably, lower amplitudes, they 
cannot be the narrow outbursts observed in dwarf novae. They should be 
treated, as pointed out by Smak (1999a, 2000), as unphysical. 
 
Outside-in outbursts have no problems with propagating. They travel down 
the surface-density gradient, so there is no danger that the post-front 
$\Sigma$ will not be larger than $\Sigma_{\rm min}$. We will say more 
about these type of outbursts in Sect. \ref{io}. In this case, after the 
heating front passage the disc's luminosity is practically at maximum. 
 
In an inside-out outburst the disc behind the front is hot and  
matter begins to diffuse inwards. Since most of the mass is located in 
the outer disc,  this leads first to a slow increase in $\Sigma$ (and in 
accretion rate) as seen in Fig. \ref{profhc} (see also Fig. \ref{rad}). 
Locally, on the \bS-curve, each ring, now in thermal equilibrium, moves 
to higher temperatures and higher surface-densities along the upper 
branch. In the light-curve this is seen as the rise-to-outburst's 
maximum. In inside-out outbursts the rise to maximum is therefore 
slower than when the outburst is outside-in, and the light curves have 
a `rounder' shape. We will come back to this in Sect. \ref{uvdel}. 
 
\subsection{The outburst maximum} 
 
Near the brightness maximum the disc is everywhere in thermal 
equilibrium on the hot branch of the \bS-curve. In most of the disc, 
except for its outermost regions, the accretion rate is constant. This 
part of the disc is well represented by a Shakura-Sunyaev (1973) 
solution. 
 
The maximum accretion rate can be estimated as: 
\begin{equation} 
\dot M_{\rm max}\approx 7.8 \times 10^{16} 
\left(\frac{\alpha_{\rm hot}}{0.2}\right)^{0.01} P_{\rm hr}^{1.79} 
\label{mdotmax} 
\end{equation} 
where $P_{\rm hr}$ is the orbital period in hours. To obtain Eq. 
(\ref{mdotmax}) it was assumed that the surface density at the  
disc's outer edge is close to the critical one $\Sigma\sim \Sigma_{\rm min}$ 
(in reality, as discussed above $\Sigma > \Sigma_{\rm min}$). Since at 
maximum the accretion rate in the disc is roughly constant one can use 
Eq. (\ref{mdmin}) to write $\dot M_{\rm max}\approx \dot{M}_{\rm 
B}\left(r_{\rm out}\right)$. Assuming that at outburst maximum $r_{\rm 
out}=0.9 R_{\rm L1}$ 
one obtains Eq. (\ref{mdotmax}). Warner (1995a) estimated $\dot M_{\rm 
max}$ by assuming that the disc before outburst is filled up to the maximum 
mass ($M_{\rm max}= 2\pi \int{\Sigma_{\rm max} r dr}$) and calculated the 
same mass at outburst's maximum assuming that the hot disc is described 
by a stationary Shakura-Sunayev solution. This gives a relation between 
$\dot M$ and $\Sigma$ from which a formula for $\dot M_{\rm max}$ 
follows. The two formulae should be roughly equivalent. The one, 
published in Warner (1995a), suffers from misprints: e.g. the power of the 
orbital period should be `1.78' and not `1.18' as printed.  
 
The observed $M_V-P_{\rm orb}$ relation can be obtained from Eq. 
(\ref{mdotmax}) by using Smak's (1989) calibration (see Warner 1995a for 
details).

\subsubsection{The decay from outburst} 
 
\begin{figure} 
\resizebox{6.8cm}{!}{\includegraphics{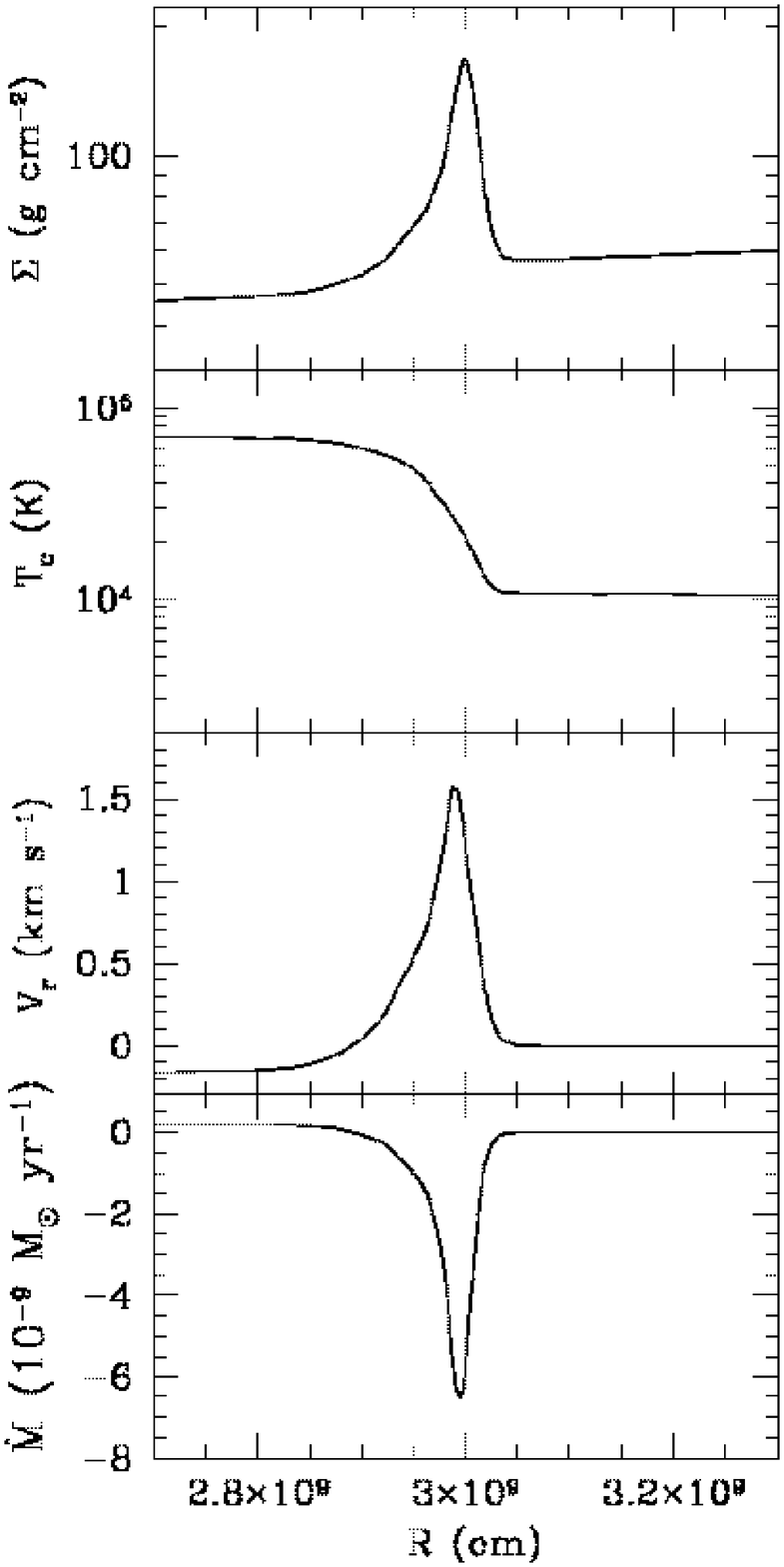}}  
\resizebox{6.8cm}{!}{\includegraphics{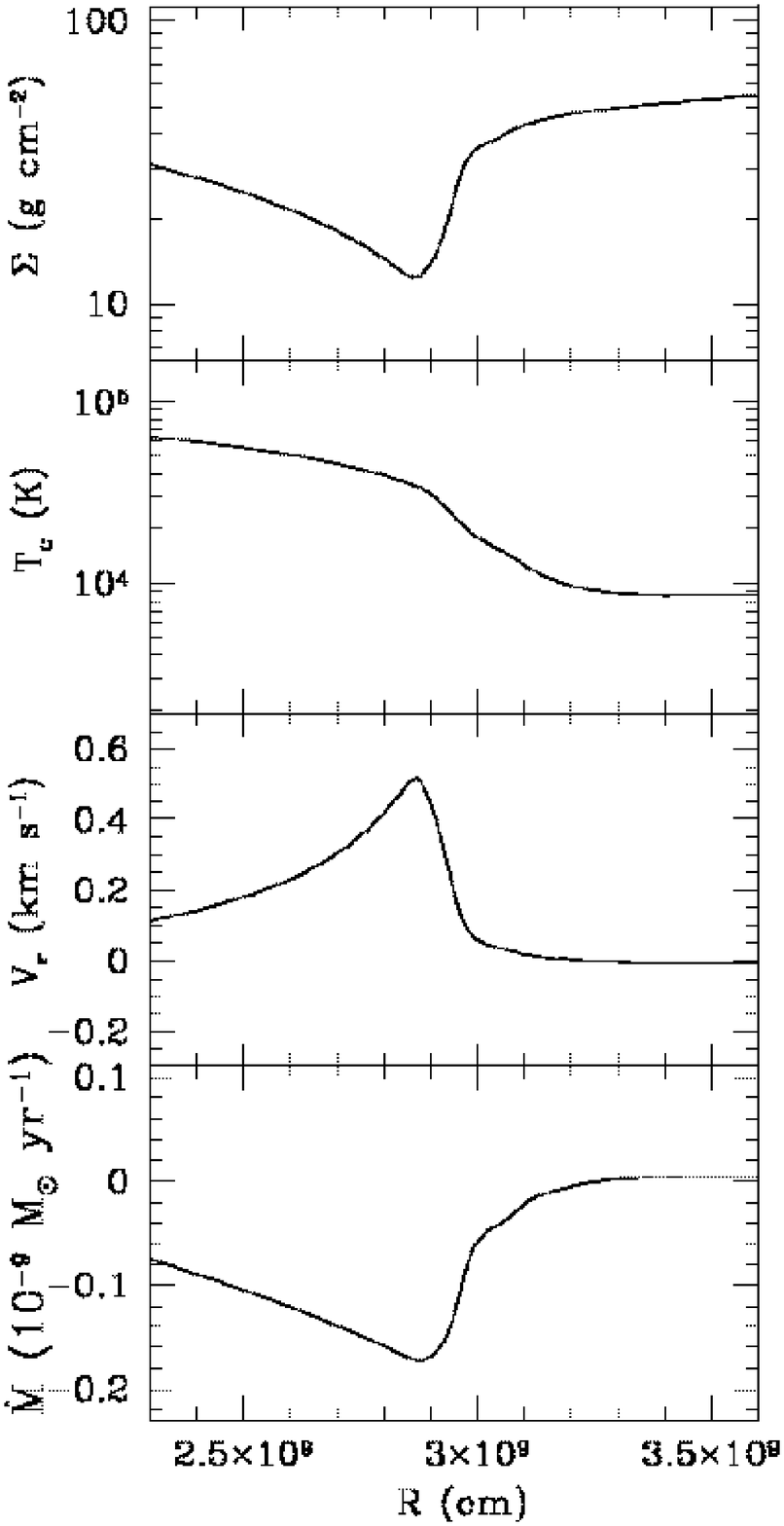}} 
 \caption{Structures of 
an inside-out heating front (left) and of a cooling front (right). The 
disc parameters are the same as in Fig. \ref{fig:std}. }  
\label{hc} 
\end{figure}

In the hot disc $\Sigma \sim r^{-3/4}$ whereas the critical 
surface-density $\Sigma_{\rm min}\sim r^{1.11}$ (Eq. \ref{sigmin}). 
Therefore somewhere in the outer disc the surface density is close to 
the critical value (see e.g. Fig. \ref{sma}). Since $\Sigma$ must now 
decrease, the outer disc will have to leave the \bS-curve. On the 
$\Sigma-T_{\rm eff}$ plane this corresponds to the region where cooling 
dominates heating. Again a steep temperature gradient is created. 
The viscosity in the inner hot part of the disc is now higher than in 
the outer ring, where $T < T\left(\Sigma_{\rm min}\right)$. This creates 
an {\sl outflow} which allows the low-temperature region (a 
cooling-front) to propagate inwards. As in the case of an inside-out 
heating front, a cooling front propagates against the density gradient. 
This creates a strong outflow, as seen in Fig. \ref{hc}. The outflow can 
also be seen `implicitly' in Fig. \ref{fig:std}, where, after getting 
through $\Sigma_{\rm min}$, the system follows for a while the middle 
branch of the \bS-curve. This corresponds to an outflow because this 
branch of the \bS-curve has a negative slope, which from Eq. 
(\ref{noneq}) corresponds to a negative accretion rate.  
The outflow through the cooling front results in an 
increased surface-density and after the front passage (but not 
immediately, see Fig. \ref{radnoill}) $\Sigma$ forms a profile parallel 
to $\Sigma_{\rm min}$. A cooling front will be able to keep propagating 
if the post-front surface density is lower than $\Sigma_{\rm max}$. If 
it is not, a heating front will form in the wake of the cooling front 
and new outburst will start (a ``reflare", see Sect. \ref{reflares}). 
This is a problem `inverse' to the one encountered by an inside-out 
heating front. There, the danger was that the post-front $\Sigma$ will 
be too small for the front to propagate; here the surface density might 
be too large. In both cases the light-curves produced do not correspond 
to what is observed in real systems, so a way  should be found of getting 
rid of these undesirable reflection features. The heating-front 
`reflection' is more of a problem for dwarf-nova models, while `reflares' 
are a deficiency of  LMXBT models.  
\begin{figure} 
\centering 
{\includegraphics[scale=1,totalheight=9cm]{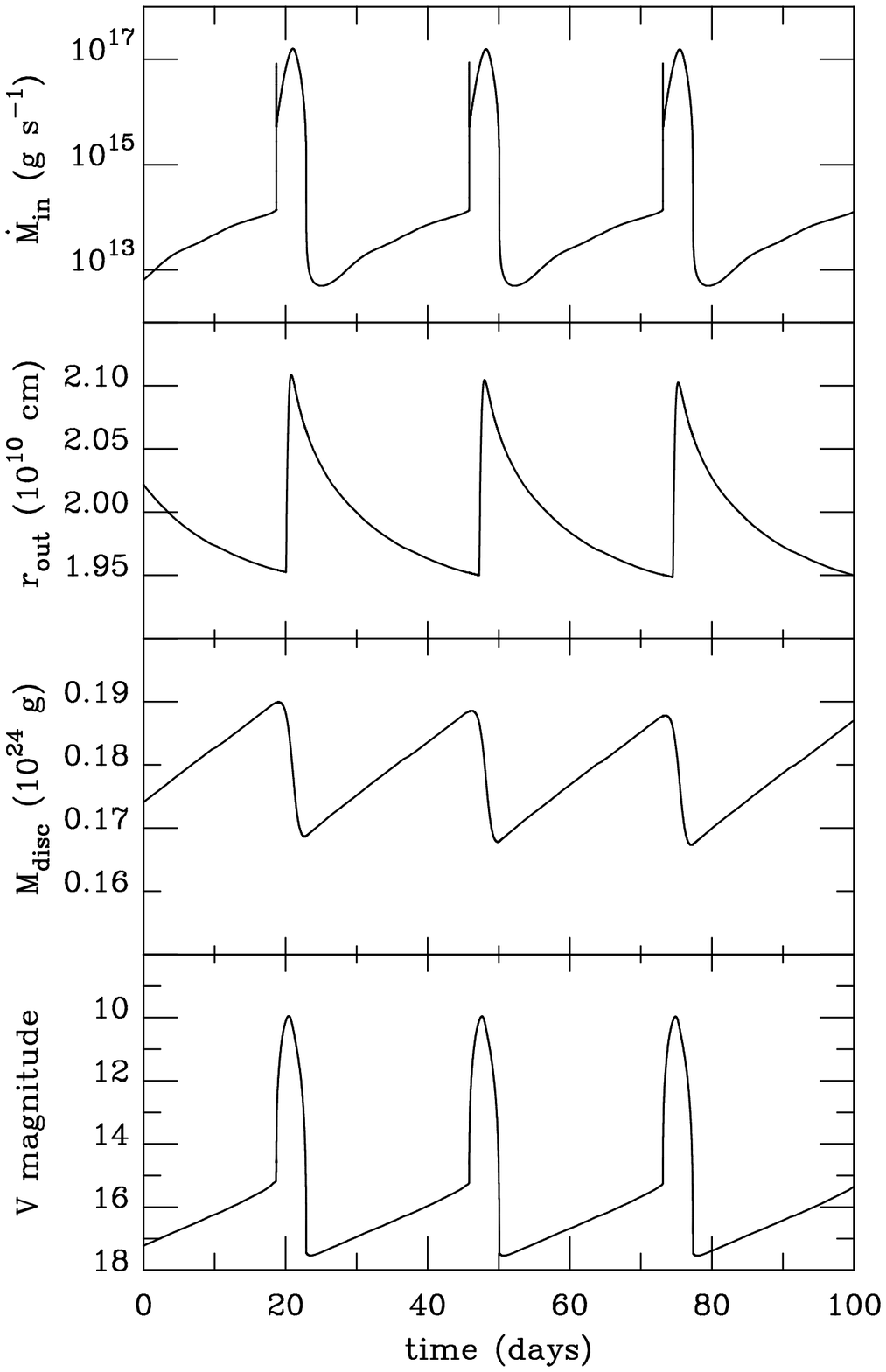}} 
\hskip 1cm 
{\includegraphics[scale=1,totalheight=9cm]{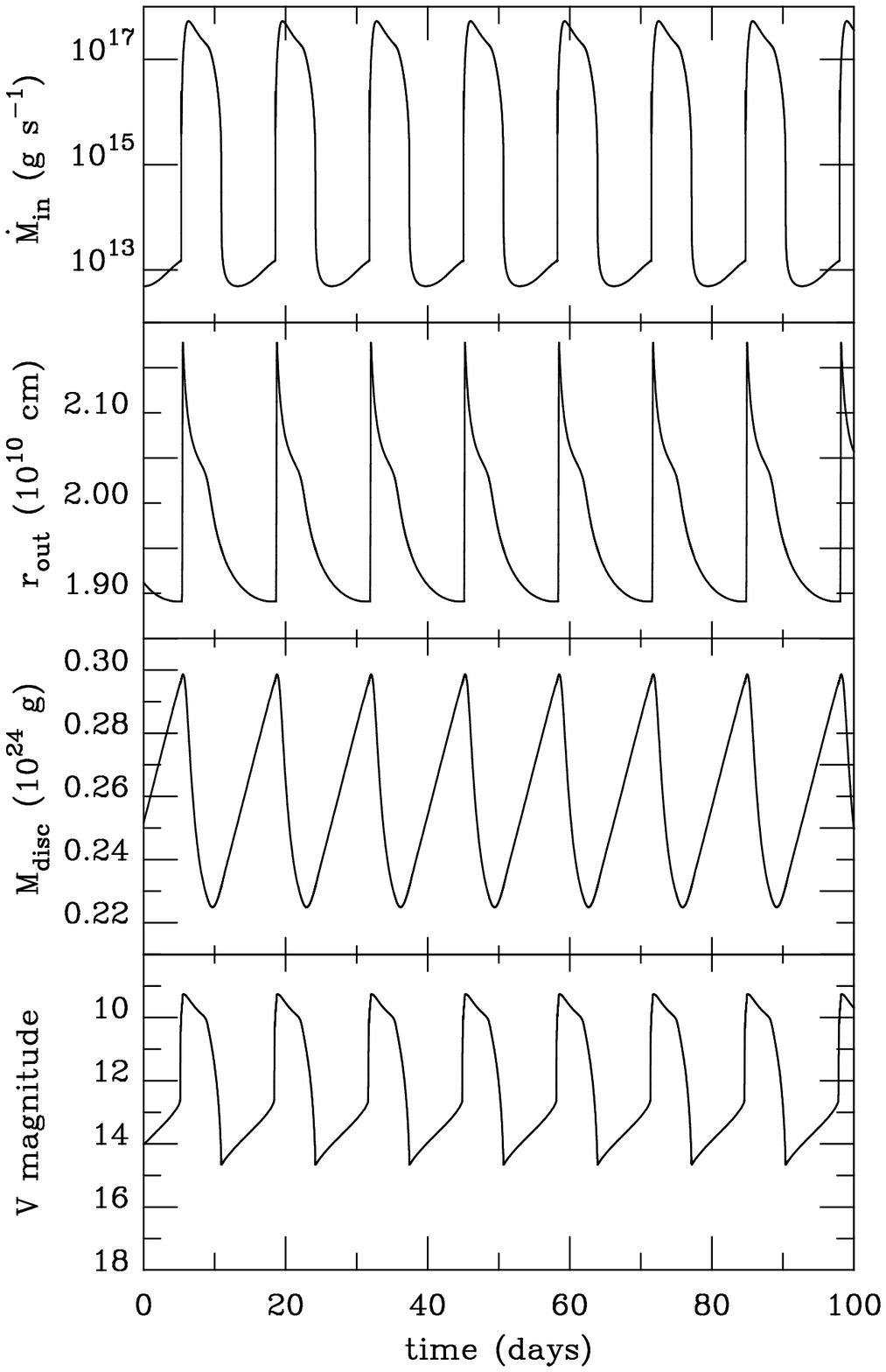}} 
\caption{Outburst 
properties for $M_1 = 0.6$ M$_\odot$, $r_{\rm in} = 8.5 \times 10^8$ cm, 
$\alpha_{\rm cold} = 0.04$, $\alpha_{\rm hot} = 0.20$, $<r_{\rm out}> = 
2 \times 10^{10}$ cm, and for $\dot{M} = 10^{16}$ g s$^{-1}$ (left) 
and for $\dot{M} = 10^{17}$ g s$^{-1}$ (right). The  
panels show (top to bottom) the mass accretion rate onto the white dwarf,  
the outer disc radius, the disc mass, and  
the visual magnitude. (HMDLH)}  
\label{ref1}  
\end{figure}

When the cooling front is allowed to propagate, as in the case shown in 
Fig. \ref{profhc}, it arrives at the inner disc edge, switching off the 
outburst. At the end of the cooling-front journey the whole disc settles 
down to a quiescent state.

\subsubsection{Fronts} 
\label{fronts}

As pointed out by Meyer (1984), heating fronts in dwarf-nova accretion 
discs are similar to ignition fronts in combustion waves. Such ignition 
fronts propagate into unburnt matter (``cool state" in a disc) leaving 
behind burnt matter (``hot state"). Matter inside the front is heated on 
the burning time-scale and then by heat diffusion ignites the matter into which 
it is propagating. During the burning time $\tau_b$ heat diffuses over a 
distance $V_{\rm front}\tau_b$ covered by the front during this time, so the 
front speed $V_{\rm front}$ can written as: 
\begin{equation} 
V_{\rm front}\approx \frac{\nu_{th}}{\tau_b} 
\end{equation} 
where $\nu_{th}$ is the coefficient of thermal diffusion. In accretion discs the 
the thermal diffusivity coefficient (which is the same as the kinematic 
viscosity coefficient) can be written as 
\begin{equation} 
\nu_{th}\approx \frac{H^2}{\tau_{th}} 
\end{equation} 
where $\tau_{th}=\left(\alpha \Omega_K\right)^{-1}$, so that the front 
speed is $V_{\rm front}\approx \alpha c_s$. 
 
These heating-front properties were confirmed in detailed numerical 
calculation by Menou et al. (1999a) with some complications 
due to the fact that two values of $\alpha$ must be used in the model of 
dwarf nova outbursts. The dependence on $\alpha_{\rm cold}$ is, 
however, rather weak and the front velocity is of the order $\alpha_{\rm 
hot} c_s$. Another ignition-front property is also confirmed by Menou et 
al. (1999). The width of the heating front is 
\begin{equation} 
\delta w \approx V_{\rm front}\tau_{th} \sim H 
\end{equation} 
as estimated by Meyer (1984). The thermal structure of the front 
proper (called a `precursor' by Menou et al. 1999a) is dominated by 
viscous radial heat diffusion, which explains the success of the 
ignition front analogy presented above. One should not forget, however, 
that the term $J$, describing radial radial energy transport, is rather 
uncertain, so the actual value of the front velocity depends on what 
is assumed about this term (Menou et al. 1999a). Especially the fact that 
$\delta w  \sim H$ could lead to a Rayleigh instability enhancing the 
radial energy transport (Lin, Papaloizou \& Faulkner 1985). In their  
observations of EX Dra, Baptista \& Catal\'an (2000) reported  
heating-front speeds a factor 2 slower that those predicted by Menou  
et al. (1999a). 
 
Cooling fronts were subject to intensive discussions after Cannizzo, 
Chen \& Livio (1995) had claimed that, according to their numerical 
calculations, the width of cooling fronts is $\delta w=\sqrt{HR}$ and 
not $\delta w\propto H$ as expected based on arguments similar to those 
used for heating fronts. Menou et al. (1999a), however, used the high 
resolution code of HMDLH to show that the front width is, as expected, 
$\delta w\propto H$, thus closing the discussion. The Cannizzo et al. 
(1995) paper had the merit of stimulating Vishniac \& Wheeler (1996) (see 
also Vishniac 1997) to try to give an analytical description of  
cooling front propagation. The formula they got for cooling front 
propagation speed 
\begin{equation} 
v_{\rm front}\approx 2 \alpha_{\rm hot} c_s\left(\frac{H}{r}\right)^{7/10} 
\label{vfc} 
\end{equation} 
 
where $c_s$ is  the sound speed at the cooling front, is very useful 
if one wishes to understand the properties of the decay from outburst. 
The factor `2' was added here following Menou et al. (1999a) who noticed 
that Vishniac \& Wheeler's (1996) assumption that gas velocity at the 
front is much larger than the front velocity is not confirmed by 
numerical calculations; the velocities are in fact comparable. 
Observations of the decay from outburst of the eclipsing dwarf nova IP 
Peg are consistent with Eq. (\ref{vfc}). This formula is a good 
approximation of the cooling front velocity in what Menou et al. (1999a) 
call the `asymptotic regime', reached after a rapid deceleration of the 
front occurring soon after its appearance. These authors note that this 
deceleration is a generic property of the DIM and could be an important 
test of the model. Recently, Baptista \$ Catal\'an (2000) reported 
observing this feature during an outburst of EX Dra as well as cooling 
front speeds comparable to those predicted by Menou et al. (1999a).  
 
However, the self-similar character of the decay from outburst found by 
several authors (Vishniac 1997; Menou et al. 1999a) is an approximation 
valid only for a restricted range of parameters, as discussed in the 
next section. 
 
Most of the study of the cooling fronts and decay from outbursts was 
motivated by the desire to obtain `exponential' light curves for LMXBTs 
in the framework of the standard DIM. As pointed out by King \& Ritter 
(1998), however, such light curves are due to the disc's irradiation 
which inhibits cooling front propagation (see Sect. \ref{sxt}) and 
models neglecting this effect are therefore only of academic interest. 
An additional motivation was provided by results of calculations in 
which the outer disc radius was kept constant. It is reassuring, 
however, that even erroneous motivations may give interesting insight 
into the structure of the DIM.

\subsection{Reflares} 
\label{reflares} 
  
We have  already mentioned the difficulties encountered by some  
types of fronts when propagating in the disc. Both inside-out heating 
fronts and cooling fronts (which are always `outside-in' and therefore  
have to move against the density gradient) may be stopped and 
`reflected'. Only outside-in heating fronts travel unhindered down the 
density slope. 
\begin{figure} 
\resizebox{\hsize}{!}{\includegraphics{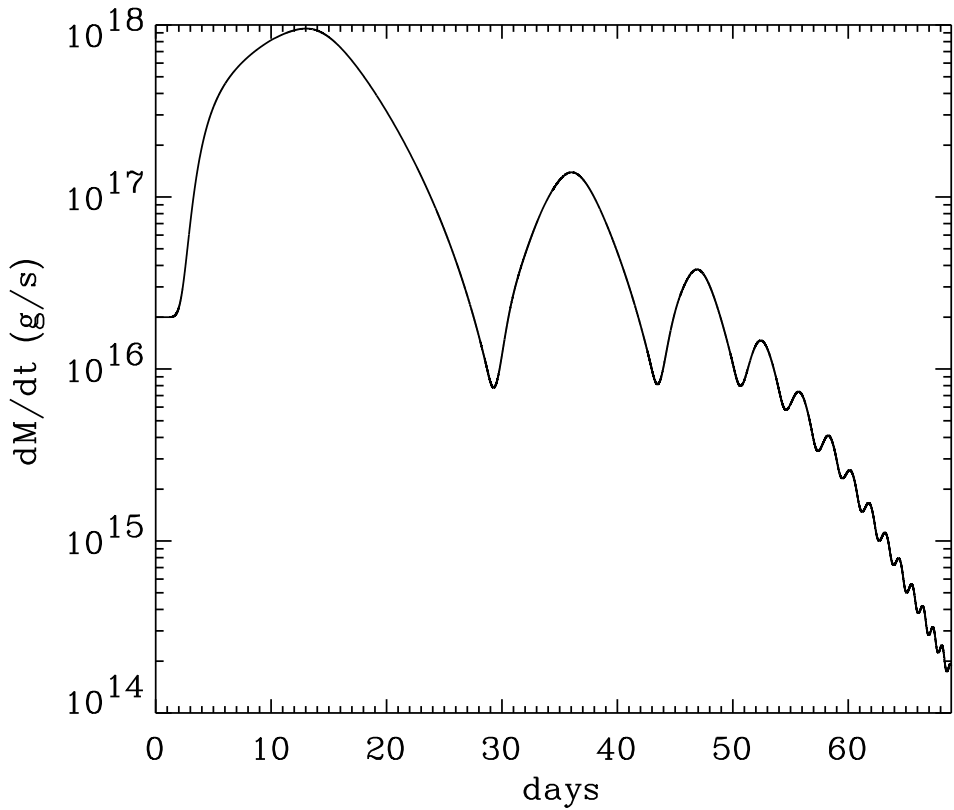}} 
\caption{The outburst light curve of a non-irradiated, non-truncated disc 
around a 1.4 M$_{\odot}$ accreting body. The mass-transfer rate is 
$2\times 10^{15}$ g s$^{-1}$, $\alpha_{\rm cold}=0.02$, $\alpha_{\rm hot}=0.1$.  
(Dubus - private communication). 
} 
\label{enlc} 
\end{figure}

\begin{figure} 
\resizebox{\hsize}{!}{\includegraphics{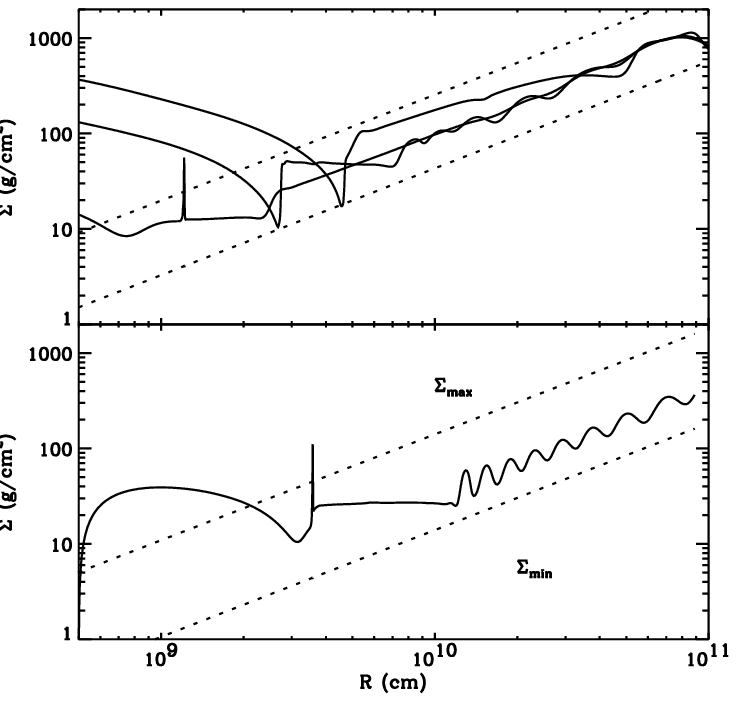}} 
\caption{Reflares in non-irradiated discs around a NS (top panel) and 
a BH (bottom panel). The full lines are the surface density $\Sigma$ 
at different times during the decay from outburst 
and the dotted lines are $\Sigma_{\rm max}$ and 
$\Sigma_{\rm min}$ (Eq.~\ref{sigmax}, \ref{sigmin}).  In the top panel, 
the situation before the onset of the first reflare is shown by the 
top curve: $\Sigma$ just behind the cooling front is close to the 
critical $\Sigma_{\rm max}$ for which the disc switches back to the 
hot state. The middle and bottom lines show the profile after several 
reflares: $\Sigma$ in the outer disc has been depleted by the 
successive passages of the heating front (which can be seen in the bottom 
curve). The lower panel shows the situation after many successive 
reflares in a disc around a BH primary. The wiggles in the density 
profile corresponding to previous reflares have not yet been smoothed 
out by viscous diffusion. The inner disc radius was taken to be $10^8$~cm  
for the neutron star case and $5\cdot10^8$~cm for an accreting black hole.  
(From DHL)} 
\label{radnoill} 
\end{figure} 
The multiple reflections of heating and cooling front produce `reflares', 
as shown, for example, in Fig. \ref{enlc}. Reflares are a natural 
property of the DIM. In particular, in a model in which $\alpha_{\rm 
hot}=\alpha_{\rm cold}$ fronts reflect indefinitely and produce only 
small amplitude variations of the disc luminosity instead of a 
dwarf-nova type light curve (see Fig. 1 in Mineshige \& Osaki 1985). 
 This is why different $\alpha$'s must be 
used in outburst and quiescence. Some authors claim that the reflares they 
found are just numerical artifacts (e.g. Cannizzo et al. 1995). 
This is possible in low-resolution calculations, but high-resolution 
calculations by Menou et al. (2000) and DHL show that 
reflares are definitely due to physical effects even when the viscosity 
parameter is different on the high and low branches of the \bS-curve. 
Menou et al. (2000) discussed the reflare problem and showed that the 
presence or absence of reflares depends not only on $\alpha$ but also on 
the mass of the accreting object. It was claimed in this article that 
for a central mass of 1.2 M$_{\odot}$ no reflares appear in the model 
while for a mass o 7 M$_{\odot}$ reflares are clearly present. This mass 
dependence was explained by an `analytical' model. 
 
Fig. \ref{enlc}, which shows a light-curve (strictly speaking the 
accretion rate onto the central object as a function of time) for a mass 
of 1.4 M$_{\odot}$, is in obvious contradiction with the conclusions 
of Menou et al. (2000). The reasons for this contradiction are very 
instructive: they show how prudent one must be when generalizing results 
of particular numerical calculations and they also show the limited 
value of `analytical' arguments in dealing with the physics of 
time-dependent accretion discs.

The argument presented in Menou et al. (2000) begins with the correct statement 
that the surface density at the cooling front is (see Sect. 
\ref{fronts} and Fig. \ref{radnoill})  
\begin{equation}  
\Sigma\left(R_{\rm front}\right)= 
\Sigma_{\rm min}\left(R_{\rm front}\right). 
\end{equation}  
 
Then, following Vishniac (1997), it is assumed that the cold, outer regions 
of the disc behind the cooling front are frozen during the cooling front 
propagation so that behind the front $\Sigma(R)=K(M_1, 
\alpha)\Sigma_{\rm min}$, where $K$ depends on the viscosity parameter 
$\alpha$ and the mass of the accreting object $M_1$. Menou et al. 
(1999a) obtained $K\approx 4$ for a central mass $M_1=1.2$ M$_{\odot}$, 
and $K \approx 6-7$ for $M_1=7$ M$_{\odot}$. Since the ratio of the 
`maximum' to `minimum' surface densities is (Eqs. \ref{sigmax} and \ref{sigmin})  
\begin{equation}  
\frac{\Sigma_{\rm max}}{\Sigma_{\rm min}}=1.26 
\frac{\left(\alpha_{\rm hot}\right)^{0.74}} {\left(\alpha_{\rm 
cold}\right)^{0.85}}  
\end{equation}  
one has  
\begin{equation} 
{\Sigma_{\rm max} \left( \alpha_{\rm cold}=0.02\right) \over \Sigma_{\rm 
min}\left(\alpha_{\rm hot}= 0.1 \right)} \approx 6.4,  
\end{equation} 
Therefore, for $M_1=1.2$ M$_{\odot}$ the post-cooling-front surface 
density $K\Sigma_{\rm min} < \Sigma_{\rm max}$, and no reflares are 
expected, while for $M_1=7$ M$_{\odot}$, the post-cooling-front surface 
density $K\Sigma_{\rm min} \gta \Sigma_{\rm max}$ and one should expect 
the presence of reflares. Menou et al. (2000) show why and how $K$ 
depends on the mass of the accreting object: $K\propto M_1^{0.35}$. 
 
The problem with this argument is that, as noticed by Menou et al. 
(1999a), just after the front the post cooling-front density is not 
proportional to $\Sigma_{\rm min}$. This is clearly seen in Fig. 
\ref{radnoill} (upper panel: just behind the front there is small 
region with a slope flatter than $\Sigma_{\rm min}$. (This region 
grows after the passage of successive reflares.  A better approximation 
of the post-front surface-density distribution would be rather: 
$\Sigma\propto R^{0.9}$ (instead of $R^{1.11}$) so that in general there 
is always a radius for which $\Sigma$ behind the cooling front becomes 
greater than $\Sigma_{\rm max}$ and reflares should always be present if 
the range of radii considered is wide enough. Menou et al. (2000) did 
not see reflares in the case of low-mass accreting objects because they 
used a code which included strong evaporation (see Sect. \ref{sxt}) of 
the inner disc during decay from outburst. As a result the inner disc 
radius was receding fast enough to prevent a reflare (see Sect. \ref{sxt}).  
 
In conclusion: the Menou et al. (2000) argument is not false, it is 
incomplete: the presence of reflares depends not only on the values of 
$\alpha$ and the central mass, but also on the disc's radial extent. 
Dwarf-nova discs which typically extend from $\sim 5\times 10^{8}$ cm to 
a few times $ 10^{10}$ cm are not prone to the presence of reflares, if 
the hot to cold $\alpha$'s ratio is large enough ($\gta 4$). 
 
The reflare problem is not just a boring technical question. Reflares 
are a fundamental part of the DIM but they seem to be observed neither 
in dwarf novae nor in LMXBTs containing neutron stars or black holes. 
`Kinks' in the light-curves of black-hole LMXBTs or `secondary' 
outbursts seen in both dwarf-novae and LMXBTs (Kuulkers, Howell \& van 
Paradijs 1996) look totally different. One must, therefore get rid of 
them for the DIM to work when applied to observed systems. For dwarf 
novae this implies a moderate ratio of hot to cold $\alpha$'s but for 
LMXBTs this ratio would have to be at least 20, but even then this will 
help only if the inner disc were truncated (Menou et al. 2000). Luckily, 
discs in LMXBTs are strongly X-ray irradiated (van Paradijs \& 
McClintock 1995), which implies lower post-front surface densities and 
less chance for reflares to appear, as showed by DHL (see Sect. \ref{sxt}).

\subsection{``Inside-out" and ``outside-in" outbursts} 
\label{io} 
 
After each outburst the disc has to fill up with matter before erupting 
once more. As discussed in Sect. (\ref{reflares}), after the passage of 
the cooling front the surface density settles down to a value 
proportional to $\Sigma_{\rm min}$, i.e at the beginning of quiescence 
the surface-density distribution is roughly parallel to the critical 
densities (see Figs. {\ref{sig} and \ref{profhc}). 
\begin{figure} 
\resizebox{\hsize}{!}{\includegraphics{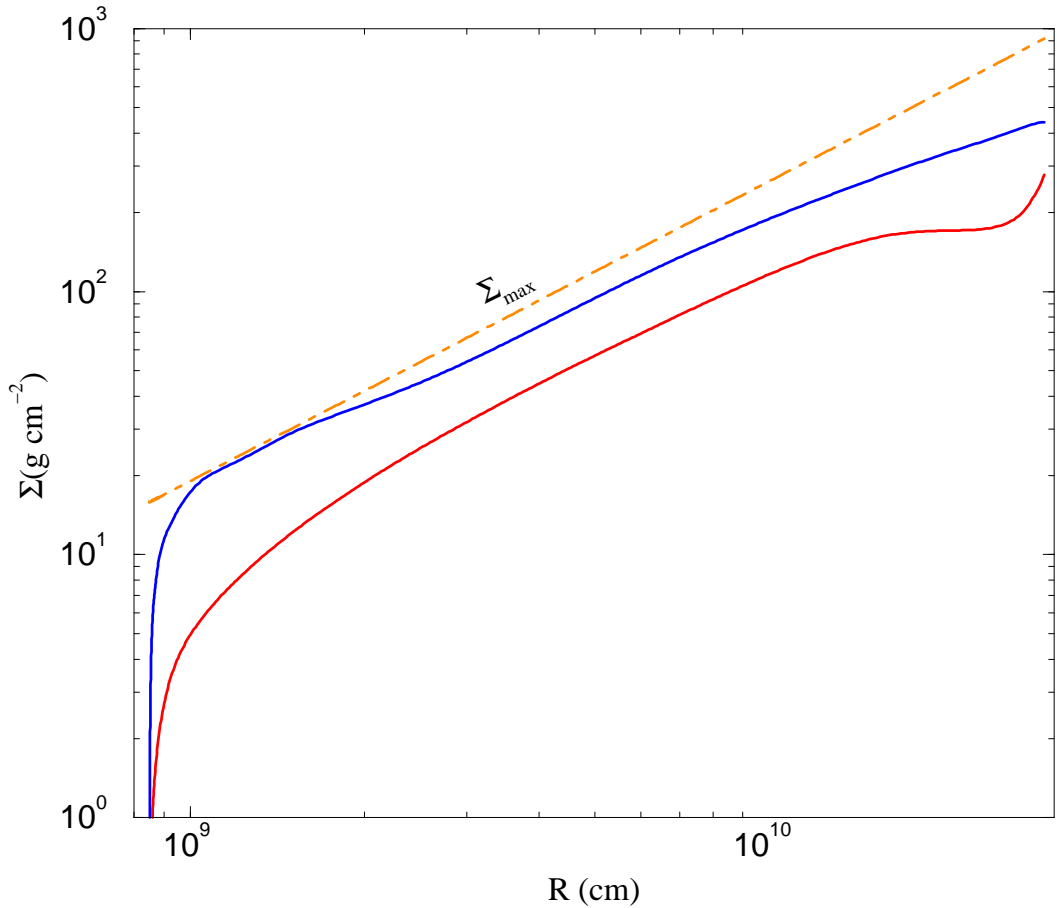}} 
\caption{Density profiles after (lower curve) and just before an outburst 
(upper curve). The 
parameters are the same as in Fig. \ref{fig:cnc}. 
} 
\label{sig} 
\end{figure} 
What happens next depends on the mass-transfer rate, the size of 
the disc and viscosity (i.e. $\alpha_{\rm cold}$).

Figure \ref{ref1} shows inside-out (left) and 
outside-in (right) outbursts for the same system in which only the mass-transfer 
rate is different. Low mass-transfer rates produce inside-out outbursts 
whereas outside-in outbursts correspond to high mass-transfer rates. 
This can be easily understood in terms of the characteristic time-scales 
of the accretion processes at the disc's outer edge. One has to consider 
two processes: accumulation and diffusion of matter. If the accumulation 
time is shorter than the (viscous) diffusion time, matter will 
accumulate near the disc's outer edge and the resulting excess of 
surface density will trigger an inward propagating outburst. This is an 
``outside-in" outburst. In the opposite case matter will never be able 
to accumulate at the outer disc regions and will diffuse inwards. Since 
in a quiescent disc the accretion rate decreases steeply with radius 
matter will have to accumulate somewhere, and eventually cross the 
critical ($\Sigma_{\rm max}$) line, starting an ``inside-out" outburst. 
This case is illustrated in Fig. \ref{sig}, which shows two 
surface-density profiles: one just after the end of an outburst and the 
second at the start of a new, inside-out outburst. 
 
We may suppose that the faster one feeds the disc (the higher the 
mass-transfer rate), the shorter the time to accumulate matter where it 
is fed to the disc, so the fact that outside-in outbursts occur at high 
transfer rates is not surprising. However, observations show (or at 
least suggest) the presence of outside-in outbursts in systems with 
quite low mass-transfer rates, so it would be useful to understand what 
is meant by `lower' and `higher' in this context.  
 
Osaki (1995) estimated the accumulation time to be 
\begin{equation} 
t_{\rm accum} = {4 \pi^2 r^2 \nu \Sigma_{\rm max}^2 \over \dot{M}_{tr}^2} \\ 
\label{acc} 
\end{equation} 
where $\dot{M}_{tr}$ is the mass-transfer rate from the secondary. 
$t_{\rm accum}$ is the time it takes to accumulate matter up to the 
critical density at which an outburst must begin. It is assumed that the 
width of the torus in which matter accumulates is governed by viscous 
spreading, i.e.  $\Delta r \sim \sqrt{\nu t_{\rm accum}}$ (it is this 
assumption which gives in $t_{\rm accum} \propto \dot{M}_{tr}^{-2}$).  
 
The diffusion (`drift') time can be estimated by using the standard 
accretion-disc diffusion equation (Eq. (5.6) in Frank et al. 1992). The 
diffusion time is then defined as 
\begin{equation}  
t_{\rm diff} \approx \left( {\partial \ln \Sigma \over \partial t} \right)^{-1},  
\end{equation} 
(e.g. Smak 1993; Osaki 1995a). We are interested in the time in takes to 
produce an outburst after the disc had settled down to a quiescent 
state. This is, therefore, the time it takes to form (or to reduce) by 
viscous diffusion a surface density excess bringing $\Sigma$ close to  
$\Sigma_{\rm max}$. Following Osaki (1995a) one can then write 
\begin{equation} 
t_{\rm diff} = \frac{r^2}{\nu}\zeta^2\delta 
\label{diff} 
\end{equation} 
where $\zeta^2 = (2/3)\left(\xi l\right)^{-1}$, $\xi= 2 {\partial \ln 
\nu \Sigma / \partial \ln r} + 1$, $l$ is the power-law index in $\nu 
\Sigma \sim r^l$ ($l\sim 2.64$) and $\delta=\ln\left(\Sigma_{\rm 
max}(\alpha_{\rm cold})/K\Sigma_{\rm min}(\alpha_{\rm h})\right)$ (typically 
$\delta \lta 2$), $K \Sigma_{\rm min}$ is the quiescent surface density (see Sect. 
\ref{reflares}). In discs with short radial sizes the quiescent surface-density 
profiles are much flatter than $\propto \Sigma_{\rm min}$ and these assumptions 
about $\zeta$, $\delta$ etc. are no longer valid (see Fig. \ref{oi2}) . 
 
From Eqs. (\ref{acc}) and (\ref{diff}) one obtains a condition for $t_{\rm 
accum} < t_{\rm diff}$ 
\begin{equation} 
\dot{M}_{\rm tr} > 2 \pi \nu \Sigma_{\rm max} \zeta^{-1} \delta^{-1/2} 
\label{moi1} 
\end{equation} 
One can now use Eq. (\ref{noneq}) to write $\nu \Sigma_{\rm max}= \dot 
M_{\rm A}/3\pi \xi$ , where $\dot M_{\rm A}$ is the accretion rate 
corresponding to $\Sigma_{\rm max}$ (Eq. \ref{mdmax}). 
Therefore, the criterion for outside-in outbursts (Eq. \ref{moi1}) 
becomes simply 
\begin{equation} 
\dot{M}_{\rm tr} \gta \frac{0.5}{\sqrt{\delta}} \dot M_{\rm A}, 
\label{moi} 
\end{equation} 
which seemed probable from the very beginning. Indeed, since in 
quiescence the accretion rate everywhere in the disc is lower than the 
critical accretion rate $\dot M_{\rm A}$, fresh matter arriving at  
higher than the critical rate will have difficulty diffusing 
inwards and will tend to accumulate at the outer edge. This 
simple picture seems to work quite well in this case. Of course in 
real systems things will be more complicated because matter is never 
brought exactly to the disc edge, etc. 
 
In terms of the system's parameters the criterion for the occurrence  
of outside-in outbursts is 
\begin{equation} 
\dot{M}_{tr} \gta \dot{M}_{\rm A} = 2.0 ~ 10^{15} \ \delta^{-1/2} ~\left( {M_1 \over   
\rm M_\odot} \right)^{-0.88} \left( {r_{\rm out} \over 10^{10} \rm cm} \right)^{2.65}  
~\rm g~s^{-1} 
\label{moir} 
\end{equation} 
where we neglect the very weak $\alpha$ dependence. The critical mass- 
transfer rate depends very strongly on the disc radius, so to obtain 
outside-in outbursts also for a low mass-transfer rate (as required by 
observations) one needs to have small discs. The problem is illustrated 
by the Figures \ref{oi1} and \ref{oi2} from Buat-M\'enard et al. (2001a). 
 
For the moment we are interested only in the lower panels of these 
figures; we will discuss the upper panels in Sect. \ref{recc}. Figure 
\ref{oi1} shows the radius at which the outburst begins as a function of 
the mass-transfer rate. The binary parameters are those of SS Cyg. Three 
model predictions are shown: the standard DIM, the DIM plus 
stream-impact heating and the DIM plus stream-impact and tidal-torque 
heating. It appears that Eq. (\ref{moir}) slightly underevaluates the 
value of the critical mass-transfer rate at which outbursts change the 
direction of propagation. The transition is quite neat: the inside-out 
outbursts all begin at $1.4 \times 10^9$ cm (very close to the inner 
radius fixed at $10^9$ cm) whereas outside-in outbursts begin from 2.14 
to $3.8 \times 10^{10}$ cm depending on the mass-transfer rate. This is 
very close to the outer disc radius. Heating of the external disc lowers 
the critical mass-transfer rate and causes the outside-in outburst to 
start closer to the outer edge. The reason is that additional heating of 
the outer disc regions lowers the value of $\Sigma_{\rm max}$ and 
$\Sigma_{\rm min}$ and shortens the distance between them as can be seen 
in Fig \ref{tistr}. This makes it easier for thermal instability to be 
triggered close to the outer disc radius, where the effects are 
strongest. It also stabilizes the disc for mass-transfer rates that are 
lower than in the standard DIM: in Fig. \ref{oi1} the highest 
mass-transfer models are close to the stability limit. 
 
Additional heating reduces the critical mass-transfer rate to 
$\dot{M}_{\rm io} = 1.8 \times 10^{17}$ g s$^{-1}$, which is 1.4 times 
smaller than in the standard case; this is reasonably close to estimates 
of SS Cyg mass transfer rate $\dot{M}_2 = 6 \times 10^{16}$ g s$^{-1}$ 
(e.g. Patterson 1984). An increase of this mass transfer rate by a 
factor 2 -- 3 is therefore sufficient to provoke outside-in outbursts.  
 
Short period systems require special attention. In two of these systems 
OY Car (Vogt 1983) and HT Cas (Ioannou et al. 1999), outside-in 
outbursts have been directly observed (see Sect. \ref{eclipses}). The 
mass-transfer rate in OY Car is $\lta 6 \times 10^{15}$g s$^{-1}$, (Wood 
et al. 1989b) and a similar or lower rate is expected in HT Cas and most 
of the other short period systems (Baraffe \& Kolb 1999). This requires 
very small discs radii if one wishes to obtain outside-in outbursts. 
Fig. \ref{oi2} from BMHL shows the transition between inside-out and 
outside-in outbursts for disc with $<r_{\rm out}>=1.3\times 10^{10}$ cm 
around a 0.8 M$_{\odot}$ white dwarf. Here the transition is at $\sim 
6\times 10^{15}$ g s$^{-1}$ for the standard DIM and slightly lower when 
heating by the stream is added. Dissipation of the tidal torques is not 
included as it should be negligible in a disc whose radius is well 
inside the tidal radius (see BMHL for details). In this case the 
transition between inside-out and outside-in is less marked than for 
larger discs. This is caused by the reduced extent of the disc ($r_{\rm 
out}/r_{\rm in}=13$) and the fact that the disc structure everywhere 
feels the influence of the boundary conditions (see BMHL). 
 
The disc radius in quiescence is $1.3\times 10^{10}$ cm in OY Car (Wood 
et al. 1989b) and $1.0\times 10^{10}$ cm in HT Cas (Horne, Wood \& 
Stiening 1991) so the results presented in Fig. \ref{oi2} are quite 
satisfactory, especially if one keeps in mind the simplistic way that 
stream heating is treated. The disc radius in OY Car was estimated when 
the system was in quiescence after a superoutburst and before a normal 
outburst. During the supercycle the mean disc radius radius increases 
and during a superoutburst reaches the 3:1 resonance radius (see Sect. 
\ref{tt}). It would be interesting to see if the type of 
normal outburst changes during the supercycle. The disc radius in the 
outburst observed by Vogt (1983) seems to be similar (Rutten et al. 
1992) to the one determined by Wood et al. (1989b). 
 
\begin{figure} 
\resizebox{\hsize}{!}{\includegraphics[angle=-90]{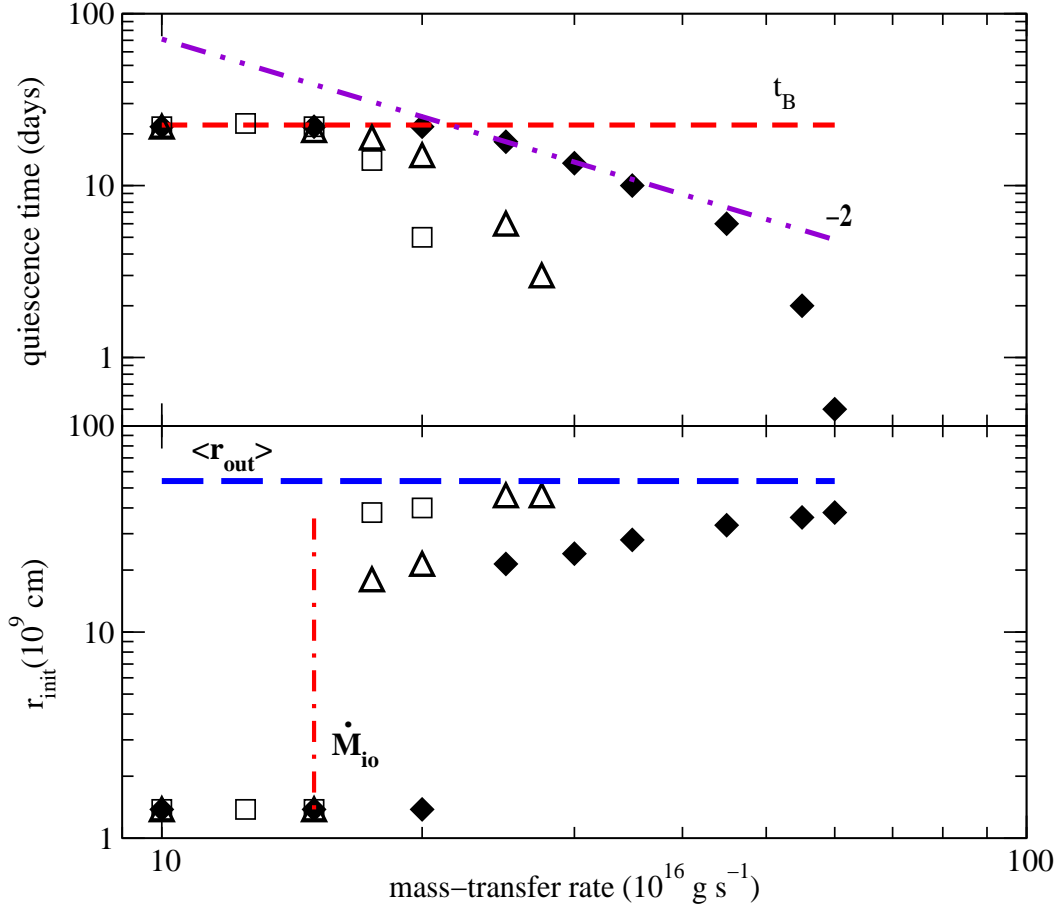}} 
\caption{Transition from inside-out to outside-in outbursts with 
increasing mass transfer rate. The mass of the accreting object is 
1.2M$_{\odot}$, the inner disc radius $10^9$ cm, the mean outer radius 
$5.4 \times 10^{10}$ cm, $\alpha_{\rm cold}=0.04$. The diamonds 
correspond to the standard model, up-triangles to the models that 
include stream-impact heating and squares to models that include both 
this effect and tidal torque dissipation. The upper panel shows the time 
spent in quiescence as a function of the mass transfer rate. The dashed 
line marked `$t_{\rm B}$' corresponds to Eq. (\ref{tb}). The lower panel 
shows the radius at which the heating front begins its propagation. The 
vertical dot-dashed line $\dot M_{\rm io}$ corresponds to Eq. 
(\ref{moir}) with $\delta=2$. The dashed line marked `$<r_{\rm out}>$' 
corresponds to the mean value of the outer disc radius.  
} 
\label{oi1} 
\end{figure} 
 
Ichikawa \& Osaki (1992) preferred to assume a special form of $\alpha$ 
in order to get rid of inside-out outbursts at low mass-transfer rates. 
It is possible that $\alpha$ has this desired form but since no 
physical argument is given to explain why and when such form should be 
assumed this solution does not seem to be very satisfactory. 
 
\subsubsection{Inside-out outburst in LMXBTs} 
 
LMXBTs have orbital periods longer than 4 hours. Their mass transfer 
rates are estimated to be less than $\sim 10^{16}$ g s$^{-1}$ (van 
Paradijs 1996; Menou et al. 1999b) so that according to Eq. (\ref{moir}) 
in all cases one can expect only inside-out outbursts. This is confirmed 
by the models (Menou et al. 2000; DHL).

Because of an observed `X-ray delay' the 1996 outburst of GRO J1655-40 
had the reputation of being of the outside-in type (Orosz et al. 1997). 
However, the successful model of the rise of this outburst by Hameury et 
al. (1997b) clearly shows an inside-out outburst. Because in this model 
the disc is truncated the outburst starts rather far from the black hole 
(and `outside' the inner `hole') but at the inner edge of the cold disc, 
and a heating front propagates {\sl outwards}. In this model the inner 
hole is filled by an ADAF (Abramowicz et al. 1995; Narayan \& Yi 1995). 
So if this model is correct this should be an inside-out outburst. A 
3-day delay between $I$ light and X-rays was observed at the rise to 
outburst of the neutron-star LMXBT Aql X-1 in August 1997 by Shahbaz et 
al. (1998). These authors call this event an `outside-in' outburst, but 
if the same reasoning as for GRO J1655-40 is applied to this system it 
is clear that what they observed was an inside-out outburst.  
 
\begin{figure} 
\centering 
\resizebox{\hsize}{!}{\includegraphics[angle=-90]{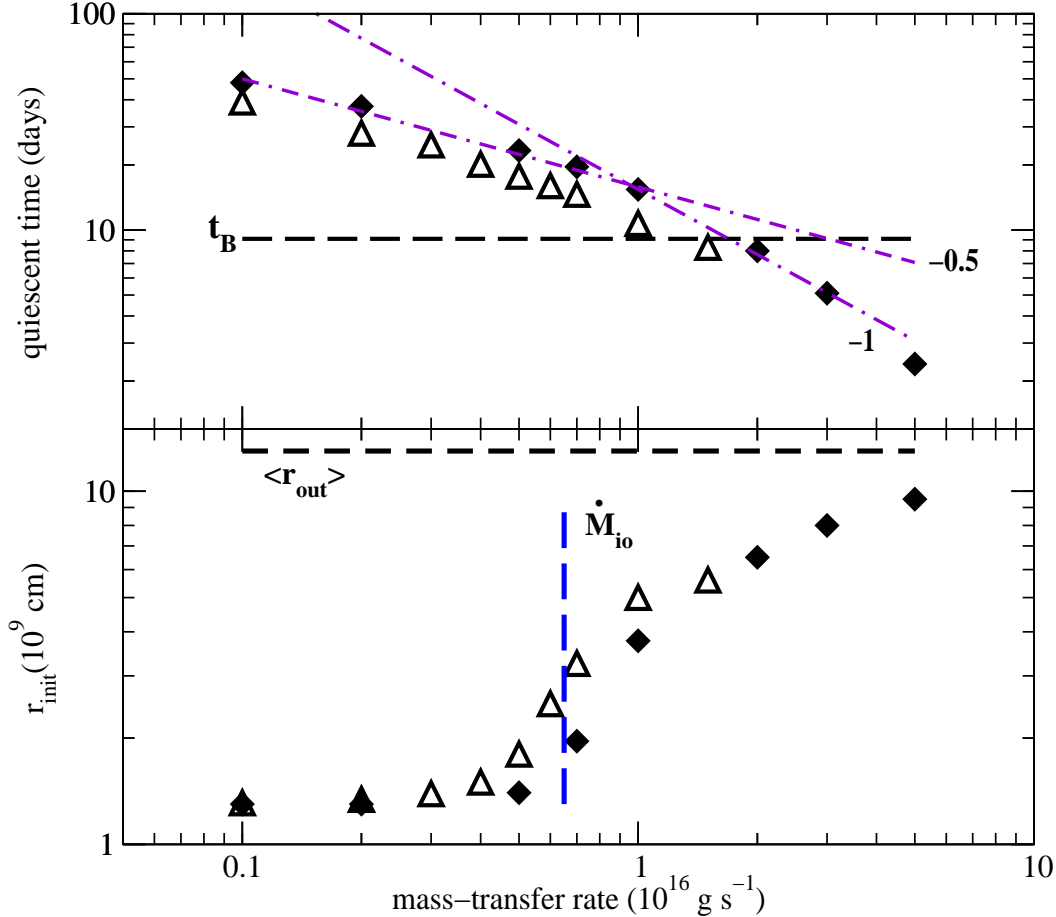}} 
\caption{The same as in Fig. \ref{oi1} but for 0.8M$_{\odot}$, and  
$<r_{\rm out}>=1.3 \times 10^{10}$ cm.  
} 
\label{oi2} 
\end{figure} 
 
This leads to the question: how can we tell from observations that an outburst 
is outside-in or inside-out? Three indicators have been used in the 
literature: 
\begin{itemize} 
\item{} The evolution of the eclipse profile during the outburst 
\item{} The shape of the outburst's lightcurve 
\item{} The length of the ``UV-delay" 
\end{itemize} 
 
We will now discuss these three methods. 
 
\subsubsection{Eclipse profiles} 
\label{eclipses} 
 
This method is the surest but by its nature is limited only to a 
minority of systems. It can also be useful, however, as a test for the 
other two methods, which can be applied, in principle at least, to all 
systems. Observations of eclipsing systems show where the outburst 
begins. Already in 1971 Smak observed an outburst of U Gem that began at 
the outer disc rim. A normal (i.e. not a superoutburst) of OY Car was 
also seen to start at the outer disc (Vogt 1983; Rutten et al. 1992). 
The same is true of a normal outburst of HT Cas (Ioannou et al. 1999), 
whereas outbursts of IP Peg (Webb et al. 1999) and EX Dra (Baptista, 
Catal\'an \& Costa 2000) were seen to begin close to the inner disc 
edge. Webb et al. (1999) confirm in this case that a slow rise usually 
associated with inside-out outbursts (see below) is indeed observed in 
the IP Peg outburst, whose type is determined from the eclipse profiles. 
 
\subsubsection{Shapes} 
 
Sequences of inside-out and outside-in outbursts are shown respectively  
in left and right panels of Fig. 
\ref{ref1}. One can see that the two types 
of outbursts differ in the shapes of their light-curves. The inside-out 
outburst light-curves are fairly symmetrical: the rise is slow since at the 
outburst's beginning only a small surface and little mass are implicated in 
the process. We will call such a shape a `type-B' shape. The outside-in 
outbursts are asymmetrical: the rise is faster than the decay and there is a 
flat-top part just after the maximum. This shape is characteristic of 
outbursts in which a large surface and most of the disc's mass take part 
in the outburst from the outset. We will call such a shape a `type-A' 
shape.  
 
This nomenclature relating to light-curve shapes is introduced to 
distinguish between the shape and the direction in which the outburst 
propagates, because in some conditions inside-out outbursts may have 
also `type-A' shapes. This is possible when the inner disc is truncated. 
Such truncated discs seem to be necessary in LMXTBs (see Sect. 
\ref{sxt}) and they are often invoked in the context of dwarf novae and 
CVs in general. In particular if the inner disc is truncated at a 
sufficiently large radius an {\sl inside-out} outburst will have a {\sl 
`type A'} shape as seen in Fig. \ref{ssrise}, where such an outbursts is 
shown (solid lines). In this model the inner disc radius is variable, 
and its value is determined by an evaporation law (slightly different 
from Eq. (\ref{evap}), see Hameury et al. 1999 for details). In 
quiescence it is equal to $3.5\times 10^{9}$ cm, i.e to about 7 white 
dwarf radii. The outburst starts near the inner disc radius and 
propagates only outwards as there is nothing to propagate into in the 
other direction. Fig. \ref{ssrise} shows also two other models of 
outbursts for a disc extending down to the white dwarf surface. The 
inner part of such a disc is irradiated by the white dwarf (dotted line) 
and by both the white dwarf and the boundary layer between the disc and 
the white dwarf. The inner part of such a disc remains hot in 
quiescence, as suggested by King (1997), but contrary to King's 
suggestion the DIM the predicts an optically thick accretion disc. Smak 
(1998) describes these outbursts as having rounded shapes typical of 
`type-B'. However, a look at Fig. 7 in Hameury et al. (1999) shows that 
this is true only of the light-curves representing the secondary 
outbursts which appear when the inner disc is irradiated.  
\begin{figure} 
\centering 
{\includegraphics[scale=0.75,totalheight=12cm,angle=-90]{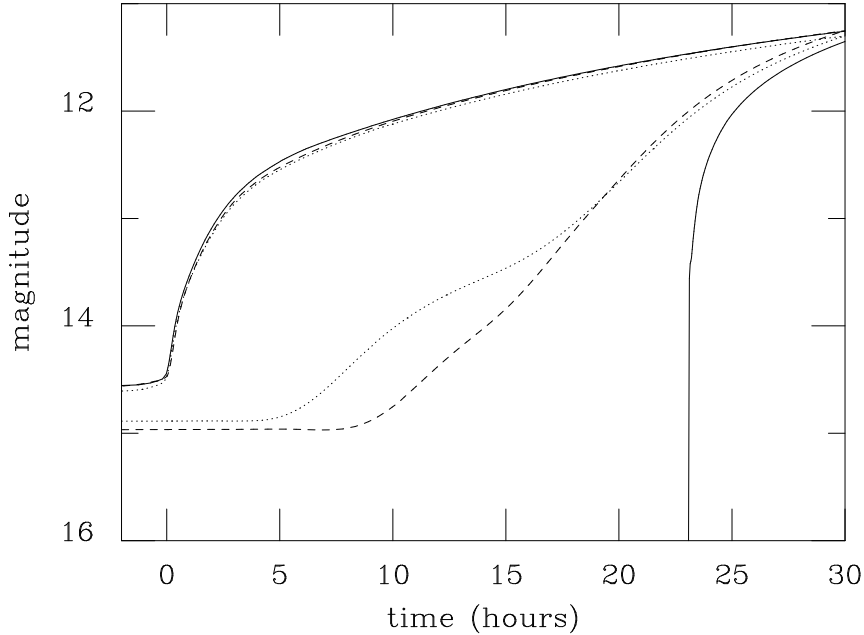}} 
\caption{Visual magnitudes (the three almost superimposed curves on the 
left) and accretion rate ``magnitudes", $m_x=27-\log \dot M_{\rm in}$ 
(the three curves on the right), for a system with 1.2 M$_{\odot}$ white 
dwarf, $R_*=5\times 10^8$ cm, $<r_{\rm out}>=4\times 10^{10}$ cm, $\dot 
M_{\rm tr}= 1.7 \times 10^{17}$ g s$^{-1}$, white dwarf temperature 
$T_*=30000K$, $\alpha_{\rm cold}=0.02$, $\alpha_{\rm hot}=0.1$. The solid curves 
correspond to a disc truncated by evaporation, the dashed line to a full 
disc irradiated by the white dwarf and accretion, and the dotted line to 
a full disc irradiated only by the white dwarf. The disc albedo is $50\%$. 
From Hameury, Lasota \& Dubus (1999).  
} 
\label{ssrise}  
\end{figure} 
 
Fig. \ref{sl}, which will be discussed in more detail in Sect. \ref{nw} 
shows another example of ambiguous shapes. Despite their different 
shapes, both types of outbursts are inside-out. A closer look shows that 
for both types of outbursts the rise-time is the same, but neither of 
them could be described as having a `rounded shape typical of type B 
outbursts', as Smak (1998) says about Fig. \ref{ssrise} (and Fig. 7 in 
Hameury et al. 1999). Also Fig. \ref{rvar} shows two types of inside- 
out outbursts (see Sect. \ref{zoo}) neither of them really `rounded'. 
 
Of course the use of the term `rounded shape' may be a question of taste 
(as when applied to human bodies) if it is not quantified. In view of 
the variety of observed and calculated outburst shapes it is not clear 
that such a quantification would make much sense. Already in 1934 
Campbell classified outbursts of SS Cyg into four types. However, each 
types contains, a variety of shapes (see Fig. 3.6 in Warner 1995a). In 
any case:``no two outbursts in a given [dwarf nova] are exactly alike" 
(Warner 1995a). As we will see in the next section, the `UV-delay' also 
does not determine the outburst type unambiguously. 
 
\subsubsection{The ``UV-delay"} 
\label{uvdel} 
 
In several dwarf-nova systems whose outbursts were observed 
simultaneously in the UV (or EUV) and in visual light, one could see 
a delay between the rise to outburst at short wavelengths and 
that at long wavelength, the visual-light flux was rising first (see e.g. 
Table 2. in Smak 1998 and reference therein). Such a delay is expected 
by the DIM (Smak 1984b). There sometimes an ambiguity in the meaning 
of the `UVs'. One certainly expects a delay in EUVs but practicaly none 
in the near UVs) . In the case of outside-in outbursts the heating 
of the outer disc regions is reduced as Fig. \ref{sma} shows, but since 
a large emitting area is available a significant increase in visual flux 
occurs very early. The UV flux will rise only when the temperature grows 
sufficiently high, which corresponds to the heating front arriving at 
the inner disc. This is especially true of the EUV flux, which can be 
emitted only very close to the white dwarf (in the `boundary layer'). In 
the case of inside-out outbursts things are more complicated. The 
heating front now forms close to the white dwarf but there, as discussed 
in the previous Section, the accretion rate is rather low at first. Also 
the emitting area is small. It is only after the arrival of the heating 
front at the outer disc that the accretion rate onto the white dwarf 
increases significantly. Then, however, the outer disc is already hot, so 
that also in this case the visual-light flux may begin to increase before the 
rise in the UV and a UV-delay can be expected. The observed UV-delays 
are between $\sim 0.5$ to $\sim 1$ day. The same range of UV-delays can 
be obtained in the standard DIM (Smak 1998). 
 
It would seem, therefore, that there is no ``UV-delay problem" in the 
DIM. It might then appear rather strange that this subject gave rise to 
many articles, debates and has been considered as one of the main 
difficulties plaguing the standard DIM. The reason was, as pointed out by 
Smak (1998), that most of the papers claiming that there is a UV-delay 
problem were using disc models with {\sl constant} and {\sl too small} 
outer disc radii. Such models produce a problem just by their unphysical 
assumptions. According to Smak there should be no UV-delay problem if 
one uses correct boundary conditions and large enough discs (large 
enough to correspond to radii expected from observations) and he 
produced a set of models to substantiate his point. However, for systems 
with large discs Smak's models did not reproduce the UV-delay {\sl and} 
the outside-in character of the outbursts. Thus the UV-delay deserves 
more discussion.

Let us look at the problem in more detail. First, the definition of the 
UV-delay is ambiguous, at least in practice. In principle one is 
concerned with the delay between the {\sl rise} to outburst in visual 
and UV (or EUV) light; the delay in the rise and not in the maximum. It 
is easy to understand that in practice it is often difficult to decide 
when the rise to outburst in a particular wavelength begins. To see this 
it is enough to have a look at the beautiful (and important) EUV and 
optical observations of SS Cyg by Mauche and his collaborators (Mauche 
1996; Mauche, Raymond \& Mattei 1995). In Mauche (1996) two outbursts of 
SS Cyg are shown. An `anomalous' 1993 outburst and a type A, 1994 
outburst. The 1993 outburst has a typical `B' shape and is anomalous in 
its length of rise to maximum: $\sim 5$ days. Because of this very slow 
rise it is difficult to determine the UV delay, or even if there is one 
at all. In Mauche et al. (1995) a 3-day delay is mentioned, but the most 
prudent statement by Mauche (1996), that the EUV `rose more quickly that 
the optical', seems more appropriate. There is no doubt about the 1994 
outburst: its typical A-type steep rise in both wavelengths shows a 
clear UV delay of $\sim$ 1 day. Therefore if one identifies `type A' 
with `outside-in' and `type B' with `inside-out', as Mauche does in his 
1996 article, everything seems to be in order: a delay in the 
outside-in, (almost) no delay in the `inside-out'. 
 
This is the problem posed by models presented in Smak (1998). Smak 
compares observations with his models, in which UV delays can be as long 
as 2 days. One of these observations is attributed to Mauche (1996) with 
a quoted (E)UV delay of 1 day. Smak's model no. 16 gives a very similar 
delay, but this model corresponds to an inside-out outburst. The 1994 
outburst, however, has the typical type-A shape attributed to outside-in 
outbursts. Therefore, although Smak is obviously right about the fact 
that correct models with large disc radii produce very long UV delays, 
he does not seem to be able to produce the right type of outburst for a 
system like SS Cyg. Indeed, he does not have models for type A outbursts 
for orbital periods larger than 4.2 hours. For this period one obtains 
an type A outburst for an outer radius of $4.2\times 10^{10}$ cm and an 
accretion rate of $1.3 \times 10^{17}$ g s$^{-1}$. The orbital period 
and radius are too short for SS Cyg but the mass-transfer rate is 
already too high. Hence, to produce type A outbursts in SS Cyg would 
require unreasonably high mass-transfer rates, as shown in Fig. 
\ref{oi1}, where (for $<r_{\rm out}>=5.4\times 10^{10}$ cm) outside-in 
outbursts appear only for $\dot M_{\rm tr} \gta 2.5 \times 10^{17}$ g 
s$^{-1}$. This figure also shows that taking into account heating by the 
accretion-stream impact and tidal forces alleviates the problem and 
allows outside-in outbursts for $\dot M_{\rm tr} \gta 1.8 \times 
10^{17}$ g s$^{-1}$ (Buat-M\'enard et al. 2001a). This is still high, but 
if the alternating sequence of type A and B outbursts in SS Cyg is due 
to mass-transfer fluctuations, an increase of mass-transfer rate by a 
factor of 2-3 would allow type A outbursts.

Of course the mass-transfer rate in SS Cyg is not really known, and one 
could decide that since type A outbursts are observed, the mass-transfer 
rate must have whatever value is necessary for their occurrence. 
However, as mentioned above, a `type A' shape is not in itself proof 
that the outburst is outside-in. There is enough disc surface available 
for the optical flux to rise quickly, whereas the EUV flux will start to 
rise much later, when the inner disc arrives close to the white dwarf's 
surface, which happens in viscous time.

Finally, it appears that information about (E)UV and optical light 
curves is not sufficient as test a of the DIM. It must be complemented 
by hard X-ray observations. Recently Wheatley, Mauche \& Mattei (2000) 
observed a SS Cyg outburst simultaneously in X-rays, EUV and optical. 
The outburst had the typical, asymmetrical type-A shape and there was a 
{\sl bona fide} EUV delay. However, hard (RXTE) X-rays began to rise 
hours before the EUV. They were shut out when the EUV started to rise. 
This multiwavelength observation is not easy to reconcile with the 
`standard' picture in which type-A outbursts are always of the 
outside-in type; it is a rather natural outcome of the truncated disc 
model. In this case the outburst rising close to the inner disc edge 
results in an enhancement of the accretion rate, which in turn increases 
the evaporation rate and thus the hard X-ray emission from the hot inner 
accretion flow.

\subsection{Recurrence time} 
\label{recc} 
  
The accumulation time defined by Eq. (\ref{acc}) is the time spent in 
quiescence between outside-in outbursts. One must make the difference 
between the recurrence time (from onset to onset)  and time spent in a  
quiescent state (from the end of an outburst to the onset of the next 
one), especially when the outburst duration is a significant part of the 
recurrence time. For typical parameters this time is 
\begin{eqnarray} 
t^{\rm A}_{\rm quiesc}  
\approx 0.5 \left(\frac{\alpha_{\rm cold}}{0.01}\right)^{-1} 
\left(\frac{\dot M_{\rm 
tr}}{10^{17}{\rm g\ s^{-1}}}\right)^{-2}&& \nonumber \\ 
\left(\frac{T_c}{3000 {\rm 
K}}\right)^{-1} &&\left(\frac{M}{\rm M_{\odot}}\right)^{- 
1.26}\left(\frac{r}{10^{10}\rm cm}\right)^{5.8} \ {\rm d}. 
\label{ta} 
\end{eqnarray} 
The quiescence time obtained with this formula is a very good estimate 
of the quiescent duration in the outside-in outburst cycle shown in Fig. 
\ref{ref1}. 
 
Eq. (\ref{ta}) is valid as long as $t_{\rm accum} < t_{\rm diff}$ 
(estimated at the outer disc radius). When the mass-transfer rate is too 
low or the outer radius too large for this inequality to be satisfied 
the quiescent time will be given by the diffusion time defined by 
Eq.~(\ref{diff}). This is the quiescence time for an inside-out outburst 
because it is the time it takes to build up a surface-density excess 
(with respect to the quiescent $\Sigma$) that will reach the critical 
surface density. Now the relevant radius is not the outer disc radius 
but the radius where a density excess forms. Since the diffusion time 
decreases with distance to the center, this radius will be, in general, 
close to the inner disc edge. From Eq. (\ref{diff}) one obtains a 
quiescence time (between inside-out outbursts) 
\begin{equation} 
t^{\rm B}_{\rm quiesc} \approx 130 \delta \left(\frac{\alpha_{\rm cold}}{0.01}\right)^{-1} 
\left(\frac{T_c}{3000 {\rm K}}\right)^{-1} \left(\frac{M}{\rm M_{\odot}}\right)^{0.5} 
\left(\frac{r}{10^{10}\rm cm}\right)^{0.5} {\rm d}. 
\label{tb} 
\end{equation} 
 
This time is {\sl independent} of the mass-transfer rate as one should 
expect of a diffusion time. Fig. \ref{oi1} illustrates this property of 
the quiescence time of inside-out outbursts (see also Ichikawa \& Osaki 
1994). Figure \ref{oi2} shows, however, that Eq. (\ref{tb}) is only an 
approximation, valid only for discs with $r_{\rm out}/r_{\rm in} \gg 1$. 
In this case most of the disc's structure is affected by the boundary 
conditions and the assumptions used to derive Eq. (\ref{tb}) do not 
apply. We will discuss this in more detail in Sect. \ref{vlrt}. 
Outbursts of irradiated discs in LMXBTs also fail, in general, to follow 
Eq. (\ref{tb}) as discussed in detail in Sect. \ref{sxt}. 
 
Menou et al. (2000) used a different formula for the quiescence time 
(recurrence time in their case, since in LMXBTs they consider the duty 
cycle is very short), using the effective instead of the midplane 
temperature. Their formula is an upper limit for the quiescence time 
because the factor $\zeta^2\delta$ is not taken into account. This 
corresponds rather to a uniform growth of surface-density. In the case 
of small truncated discs where the $\Sigma(r)$ profile is flat, 
especially if they were irradiated during outburst (see Fig. 
\ref{reclc}), a formula without the `slope' factor $\zeta^2\delta$ can 
be useful for comparisons with model calculations. The drawback of the 
formula used by Menou et al. (2000) is its very strong dependence on the 
effective temperature.  
 
\subsubsection{Very long recurrence times} 
\label{vlrt} 
 
For the parameters of U Gem-type dwarf novae, quiescence times given by 
Eqs. (\ref{ta}) and (\ref{tb}) are very close to those calculated in the 
models and, luckily, close to the observed ones. Problems arise when one 
wishes to make a DIM model of long recurrence times systems as, for 
example WZ Sge (Smak 1993; Osaki 1995a). The recurrence time for this 
system is $\sim 30$ years. In this case the difference between 
quiescence and recurrence time is not important. It is clear from Eq. 
(\ref{tb}) that the DIM will give such long recurrence time only if 
$\alpha_{\rm cold}$ is much lower than the $\sim 0.01$ used in the 
standard approach; the value of the viscosity parameter would have to be 
$\lta 10^{-4}$ to give a recurrence time of about 30 years. This 
conclusion about the very low value of $\alpha$ is confirmed by a second 
argument (Smak 1993; see also Hameury et al. 1997) which we will now 
present. 
 
The mass in the disc can be estimated if we assume that the quiescent disc 
just before the outburst is filled up to the critical surface density. 
The mass (the maximum disc mass) is  
\begin{equation} 
M_{\rm D, max}=2.7 \times 10^{21} \alpha^{-0.83} \left( {M_1 \over \rm  
M_\odot} \right)^{-0.38} \left( {r \over 10^{10} \; \rm cm} \right)^{3.14}  
\ {\rm g}, 
\label{diskmass} 
\end{equation} 
(see Eq. (\ref{sigmax})). 
The mass transfer rate in WZ Sge is estimated to be $\sim 2\times 
10^{15}$ g s$^{-1}$, a value expected at such a short orbital period 
($\sim$ 80 min). The outer disc radius is $\sim 10^{10}$ cm and the 
white-dwarf mass $\sim 0.5$ M$_{\odot}$ (see Lasota, Kuulkers \& Charles 
1999 and references therein). The quiescence (`recurrence') time can be 
defined as the time it takes, after an outburst, to refill the disc up 
to the critical level. Therefore: 
\begin{equation} 
t_{\rm recc}\approx \frac{\epsilon M_{\rm D, max}}{\dot M_{\rm tr}- 
\dot M_{\rm in}}= 
2.5 \left(\frac{\alpha_{\rm cold}}{0.01}\right)\left(1 - \frac{\dot M_{\rm 
in}}{\dot M_{\rm tr}}\right)^{-1} \ {\rm y} 
\label{treccu} 
\end{equation} 
where the fraction of the disc's mass lost during outburst is $\epsilon= 
\Delta M_{\rm D}/M_{\rm D, max}$ and  $\dot M_{\rm in}$ is the accretion 
rate at the disc's inner edge. 
 
Since in the standard DIM $\dot M_{\rm in} \ll \dot M_{\rm tr}$, getting 
the observed recurrence time for WZ Sge requires very small viscosity 
parameter in quiescence: $\alpha_{\rm cold} \sim 4.5 \times 10^{-4} 
\epsilon^{1.2}$. In the standard case $\epsilon \sim 0.1$, so that the 
required value of $\alpha_{\rm cold}$ ($\sim 3 \times 10^{-5}$) would be 
very small indeed.  
 
WZ Sge is very special: it shows only superoutbursts. Its superoutbursts 
are also special: they last more than a month whereas usual 
superoutbursts of SU UMa systems last for about two weeks. Once more 
this could suggest very low $\alpha$'s in quiescence. The duration of 
the outburst can be estimated as (Hameury et al. 1997) 
\begin{equation} 
t_{\rm dur}\approx \frac{\epsilon M_{\rm D, max}}{\dot M_{\rm in, max}- 
\dot M_{\rm tr}}= \epsilon 
2 \left(\frac{\alpha_{\rm cold}}{0.01}\right)\left(1 - \frac{\dot M_{\rm 
tr}}{\dot M_{\rm in, max}}\right)^{-1} \ {\rm d} 
\label{tdur} 
\end{equation} 
where we took $\dot M_{\rm in, max}=10^{18}$ g s$^{-1}$ (Smak 1993). 
Therefore, to obtain a duration of $\sim 30$ days, one needs $\alpha_{\rm cold} < 
8 \times 10^{-4} \epsilon^{1.2}$, if $\dot M_{\rm in, max} \gg \dot 
M_{\rm tr}$. Finally, an equivalent argument shows that since for 
$\alpha_{\rm cold}=0.01$ the maximum mass of the disc is $M_{\rm D, max}\approx 
1.6 \times 10^{23}$ g while $\gta 10^{24}$ was accreted during the 
outburst, either $\alpha_{\rm cold} < 8 \times 10^{-4}$ or mass was added to the 
disc during the outburst. This last possibility means in practice that 
during the outburst the mass-transfer rate increased to a value close to 
$\dot M_{\rm in, max}$, i.e. by more than two orders of magnitude. 
 
WZ Sge outbursts pose two problems: the recurrence time and the outburst 
duration are much too short when calculated in the standard DIM. Both 
problems can be solved by assuming that for this system the viscosity 
parameter in quiescence is much lower than in other dwarf-nova discs. We 
will discuss the quiescent disc viscosity problem in Sect. \ref{quidn}. 
Even with these small $\alpha$'s not all problems are solved: one must 
still explain the unusual duration of the (super)outburst and 
its `tremendous' amplitude. These two properties of WZ Sge cannot be 
explained by the DIM. The tidal-thermal instability model (see Sect. 
\ref{tt}) is more successful (Osaki 1995a, 1996). 
 
There is another possibility. As suggested by Hameury et al. (1997, see 
also Lasota, Hameury \& Hur\'e 1995) the disc in WZ Sge could be 
truncated at a radius sufficiently large to make what is left of the 
disc stable or marginally stable. This, however, would solve only part 
of the problem, because although this would allow us to keep `normal' 
$\alpha$'s in quiescence, the outburst itself would be a flop 
(see Warner, Livio \& Tout 1996). And no tidal-thermal process would 
help because there is simply not enough mass in the disc. If one wishes 
to keep not extravagantly low $\alpha$'s one {\sl must} add mass to the 
disc {\sl during outburst} as proposed by Hameury et al. (1997). 
 
The secondary star in WZ Sge is a very low-mass cold star ($\lta 2000$) K  
but during outbursts it gets heated up to $\sim 20000$ K (Smak 1993). The 
irradiation flux, $\gta$ 1000 times larger than the star's intrinsic flux must 
influence the rate at which matter is transfered. We will discuss the 
secondary's irradiation effects in Sect. \ref{zoo}. Here it suffices to 
mention that the increase of mass-transfer rate by a factor of 100 that  
would be required in WZ Sge is not unreasonable as shown in Hameury et 
al. (2000). 
 
It is difficult to decide at present what is the solution of the WZ Sge 
puzzle. On the one hand the fact that in WZ Sge $ \dot M_{\rm tr} t_{\rm 
recc} \sim 2 \times 10^{24}$ g is equal to the mass accreted during 
outburst speaks strongly in favour of the DIM, and hence in favour of 
very low $\alpha$'s (as pointed out by A.R. King - private 
communication); on the other, there are reasons independent of the 
dwarf-nova character of WZ Sge for supposing that its quiescent disc is 
not an accretion disc at all, but a magnetically controlled (by the 
white dwarf's magnetic field) flow (Lasota, Kuulkers \& Charles 1999). 
In such a case the quiescent ``disc" would be stable (as originally 
proposed by Lasota et al. 1995). The outbursts would occur only when an 
upward fluctuation of the mass-transfer rate reconstructs the disc.  
The idea that dwarf nova systems become stable 
close to the CV's minimum period (thus ceasing to be dwarf novae) may 
help to solve the discrepancies between evolutionary model predictions 
and observations (Baraffe \& Kolb 1999). Models predict a CV minimum 
period at $\sim 70$ min while the observed minimum is at $\sim 80$ min. 
Also an excess of systems with the minimum orbital period is expected 
but not observed. Mass-transfer rates at short periods ($\sim 10^{15} 
$ \gs) should be such that all CVs harbouring full fledged accretion 
discs should be unstable with respect to the dwarf-nova instability. 
Such faint systems can be identified only if they undergo substantial 
outbursts if not they would be `invisible'.

Because of the very long recurrence times of LMXBTs a similar problem 
arises. Similar but not the same, because in this case there is enough 
mass in the disc to account for the mass accreted during the outbursts. 
Also disc irradiation allows for longer recurrence times. We will 
discuss this problem in Sect. \ref{sxt}. 
 
\section{The generalized DIM} 
 
\subsection{Bimodality of outburst duration: a solution} 
\label{nw} 
 
As mentioned at the beginning of Sect. \ref{dn}, outburst durations 
(widths) fall into two classes: short (narrow) and long (wide). 
(Superoutbursts are different from long outbursts, e.g. Warner 1995a; 
Smak 2000, and will be discussed in Sect. \ref{tt}). The bimodality is 
best seen in SS Cyg. It is interesting that the article by Bath \& van 
Paradijs (1983) in which this effect was analyzed was devoted to the 
support of the mass-transfer instability model and to a critique of the 
DIM. Nowadays the bimodality of SS Cyg and co. outbursts is still seen 
as the DIM's weakness, and mass-transfer fluctuations are called at its 
rescue (Smak 1999a, 2000, Hameury et al. 2000). However, nobody any 
longer believes in the mass-transfer instability model. The bimodality 
was observed in over a dozen of U Gem and Z Cam type stars (Warner 1995a 
and references therein). It seems that no bimodality of {\sl normal} 
outbursts was observed in SU UMa stars (the exception being TU Men, 
which has the longest orbital period among SU UMa stars: the only such 
star above the `period gap'). 
 
As mentioned before, the main deficiency of the DIM is its inability to 
reproduce a sequence of alternating narrow and wide outbursts with 
roughly the same amplitude. If one could superimpose the two panels of  
Fig. \ref{ref1} 
one would get something close to the required result. Of 
course these figures cannot be superimposed; they represent models in 
which the mass-transfer rate differs by an order of magnitude. Varying 
the mass-transfer rate by a factor of 10 would not give the right 
result. But mass-transfer variations could help, as shown by Smak 
(1999a). 
 
The i1dea is to repair what the middle panel of Fig. \ref{comp_bc} shows: 
the narrow outbursts result from heating fronts unable to reach the 
outer disc regions. In his numerical experiments Smak used a factor of 
2, instant enhancement of the mass-transfer rate. As a result he 
obtained a sequence of outbursts of the same amplitude but also of the 
same duration. To obtain wide outbursts Smak (1999a) used `major' mass- 
transfer rate enhancements which brought the disc to a steady state. We 
will see that a variant of this idea explains the standstills of Z Cam 
stars. It could also produce SS Cyg type wide outbursts but the width of 
the outburst would be given by a `clock' in the mass-losing star. There 
is nothing wrong with such a solution: after all, the duration of Z Cam 
stars standstills is supposed to be determined by such a clock. But here 
a different solution is possible. In fact several solutions, all 
involving mass-transfer fluctuations, are possible. 
\begin{figure} 
\resizebox{\hsize}{!}{\includegraphics{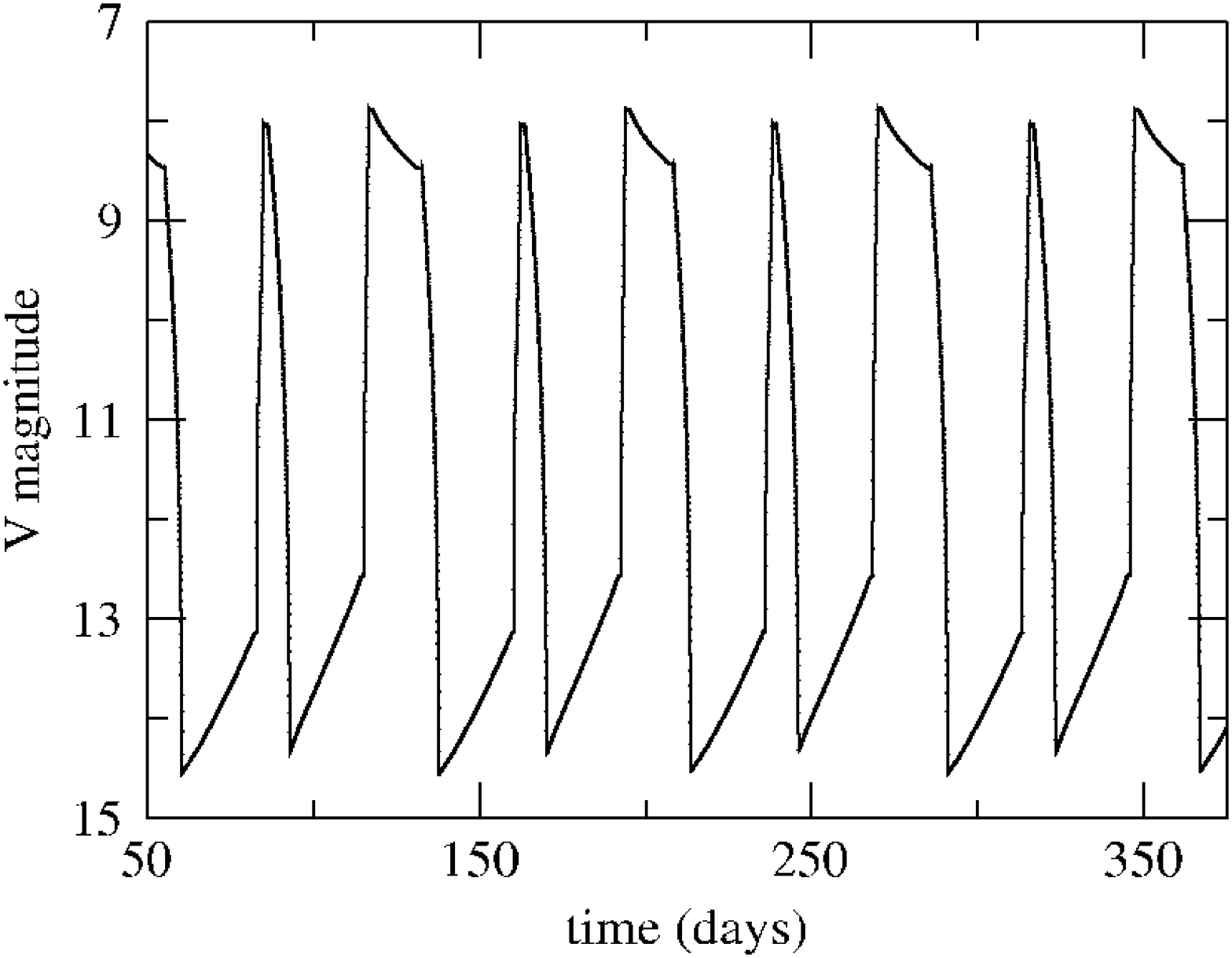}}
\caption{The visual light curve for the DIM plus tidal dissipation 
(without stream impact) with $M_1$ = 1.2 M$_\odot$, $\dot{M}_{\rm tr}  
= 1.5 \times 10^{17}$g s$^{-1}$, the average disc radius $5.4  
\times 10^{10}$. Both narrow and wide outbursts are inside-out 
($\dot M_{\rm A}\lta 2.4 \times 10^{17}$ g s$^{-1}$, see Eq. \ref{moi}).} 
\label{sl} 
\end{figure} 
 
We shall see in Sect. \ref{zoo} how mass-transfer fluctuations treated 
in a slightly more sophisticated way than Smak's may give narrow and 
wide outbursts without assuming enhancements to bring the disc to a 
steady state. 
 
Here, however, we will present a different although related model. BMHL 
showed that including the heating of the outer disc by tidal torque 
dissipation and by the stream impact naturally produces a sequence of 
narrow and wide outbursts for a rather narrow range of mass-transfer 
rates $\dot M_{\rm SL} (P_{\rm orb})$ , which depend on the orbital 
period. They are close to the values dwarf novae are supposed to have 
above the period gap ($\gta 3$ hr) but much larger than the secular 
mass-transfer rates of short period systems such as SU UMa stars. At 
$P_{\rm orb}=2$ hr $\dot M_{\rm SL}\approx 3\pm 0.5\times 10^{16}$ g 
s$^{-1}$ and it increases to $\dot M_{\rm SL}\approx 1\pm 0.5\times 
10^{17}$ g s$^{-1}$ at $P_{\rm orb}=7$ hr. 
 
Fig. \ref{sl} shows an example of a sequence of narrow and large 
outbursts for a system with SS Cyg parameters; $M_{\rm SL}=1.5\times 
10^{17}$ g s$^{-1}$. Only tidal-torque dissipation is taken into 
account. Including heating by stream impact reduces $M_{\rm SL}$ by 
$\sim 15\%$. Outer disc heating helps inside-out propagating 
heating-fronts to reach the outer rim. As explained in Sect. 
\ref{fronts} heating fronts get reflected when they cannot bring the 
surface density above $\Sigma_{\rm min}$. Heating by the stream and 
tidal torques lowers the value of this critical surface density, which 
allows the fronts to reach the outer rim and all the outbursts have a 
decent maximum brightness. The width of the outburst depends on the 
mass-transfer rate. For $\dot M_{\rm tr}< \dot M_{\rm SL}$ only narrow 
outbursts are present while in the opposite case only wide outbursts 
appear.

Long outbursts are slightly brighter than the short ones; a property 
observed in real systems (Oppenheimer et al. 1998). Near the peak of the 
outburst, the outer radius increases and with it the tidal dissipation. 
The tidal dissipation lowers the value of $\Sigma_{\rm min}$ so it takes 
longer for the disc to empty down to the critical density where a 
cooling front will start to propagate shutting off the outburst. 
Outbursts therefore last longer than in the standard case and have the 
characteristic ``flat top" shape. 
 
Narrow outbursts are always of the inside-out type, whereas wide 
outbursts can be of either type because the critical mass-transfer rate 
above which they occur $\dot{M}_{\rm SL} < \dot{M}_{\rm oi}$. In Fig. 
\ref{sl} all outbursts are inside-out, but if moderate (factor 2--3) 
fluctuations of the mass-transfer are present, even at $\dot{M}_{\rm tr} 
= \dot{M}_{\rm SL}$ outbursts will be of both propagation-direction 
types because $\dot{M}_{\rm SL}$ is usually very close to $\dot{M}_{\rm 
oi}$. For example for the binary parameters of SS Cyg $\dot{M}_{\rm 
SL}\approx 1.3 \times 10^{17}$ \gs while $\dot{M}_{\rm oi}\approx 1.8 
\times 10^{16}$ \gs. 
 
Including the heating by tidal-torque dissipation and stream-impact 
heating in the DIM may therefore solve one of the main `unsolved 
problems' (Smak 2000) of this model: the origin of the bimodal 
distribution of the outburst widths.  
 
van Paradijs (1983) found a correlation between outburst width and the 
orbital period. The BMHL models reproduces this correlation for short 
outbursts  but this is true for everybody . Their duration is 
independent of the mass transfer rate and depends on the disc's size 
(i.e. on the orbital period) and $\alpha_{\rm hot}$ (Smak 1999b). It is 
more difficult to reproduce this alleged correlation for long outbursts, 
since their duration is quite sensitive to $\dot{M}_2$, whose variation 
with the orbital period is not really known. And it is not well 
established, as (i) the sample is small, and (ii) the correlation may 
result in part from the very definition of ``long" outbursts: they must 
last longer than short outbursts, whose duration does increase with the 
orbital period (see above).  
 
Moderate mass-transfer fluctuations play only a auxiliary role here: 
they are necessary to produce a sequence of inside-out and outside-in 
outbursts. A model combining both the additional heating effects and 
mass-transfer fluctuation has still to be calculated, but it will 
certainly give the desired result. Of course, one must always keep in 
mind that the way additional heating (especially by the stream impact) 
is included into the DIM is rather rough.

\subsection{Z Cam stars} 
\label{zcams} 
 
\begin{figure} 
\centering 
{\includegraphics[scale=0.75,totalheight=10cm,angle=-90]{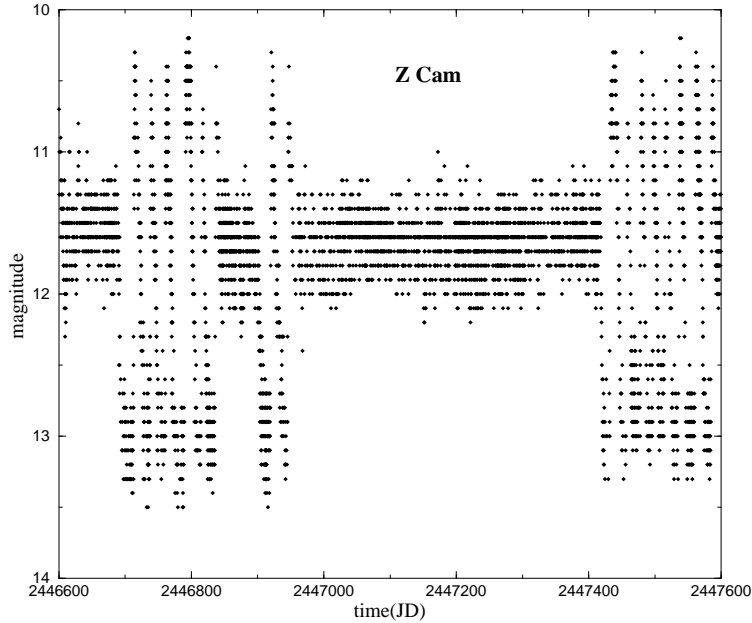}} 
\caption{The light curve of Z Cam. Data from AAVSO} 
\label{zclc} 
\end{figure} 
Z Cam stars are characterized by a `standstill' phenomenon: the decline 
from normal outburst maximum is interrupted and the luminosity of the 
systems settles down to a value $\sim 0.7$ mag lower than the peak 
luminosity (see Fig. \ref{zclc}. In some cases the magnitude difference 
is smaller (see Warner 1995a). Such standstills may last from ten days to 
years. After that the system luminosity declines to the usual quiescent 
state. Osaki (1974) interpreted standstills as stable phases of 
accretion in the framework of his disc instability model. In his 
disc-radius and mass-transfer rate diagram, Smak (1983) described these 
stars as an intermediate case between stable nova-like stars and 
unstable dwarf novae. Meyer \& Meyer-Hofmeister (1983b) proposed that Z 
Cam stars are dwarf novae with a mass transfer rate that fluctuates 
about the critical rate. Also Lin et al. (1985) concluded that such 
moderate fluctuations of the mass-transfer rate can produce Z Cam-type 
light curves. Despite this early and generally accepted diagnosis no 
model has been able to reproduce the observed light curves of Z Cam stars. 
The latest attempt by King \& Cannizzo (1998) was not very successful 
(for reasons explained below). 
\begin{figure} 
\centering 
{\includegraphics[scale=0.75,totalheight=12cm]{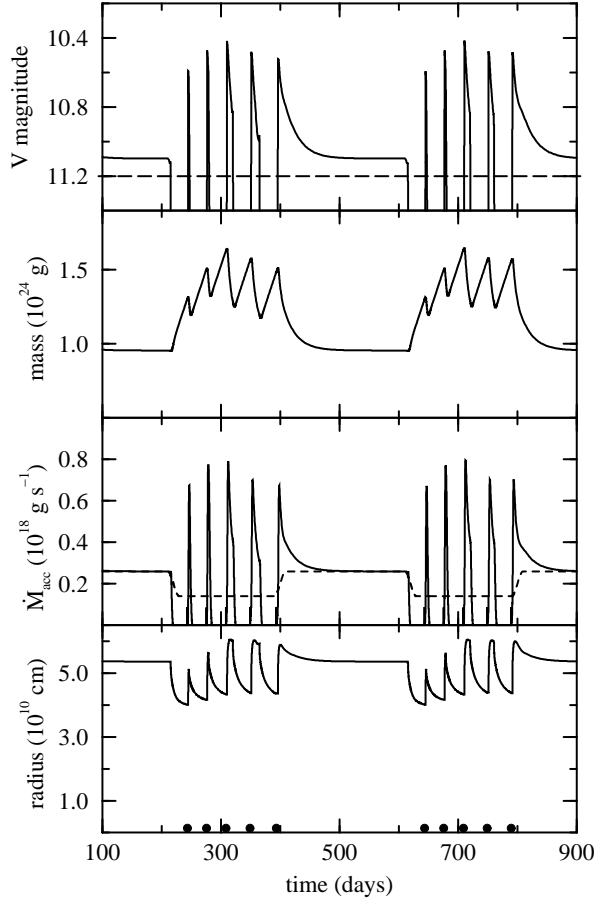}} 
\caption{Z Cam modelled by the DIM including stream impact and 
tidal-torque dissipation heating for $\dot{M}_{\rm tr} = 2.0 \times 10^{17} 
\pm 30 \%$g s$^{-1}$. The upper panel shows the visual magnitude (solid 
line). The dashed line represents the V magnitude (11.2) of a disc which 
accretes at exactly the critical value: $\dot{M}_{\rm B} = 2.0 \times 
10^{17}$g s$^{-1}$. The next panel gives the disc mass and the third 
represents the mass accretion rate onto the white dwarf (solid line), as 
well as the mass transfer rate from the secondary (dashed line). The 
bottom panel shows the variations of the outer disc radius. The dots 
show points where the instability is triggered at the onset of an 
outburst. All outbursts are of the inside-out type. From Buat-M\'enard 
et al. (2001b)} 
\label{zcam} 
\end{figure}

Buat-M\'enard, Hameury \& Lasota (2001b) showed that in this case the 
outer-disc heating by  stream impact and dissipation of the tidal 
torques play an crucial role. Although as they show, it is possible to 
obtain Z Cam-type light curves in the framework of the standard DIM, the 
parameters required to obtain this result are not realistic; getting 
the often observed $\Delta_{\rm mag} \gta 0.7$ would require extreme 
fine tuning of the parameters or might not be possible at all.  
 
Additional heating helps by lowering the critical parameters of the 
system. Fig. \ref{zcam} shows the model applied to Z~Cam. The results 
are very satisfactory. One obtains a magnitude difference between peak 
and standstill $0.6 < \Delta_{\rm mag} < 0.8$ mag for a fluctuation of 
30 \% around $<\dot{M}_{\rm tr}> = 2.0 \times 10^{17}$g s$^{-1}$. Having 
the outer disc heated by the stream impact and tidal torques bestows a 
bonus: narrow and wide outbursts which are observed in Z Cam 
(Oppenheimer et al. 1998). The duration of the narrow outbursts is 8 -- 9 
days, of the wide outbursts 16 -- 20 days. The recurrence time is 32 
days between narrow outbursts and 40 days between wide outbursts. These 
compare quite well with observed values: the average duration of the 
narrow outbursts is 10 days, of the wide outbursts 17. The average 
cycle length of narrow and wide outbursts is 23 and 31 days 
respectively.  
 
This is not all. When one calculates Z Cam-type light curves with the
standard DIM and requires the mass-transfer rate fluctuations to be
{\sl  moderate} one always obtains {\sl outside-in} outbursts. The
reason is  obvious: the mean mass-transfer rate has to be close to the
stability  limit, i.e. it is so high that only outside-in outbursts
can take place  (see Fig. \ref{oi1}) as e.g. in RX And. However, Z Cam
stars show also inside-out (type B)  outbursts e.g. in Z Cam and AH
Her(Warner 1995a, Table 3.6)!  This can be understood as the result of
stream-impact and  tidal-torque heating of the outer disc lowering not
only the values of   mass-transfer rates critical for stability -
$\dot M_{\rm B}$ and for outburst type - $\dot{M}_{\rm AB}$
(Fig. \ref{oi1}) but also the value of $(\dot{M}_{\rm B} -
\dot{M}_{\rm oi})/\dot{M}_{\rm B}$ which makes possible also
inside-out outbursts during the unstable phase of Z Cam systems even
for moderate fluctuations of the mass-transfer rate.  In Fig.
\ref{zcam} all the outbursts are of the inside-in type, which tallies
with observations of Z Cam and AH Her.  

The conclusion that in Z Cam $\dot{M}_{\rm oi}$ is close to
$\dot{M}_{\rm B}$ has interesting implications for the relation
between Z Cam and U Gem stars (Osaki, private communication). Indeed,
also  for SS Cyg  the two critcal accretion rates should be close  and
the transfer rate close to $\dot{M}_{\rm oi}$, so that a small
increase of the {\sl average} mass-trasnefr rate should transform SS
Cyg into a Z Cam stars. This would be a confirmation of the old
suspicion (Warner 1995a) that ``all U Gem stars are unrecognized Z Cam
stars".
 
Therefore, modeling Z Cam-type outbursts by varying the mass-transfer 
rate around the value critical for disc stability is very successful 
when effects of heating by the stream-impact and tidal torque 
dissipation are included in the DIM. In Fig. \ref{zcam} the outburst 
preceding the standstill is longer than observed which suggests that 
mass-transfer variations do not follow the prescription assumed 
in this particular model and the mass-transfer increase should begin
around 5 days after the outburst maximum.
(see Buat-M\'enard et al. 2001b). 
 
One should note that the range of parameters producing standstills is 
rather restricted. This is because close to the disc's outer edge 
non-local effects of the outburst bring the surface density very close 
to the critical value $\Sigma_{\rm min}$ (see Fig. \ref{sma}), i.e. 
$\dot M$ close to $\dot{M}_{\rm B}$. Since at maximum the accretion rate 
in the disc is roughly constant $\dot{M}_{\rm max} \sim \dot{M}_{\rm 
c}$. In Esin, Lasota \& Hynes (2000) it is assumed that $\dot{M}_{\rm 
max} = \dot{M}_{\rm B}$ and deduced that an enhancement of mass transfer 
above $\dot{M}_{\rm B}$ will always produce a standstill brighter than 
the outburst maximum. This is not true because in reality the peak 
accretion rate is larger by a factor of about 2. Therefore, King \& 
Cannizzo (1998), who obtained `standstills' brighter that the outburst 
maximum, guessed correctly that they had increased the mass-transfer 
rate by an amount too large to get a Z Cam-type standstill (they 
increased the mass-transfer rate by a factor 6).  

\subsection{Irradiation and fluctuations} 
\label{zoo} 
 
It seems that some of the basic outburst properties of U Gem and Z Cam 
type dwarf novae can be reproduced by the DIM if effects of outer disc 
heating are taken into account and if moderate fluctuations of the mass- 
transfer rate occur during the outburst cycle. These fluctuations can be 
due to various effects. The stochastic fluctuations leading to the 
bimodal distribution of outburst duration and heating-front propagation 
direction could be due to stellar spots (e.g. Livio \& Pringle 1994). This 
could also be the origin of the fluctuations generating Z Cam behaviour. 
Not much is known about these stellar spots. Stellar spots might be 
responsible for the low states of AM Her and other magnetic systems 
(Hessman, G\"ansicke \& Mattei 2000) but contrary to the assertion by 
Schreiber et al. (2000) mass-transfer fluctuations in dwarf-nova systems 
cannot be of the same type, as shown by BMHL. 
 
Another type of mass-transfer fluctuation (enhancement) is due to 
irradiation of the secondary during the dwarf-nova outburst, as argued by 
Smak (1996,1999a,2000 and references therein; see also Hameury et al. 
1997). Secondary stars in dwarf-nova binaries are observed to be 
irradiated by the accretion flow (the irradiating source is usually 
called the ``boundary layer") and the resulting enhancement of the 
mass-transfer rate is observed as a brightening of the `hot spot' where 
the stream of transferred matter hits the disc rim (Smak 1996).  
Also Warner (1998) has suggested that a large variety of observed dwarf-nova  
outbursts could be explained by the DIM if this model included 
effects of irradiation of both the secondary and the inner 
disc.  
 
The importance of the irradiating flux depends on the binary parameters. 
Smak (2000) plotted the ratio of the mean irradiating to the secondary's 
intrinsic flux $<F_{\rm irr}>/ F_2$ for several dwarf novae with well 
established properties. For dwarf novae not showing superoutbursts this 
ratio is less than 10, while dwarf novae with superhumps (which we 
deliberately do not call `SU UMa stars' for reasons that will be clear in a 
moment) have much larger values $<F_{\rm irr}>/ F_2 >20$. For WZ Sge the 
ratio may be larger than $10^3$ (Smak 1998). It is therefore clear 
that in superoutburst systems irradiation of the secondary must play a 
crucial role.  
 
U Gem, the prototypical dwarf nova, exhibited a very  
long (45 days)large amplitude outburst in 1985 (Mattei et al. 1987; 
Mason et al. 1988).  
Applying Eq. (\ref{tdur}) to U~Gem, one gets the maximum duration  
of an outburst in this system: 
\begin{equation} 
t_{\rm max} = 26 \left( {\alpha_{\rm cold} \over 0.01}\right) ^{-0.83} M_1^{0.51}  
\left( {r_{\rm out} \over 4.1 \; 10^{10} \rm cm} \right)^{0.47} \; \rm d 
\end{equation} 
Contrary to WZ Sge, U Gem shows mainly normal outbursts with a 
recurrence time of 100 days, so that $\alpha_{\rm cold}$ cannot be 
smaller than 0.01; in fact it should be greater (e.g. Livio \& Spruit 
1991). Therefore, $t_{\rm \max}$ could never be as high as 45 days. This 
means that the total amount of mass accreted during this very long 
outburst is larger than the mass of the disc in quiescence; this is 
possible only if the mass transfer rate from the secondary has increased 
to a value close to the mass accretion rate onto the white dwarf (see 
Sect. \ref{vlrt}). Such an increase of the mass transfer rate is very 
likely caused by irradiation, since U Gem has a borderline ratio 
$<F_{\rm irr}>/ F_2 \approx 20$ (Smak 2000). 
 
The large outburst of U Gem is not usually called a `superoutburst', 
because it did not exhibit the `superhump' characteristic of 
superoutburst light-curves of SU UMa stars. The period of superhumps is 
a few percent longer that the orbital one. Superhumps are believed to 
result from disc deformations due to a 3:1 resonance between the disc 
flow and the binary motion (Whitehurst 1988, Whitehurst \& King 1991; 
Hirose \& Osaki 1990; Lubow 1991,1994). This resonance is possible only 
if the mass ratio (secondary/primary) is $\lta 0.33$ (see e.g. Frank et 
al. 1992). For U Gem $0.46\pm 0.03$ (Friend et al. 1990) so no superhump 
is expected, nor was one observed.  
 
Is a superhump a necessary ingredient of a superoutburst? Osaki (1989, 
see 1996 for a review) and his collaborators answer `yes'. They 
generalized the DIM, adding a {\sl tidal} instability to the original 
thermal thus creating the tidal thermal instability model (TTI) which 
would apply to SU UMa stars. In the TTI the mass-transfer rate is 
assumed to be constant during the whole cycle of normal outbursts and 
superoutbursts. Vogt (1983) and Smak (1984c), however, suggested that 
superoutbursts result from mass-transfer rate enhancements. Osaki (1985) 
proposed a model according to which these enhancements are due to 
irradiation of the secondary by the accretion flow. In these models the 
superoutburst was supposed to be due an irradiation-triggered 
instability of the {\sl secondary}. Hameury et al. (1986) showed that 
this not possible in a binary in which the accreting body is a white 
dwarf. The later evolution of the respective opinions on this subject is 
quite interesting: Osaki devised the TTI model while Hameury and the 
author of this review studied the role of irradiation of the secondary 
in various classes of dwarf nova stars (e.g. Hameury et al. 2000). 
 
In the next section we will discuss SU UMa stars and superoutburst 
models.

\subsection{SU UMa stars} 
\label{tt} 
 
SU UMa stars are dwarf novae having superoutbursts. This is the traditional 
definition; Warner (1985a) says that the current ``operational 
definition" of SU UMa stars involves the further requirement that they exhibit  
superhumps. He adds that if dwarf novae are ``ever found with 
superoutbursts lacking superhumps they will define a class of their 
own". Therefore U Gem is not an SU UMa star, which is rather reassuring. 
With one exception all the SU UMa stars have orbital periods of less than 
2.1 hours, i.e. below the CV period gap. The exception is TU Men whose 
period is very close to the upper edge of the period gap. Superoutbursts 
are $\sim 0.7$ mag brighter than normal outbursts; their light-curve at 
maximum has a form described as `flat-top' or `plateau'. Superoutbursts 
last 5--10 times longer than normal outbursts. 
 
In order to produce such superoutbursts in the standard DIM, two conditions must 
be fulfilled: 
\begin{itemize} 
\item{} There must be enough mass in the disc for the outburst to last 
        the required time 
\item{} The cooling-front propagation must be withheld for most of the 
        outburst duration to allow more mass 
\end{itemize} 
 
\subsubsection{Tidal-thermal instability} 
\label{tti} 
 
These two conditions cannot be satisfied in the standard DIM. Osaki 
(1989), Ichikawa et al. (1993) (see Osaki 1996 for a review) satisfied 
this them by having a sequence of normal outbursts during which only a 
very small fraction of the disc mass is accreted onto the white dwarf. 
The disc mass therefore increases together with the disc outer radius. 
Once the outer radius reaches a critical radius $r_{\rm critical}=0.46 
a$, the tidal torque is increased by a factor of 20. This amounts to 
changing the value of $c\omega$ in Eq. (\ref{eq:defT}). Therefore, in 
the TTI model the superoutburst is due to a `tidal instability'. The 
critical radius is the 3:1-resonance radius and the `tidal 
instability' is somehow related to the instability responsible for the 
superhump. The factor by which the tidal torque is increased is a free 
parameter, as is the radius at which the torque changes back to its 
normal value. In usual SU UMa stars this radius is assumed to be 0.35 
$a$, but in order to explain the properties of the rapid system RZ LMi, 
Osaki (1995b) has to assume that the tidal instability stops when the 
radius has shrunk by less than 10\%. In the TTI model with the standard 
$\alpha$ prescription according to which the viscosity parameters in the 
hot and cold states are constant normal outbursts are always inside-out 
for mass-transfer rates typical of SU UMa stars (see Sect. \ref{io}). 
Osaki and his collaborators prefer, however, to have outside-in normal 
outbursts and they use a radius-dependent $\alpha_{\rm cold}$ (see 
below).  
 
The TTI model is different from the standard DIM in two 
respects: it introduces a new free parameter whose value depends on the 
system to which the model is applied, and it modifies the viscosity 
prescription.  
 
Vogt (1981) and Warner (1987, see also 1995a,b) found a correlation 
between the recurrence times of normal outbursts, $t_{\rm N}$ and  
superoutbursts $t_{\rm S}$: 
\begin{equation} 
t_{\rm S} \propto t_{\rm N}^{0.5}, 
\label{tstn} 
\end{equation}  
which means that the number of normal outbursts in a supercycle ($t_{\rm 
S}/t_{\rm N}$) is inversely proportional to its duration. Of course 
$t_{\rm S} \geq t_{\rm N}$. According to Warner (1995b) $660 {\rm d} \leq t_{\rm 
S}=t_{\rm N} \leq 1450$ d, whereas observations suggest the absence of 
normal outbursts for $t_{\rm N} \gta 400$ d. Eq. (\ref{tstn}) reflects, 
therefore, an important property of the supercycle mechanism. 
 
Osaki (1994,1995a,1996) argues that Eq. (\ref{tstn}) has a simple 
interpretation in the framework of his tidal-thermal DIM. If the 
recurrence time for normal outbursts is given by Eq. (\ref{ta}) then 
$t_{\rm N}\propto 1/\dot M_{\rm tr}^2$. Since superoutbursts are 
supposed to occur when sufficient mass is accumulated in the disc 
$t_{\rm S}\propto 1/\dot M_{\rm tr}$ and Eq. (\ref{tstn}) follows. It 
was on this argument that Osaki based his ideas about the ``SU UMa/WZ 
Sge connection" and in general about the relation between the 
mass-transfer rate and the dwarf-nova outburst type (1995, 1996). One 
should note, however, that this argument can also be used in favour of 
models in which superoutbursts are not produced by a tidal viscosity 
enhancement. It is sufficient for the normal outburst recurrence time to 
satisfy Eq. (\ref{ta}) and for the superoutbursts to result from mass 
accumulation (Hameury et al. 2000). 
 
This apparently convincing argument, however, is based on very strong 
assumptions which, when looked at more closely, are not satisfied by the 
model. First, Eq. (\ref{ta}) is satisfied only by outside-in outbursts 
and as we saw in Sect. \ref{io} it is very difficult to have this type 
of outburst for small discs and low mass-transfer rates typical of SU 
UMa stars. Osaki uses a form of the viscosity parameter $\alpha_{\rm 
cold}\propto \left(r/R_{\rm tidal}\right)^{0.3}$ specially designed to 
suppress inside-out outbursts (Ichikawa \& Osaki 1992). However, in 
general, the relation $t_{\rm N}\propto 1/\dot M_{\rm tr}^2$ is 
satisfied only for outside-in outbursts in systems with large discs (U 
Gem-type) as clearly seen in Figs. \ref{oi1}, \ref{oi2}, which considerably 
weakens the argument.  
 
Second, and more important, Osaki's argument assumes that recurrence 
times depend only (or mainly) on the mass-transfer rate. This is clearly 
not the case. It is clear that the recurrence time depends also on the 
viscosity parameter, as it is evident in Osaki's explicit assumption 
about $\alpha_{\rm cold}$. In addition one should expect the recurrence 
times to depend on the binary parameters. In fact, Menou (2000a) showed 
that the recurrence times of both normal outbursts and superoutbursts 
are strongly correlated with binary component mass ratio. We will 
discuss his interpretation of this correlation in Sect. \ref{quidn}. 
Here it suffices to say that the existence of this correlation casts 
serious doubt on the interpretation of the $t_{\rm S}/t_{\rm N}$ 
correlation as reflecting a mass-transfer dependence of the recurrence 
times. 
\begin{figure} 
\centering 
{\includegraphics[scale=0.75,totalheight=12cm]{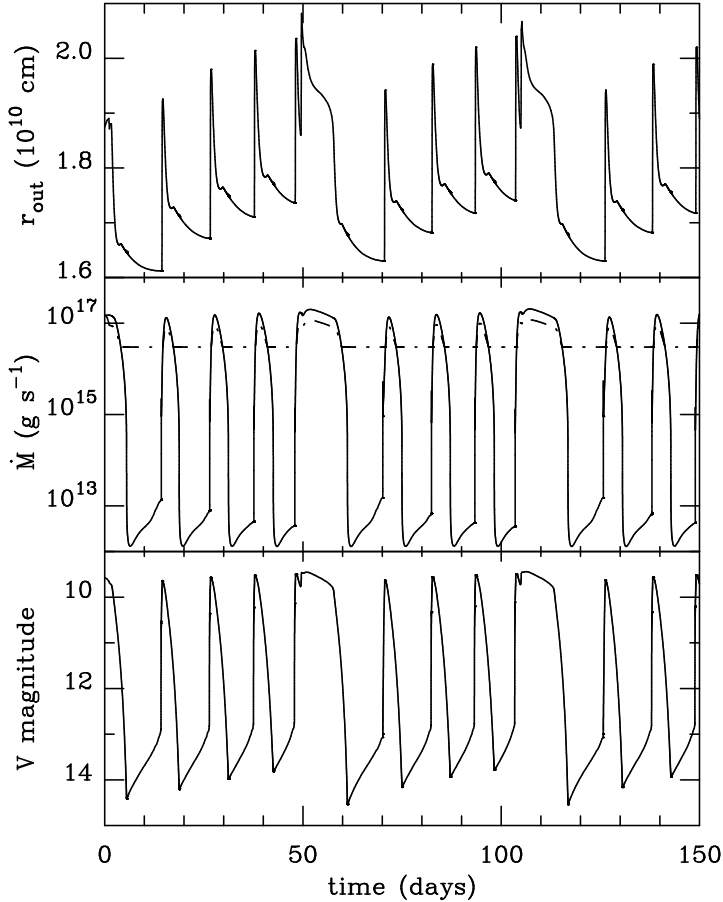}} 
\caption{The predicted light curve when the secondary is irradiated. 
The parameters are $M_1$ = 0.6 M$_\odot$, $\dot M^0_{\rm tr} =3 \times 
10^{16}$ g~s$^{-1}$, the average disc radius $1.9 \times 10^{10}$ cm and 
the irradiation parameter is $\gamma$ = 0.5. The upper panel shows the 
outer disc radius, the middle panel shows the accretion rate onto the 
white dwarf (solid line) and the mass transfer rate from the secondary 
(dashed line); the lower panel shows the visual magnitude of the disc. 
(Hameury et al. 2000)} 
\label{illsec} 
\end{figure} 
 
But the main deficiency of the TTI model is that it neglects a property 
of SU UMa stars and dwarf-novae outbursts in general: the {\sl observed} 
enhancement of the mass-transfer rate during outbursts (Smak 
1991,1995,1996,2000; Vogt 1983). This enhancement must change the 
sequence of events because it strongly influences the size of the disc. 
As indicated by Smak (1996) the weakest point of the TTI model is the 
sequence of events it predicts in the superoutburst: everything begins 
with the tidal instability leading to the formation of an eccentric disc 
and causing a major enhancement of the mass-transfer rate whereas the 
superhump is observed to appear well after the beginning of the 
superoutburst. Smak (1996,2000) therefore suggested that both the 
mass-transfer rate enhancement and the tidal instability could be 
combined into a ``hybrid" model in which they would play a role at 
different phases of the superoutburst. The superoutburst would start 
with a major enhancement of the mass-transfer rate due to irradiation by 
a preceding normal outburst. The resulting contraction of the disc would 
screen the secondary from irradiation, thus decreasing the mass-transfer 
rate. This in turn would cause the disc expansion and bring the disc 
into the 3:1-resonance radius. Only then would the tidal instability 
enter the game. This is an attractive scenario but the `hybrid' model 
might be slightly different. For example it is not clear that shrinking 
the disc changes the secondary's irradiation.  
 
Hameury et al. (2000, hereafter HLW) presented a series of numerical 
experiments in which irradiation of the secondary and other non-standard 
effects were examined. Their results make it easier to understand what 
the hybrid model of superoutbursts could look like. 
 
The response of the secondary to sudden irradiation is rather complex 
(see e.g. Hameury et al. 1997) and HLW preferred to use a simpler 
approach in which a linear relation between the mass transfer rate from 
the secondary $\dot{M}_{\rm tr}$ and the mass accretion rate onto the 
white dwarf $\dot{M}_{\rm acc}$ is assumed: 
\begin{equation} 
\dot{M}_{\rm tr} = \max (\dot{M}_0,  \gamma \dot{M}_{\rm acc}) 
\label{eq:ill_sec} 
\end{equation} 
where $\dot{M}_0$ is the mass transfer rate in the absence of 
irradiation. Although it is an extremely crude approximation, it has at 
least the advantage of having only one free parameter $\gamma$. Its 
value must be in the range $[0-1]$ for stability reasons. Such an 
approach assumes that irradiation of the secondary can increase the 
mass-transfer rate by a substantial amount. As discussed by HLW, in 
principle irradiation during dwarf nova outbursts could bring the mass- 
transfer rate up to $10^{18}$ \gs, i.e. increase it by 3 orders of 
magnitude. Meyer-Hofmeister, Meyer \& Liu (1998) expressed doubts about 
the possibility of the increase of about 300 required in the model of 
of WZ Sge by Hameury et al. (1997), but such values are not excessive 
considering the very large irradiating fluxes (see above). 
\begin{figure} 
\resizebox{\hsize}{!}{\includegraphics{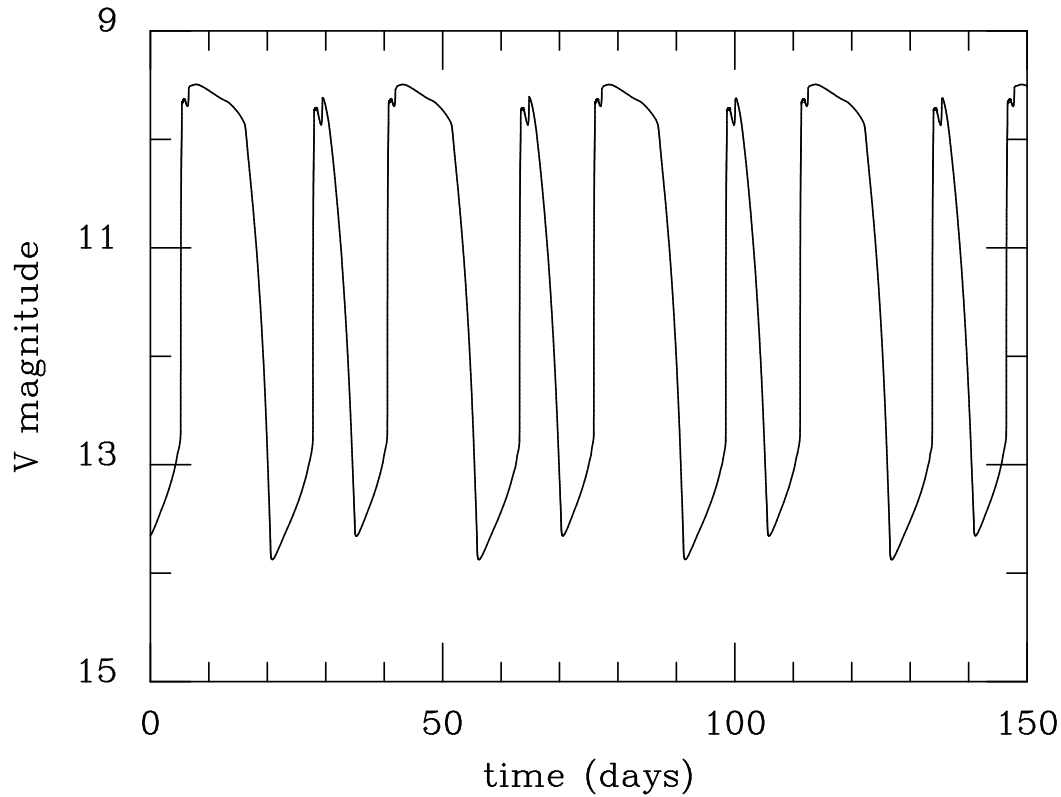}} 
\caption{Lightcurve obtained when the inner disc is truncated at a 
radius given by Eq. \ref{rmag} where $\mu_{30}=2$. The 
other parameters are $\gamma$ = 0.5, $\dot{M}_0$ = $4 \times 10^{16}$ 
g~s$^{-1}$, $M_1= 1 {\rm M_{\odot}}$. 
} 
\label{rvar} 
\end{figure} 
 
Figure \ref{illsec} shows a light curve (lowest panel) similar to those 
observed in SU UMa stars. In the model $\gamma=0.5$ was assumed. The 
uppermost panel shows that during the supercycle (between two large 
outbursts) the disc radius increases. Smak (1991,1996) studied the 
observed disc radius evolution in two SU UMa stars: Z Cha and OY Car, 
the only two systems for which sufficient data is available. In Z Cha 
observations suggest that the disc radius is actually decreasing, but 
the data scatter is rather large. In observations of OY Car no trend can 
be seen. The radius increase is one of the fundamental elements of the 
the TTI model. The difference between the HLW and TTI models is that in 
the former the disc radius varies during normal outbursts (because of 
the mass-transfer rate enhancements) whereas in the latter it is almost 
constant. In the HLW model during a large outburst the disc expands 
after an initial contraction and can reach the 3:1--resonance radius. So 
in this model only the superhump, {\sl not} the superoutburst, would be 
due to a tidal instability. It worth noting that the initial contraction 
of the disc followed by an expansion to a larger radius could postpone 
the appearance of the superhump. Smak (1991) also considered the effect 
of mass-transfer rate enhancements in a SU UMa star, but obtained a 
different radius behaviour. The reason for this difference is that while 
HLW assume that $\dot{M}_{\rm tr}$ depends on $\dot{M}_{\rm acc}$ (Eq. 
\ref{eq:ill_sec}), in Smak's model $\dot{M}_{\rm tr}$ is increased by 
about one order of magnitude after the maximum of a normal outburst, 
during a fixed time. As a consequence, at the end of a large outburst 
HLW obtain a much smaller radius than Smak in his model. In HLW, when 
the cooling wave starts propagating, the accretion rate onto the white 
dwarf, and hence the mass transfer from the secondary, is unaffected, 
and the disc contracts as in the unilluminated case, whereas in Smak's 
model, the disc expands rapidly when mass transfer is reduced by a 
factor of 10; the surface density at the outer edge then drops below the 
critical value, and a cooling wave starts in quite a large disc. 
\begin{figure} 
\resizebox{\hsize}{!}{\includegraphics[angle=-90]{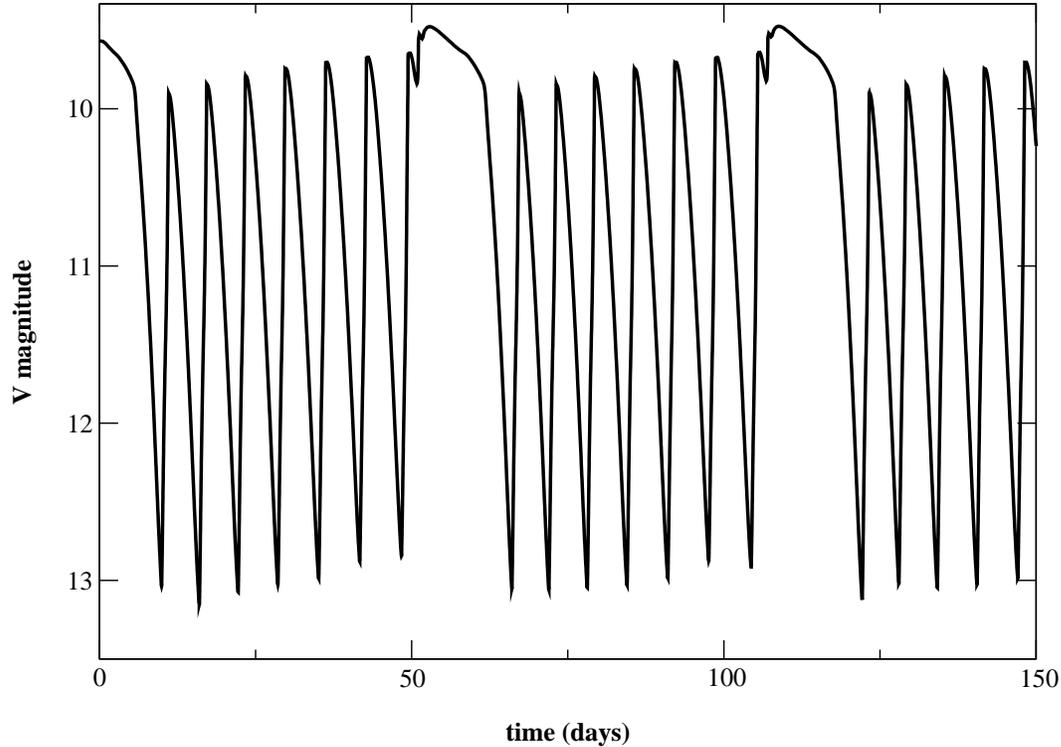}} 
\caption{Lightcurve obtained when the inner disc is irradiated 
by a 35000 K white dwarf, other parameters are the same as  
in Fig. \ref{rvar}. 
} 
\label{rzlmi} 
\end{figure} 
 
Fig. \ref{illsec}, however, does not represent a model of an SU UMa 
star: the mass transfer rate, $3 \times 10^{16}$ \gs is too large for 
such a system. For lower mass-transfer the model fails to produce large 
outbursts. The reason is that too much mass is lost during normal 
outbursts: the enhanced mass-transfer rate increases the burst's 
duration. This failure of the model is due to the linear relation 
between the accretion and the mass-transfer rates. This relation was 
assumed for the sake of simplicity but in real systems things are more 
complicated. One can nevertheless expect that Fig. \ref{illsec} 
represents a more general class of models. 
 
Fig. \ref{rvar} and \ref{rzlmi} show models in which the secondary is  
irradiated and in addition the inner disc is either truncated or 
irradiated by a hot white dwarf. The binary parameters and the mass- 
transfer rates are the same in both cases.

Fig. \ref{rvar} shows a different way of obtaining sequences of `narrow' 
and `wide' outbursts, similar to what was suggested by Smak (1999a), 
the difference being once more that in the present case there is a 
relation between the mass-transfer and accretion rate while in Smak's 
case the mass transfer rate is independent of the accretion rate. 
 
When the disc extends down to the white dwarf but is irradiated as 
suggested by King (1997) one can obtain very curious light-curves (HLW). 
Fig. \ref{rzlmi} is very reminiscent of the light curves of a subclass 
of frantic SU UMa stars: the ER UMa stars whose supercycles are very 
short. They range from 19 to 45 days (instead of the usual 130-160 days) 
(see e.g. Kato et al. 1999 and references therein). These systems are 
expected to have higher mass-transfer rates than SU UMa stars, so in 
this case $4\times 10^{16}$ \gs could be the appropriate mass-transfer 
rate. Osaki (1995b,c) obtained almost identical light-curves in the 
framework of the TTI model. But in the case of RZ LMi, the system with the 
shortest supercycle he had to assume a significantly lower 
strength of the tidal torques. HLW, on the other hand, use a very simple 
irradiation law. Perhaps also in this case a hybrid model is the answer.

So far so good: with the help of additional ingredients we were able to 
use the DIM to explain many properties of dwarf nova outbursts. The real 
trouble begins when one tries to describe the quiescence in dwarf nova 
outburst cycles.

\subsection{Quiescence} 
\label{quidn} 
 
According to the DIM a quiescent accretion disc is cold everywhere, i.e. 
its effective temperature is everywhere lower than the critical 
temperature given by Eq. (\ref{tecc}): $T_{\rm eff} < T_{\rm eff~A}\sim 
5800$ K. In fact, the effective temperature is much lower because after 
the outburst the disc ends up well below the critical values of $T$ and 
$\Sigma$ (see e.g. Fig. \ref{fig:std}). In the models the {\sl midplane} 
temperature may be as low as 2000-3000 K, whereas observations suggest 
$T_{\rm eff} \sim 4000-5000$ K (e.g. Wood et al. 1986,1989a). One 
encouraging fact (maybe the only one) is that the DIM predicts flat 
radial temperature profiles and such profiles are indeed observed. At 
first the flatness of temperature profiles elicited surprise among some 
authors, who seemed to expect a $r^{-3/4}$ profile despite the very clear 
prediction of the model (Smak 1984b; Smak, presumably irritated by the 
surprise constantly elicited by the flat temperature profiles, reminds 
the reader that in Fig. 4 of his earlier paper such profiles are clearly 
seen). 
 
The low quiescent temperatures predicted by the DIM cause several 
problems. First, they do not, of course, correspond to observations. 
However, since temperature profiles are determined only for eclipsing 
systems, i.e. for high inclination systems, it is not clear what the 2D 
modeling (Horne 1991; Smak 1994) refers to. A corona above the disc could 
spoil the whole procedure (but see Vrielmann 2000). In any case simple 
(one zone) models of the observed emission from quiescent disc require 
$\alpha_c \gta 100$ (Wood, Horne, Stiening 1992), while more refined 
modeling (Idan et al. 1999) is only ostendibly better at lowering the 
values ($\alpha \gta 1$). Observations of the quiescent state of EX Dra 
by Baptista \& Catal\'an (2000) contradict the DIM, independent of the 
version the data they use (contrary to what they seem to believe). 
 
It is important to determine the exact value of the temperature in 
quiescence is important because it will provide us with crucial 
information about the viscosity mechanism operating during this phase of 
the outburst cycle. The most recent calculations by Hawley (2000) seem 
to indicate that turbulent viscosity in real accretion discs is produced 
by the Balbus-Hawley mechanism. Since this mechanism would not operate 
in a quiescent dwarf nova disc for $T \lta 4000$ K (Gammie \& Menou 
1998; the real limit is on the magnetic Reynolds number: $\lta 10^4$), 
viscosity in quiescent discs would have to be attributed to a different 
mechanism if one wants to preserve the DIM. The DIM does not assume a 
specific viscosity mechanism but does assume that the angular-momentum 
transport mechanism is {\sl local}. In other words, the disc in the DIM 
is an $\alpha$-{\sl disc}. 
 
The DIM predicts quiescent temperatures which are below the critical 
value needed for the Balbus-Hawley instability to operate. This 
statement is not based on a self-consistent calculation because in 
calculating the critical numbers one {\sl assumes} a value of $\alpha$, 
but let us accept this conclusion for the sake of the argument. Then, 
for the DIM to be valid a different {\sl viscosity mechanism} must be at 
work in quiescence. For example, if at low temperature turbulence were 
of hydrodynamical origin (as proposed by Zahn 1991), the DIM could be 
rescued. (The interpretation of the Gammie-Menou `effect' by Livio 1999 
is nonsense because it uses the hot Shakura-Sunayev solution 
to describe the cold branch od the \bS-curve -- see Lasota 2000b). 
 
Of course one should keep in mind that the main assumption made in the 
DIM about viscosity -- the increase of $\alpha$ in outburst -- has yet to be 
confirmed by calculations. Even the relevance of $\alpha$ in the context 
of the Balbus-Hawley mechanism (or the relevance of the Balbus-Hawley 
mechanism to $\alpha$-discs) has still to be demonstrated. 
 
Non-$\alpha$-disc solutions are also possible. It has been proposed that 
there is no viscosity in quiescent dwarf-nova disc (e.g. Paczy\'nski 
1977; R\'o\.zyczka \& Spruit 1993; Gammie \& Menou 1998; Kornet \& 
R\'o\.zyczka 2000). In such a case matter would have to accumulate where 
it arrives from the secondary: at the outer disc edge. Therefore {\sl 
all} outbursts would have to be of the `outside-in' type. And they are 
not. Of course one does not have to assume that all quiescent discs are 
the same. For example, the very low $\alpha$-value interpretation of the 
long recurrence time of WZ Sge assumes that for some reason the disc in 
this system is different. If it is confirmed that all normal outbursts 
in SU UMa stars are `outside-in', one could argue that in this systems 
there is no viscosity in the disc during quiescence. One would the have 
to explain wher the observed X-rays come from (they clearly come from 
the accretion flow), but the popular {\it panacea} of accretion 
astrophysics, disc coronae, could presumably save the day. In any case 
such models would be fundamentally different from the DIM and a global 
hydrodynamical instability would have to be at the origin of the 
outbursts (e.g. R\'o\.zyczka \& Spruit 1993). Also other m echanisms 
(e.g. Caunt \& Tagger 2001 and references therein) cannot be ruled out.
 
Menou (2000a) argued that spiral waves (or shocks) tidally induced by 
the secondary could be the main physical process responsible for 
accretion in quiescence. Spiral patterns are indeed observed in some 
dwarf nova systems (see Steeghs 2000 for a review). To support his 
suggestions Menou shows that there exists a correlation between 
recurrence times of SU UMa stars and their mass- ratios (see Sect. 
\ref{tti}). This could indeed suggest that recurrence times have 
something to do with tidal torques. However, as Terquem (2000) has 
pointed out, tidal torques would be effective only in the outer disc, so 
the problem of the impossibility of `inside-out' outbursts would persist 
(Menou expects the mechanism to apply to all dwarf-nova types). In any 
case the `tidal-spiral' mechanism cannot be described in the framework 
of $\alpha$-discs (Balbus \& Papaloizou 1999). 
 
Last but not least, while all versions of the DIM predict {\sl 
increasing} quiescent fluxes, observations show that they are {\sl 
constant} or {\sl decreasing} (see e.g. Smak 2000). This may require a 
major revision of the DIM. 
 
Quiescence is therefore the biggest unsolved problem of the DIM. 
 
\subsubsection*{X-rays} 
 
According to the DIM the disc should be neutral. We do not expect to 
observe X-rays emitted by such a system. But we do observe them. 
Quiescent dwarf novae and LMXBTs both emit quite an impressive quantity 
of X-rays with luminosities reaching $10^{32}$ erg s$^{-1}$ (Eracleus et 
al. 1991, van Teeseling et al. 1996; Richards 1996). Observations of 
quiescent X-rays in eclipsing systems show clearly that they are emitted 
by the accretion flow {\sl close to the white dwarf}. If the quiescent 
disc were to extend down to the white dwarf surface and the X-rays 
emitted by a hot boundary layer (Pringle \& Savonije 1979; Tylenda 1981; 
Patterson \& Raymond 1985), the required accretion rates would be about 
two orders of magnitude higher than those allowed by the DIM (e.g. Meyer 
\& Meyer-Hofmeister 1994). Hence the idea that the disc does not extend 
down to the white-dwarf surface. The consequences of `holes in discs' 
for the DIM have been discussed above. What fills the holes is of less 
importance for the DIM (in practice, however, it adds one or two free 
parameters) as long as fronts do not propagate into it. 
 
Holes can be due to magnetic fields (e.g. Livio \& Pringle 1992; Lasota 
et al. 2000) or to evaporation into a `siphon flow' (Meyer \& 
Meyer-Hofmeister 1994, Meyer, Meyer-Hofmeister \& Liu 1996, 1998) or to 
an ADAF (Menou 2000b). The main challenge for last the two classes of 
models (especially for the ADAF model) is to explain why emission is 
observed to come from a source the size of a white dwarf radius (e.g. 
van Teeseling 1997; Wood et al. 1995) while the hot plasma forms a much 
more extended structure.

\section{Low-mass X-ray binary transient systems} 
\label{sxt} 
\begin{figure} 
\centering 
{\includegraphics[scale=1,totalheight=7cm]{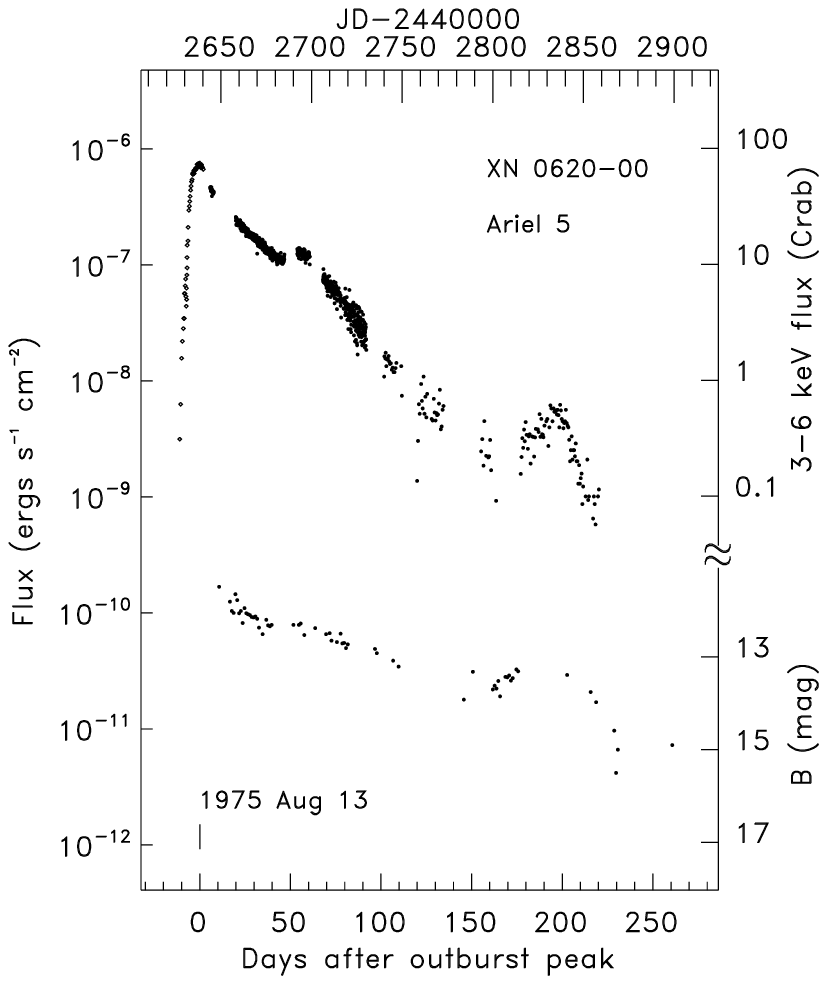}} 
\vskip 0.5cm 
\centering 
{\includegraphics[scale=1,totalheight=7cm,angle=-90]{fig27b.ps}} 
\caption{X-ray and optical light curves of two black-hole LMXBTs: 
A0620-00 (from Chen et al. 1997) and XTE J1118+48 (Kuulkers, private communication). 
In the case of XTE J1118+48 the `black-hole' classification requires 
confirmation.} 
\label{typic} 
\end{figure} 
 
Compared to dwarf nova eruptions, outbursts of low-mass X-ray binary 
transients are both easier and more difficult to describe. Easier 
because fewer systems are known and recurrence times are much longer, so 
there is less information about outbursts to deal with. There are also 
fewer multi-wavelength observations, so much of the data is of rather 
limited value to theorists (or is of unlimited value to those who use them 
to support their suspect models). Most of the bolometric luminosity in 
outburst is emitted in soft X-rays, but often only hard X-ray data are 
available, etc. In the case of dwarf novae, one can always count on an 
optical telescope to observe the outburst. In fact, this is a field in 
which amateur astronomers play a very important role. In the case of 
LMXBTs we are at the mercy of the X-ray satellites that happen to be in 
orbit during the outburst. In addition, simultaneous multiwavelength 
observations are often difficult to arrange. The systems themselves are 
also not so well observed: only four neutron-star LMXBTs have well 
determined orbital periods and only one such system is eclipsing; nine 
orbital periods of black-hole LMXBTs are known and only one is 
eclipsing. Despite this relative scarcity of data, constructing a model 
is rather difficult because the systems themselves are more complicated 
physically.

Discs in low-mass X-ray binaries are strongly X-ray irradiated; this 
influences their stability properties (Sect. \ref{irrsc}) and affects 
the propagation of heating and cooling fronts during outburst. The 
optical flux during eruptions is due to X-ray reprocessing. There is 
also a fundamental problem with the geometry of irradiation. The 
simplest model, in which a point-source located in the midplane at the 
center irradiates a planar disc, cannot work (see e.g. DHLC). Also 
eclipse statistic suggests values of the disc's aspect ratio that are 
much too large (Milgrom 1978) to be accommodated by a vertically static 
planar disc. Therefore extended sources and/or warped discs have to be 
considered, and these, especially the latter, may completely change the 
character of the outburst models. In addition, the DIM for LMXBTs 
suffers from all the deficiencies of the dwarf-nova version, although 
some of them disappear when the effects of irradiation are included in 
the model.

Fig. \ref{typic} shows lightcurves of two LMXBTs believed to contain 
accreting black holes. A0620-00, the veteran prototype X-ray transient 
system has a typical FRED-type lightcurve. But is it a typical LMXBT 
lightcurve? The answer is no since even among FREDs it is 
exceptional, in with a fast quasi-exponential decay over  
almost 2.5 decades and XTE J1118+48 shows that completely different  
lightcurves are observed in LMXBTs. Attempts to model LMXBT outbursts 
with the DIM try to reproduce A0620-00, for obvious reasons: this model 
`naturally' produces fast-rise/slow-decay lightcurves.

Van Paradijs \& Verbunt (1984) and Cannizzo, Wheeler \& Ghosh (1985) 
suggested that LMXBT outburst could have the same origin as dwarf nova 
eruptions. Time-dependent models of LMXTB outbursts, however, have been 
even less successful than models of dwarf novae. Models by Huang \& 
Wheeler (1989) and Mineshige \& Wheeler (1989) were the first heroic 
attempts to obtain LMXBT's light curves, but the result was not really 
compelling: neither the amplitudes nor the recurrence times corresponded 
to reality. Several later attempts concerned only some of the phases of 
the outburst cycle, e.g. the decay from outburst (Cannizzo et al. 1995) 
or evolution in quiescence (Meyer-Hofmeister \& Meyer 2000). In 1998 
Cannizzo presented LMXTB models for truncated discs, in which a 
non-standard viscosity law was assumed. In addition Cannizzo (1998a) 
used an incorrect fixed-radius boundary condition. It was an interesting 
and useful contribution to the problem but not a viable model. Menou et 
al. (2000) calculated a series of LMXBT models in which the disc is 
truncated and used a correct outer boundary condition. They used the 
conventional viscosity prescription in which $\alpha$ is constant on the 
thermally stable branches of the \bS-curve. They considered two classes 
of models: with a `neutron star' ($M_1=$1.4 M$_{\odot}$) and a `black 
hole' ($M_1$=6 M$_{\odot}$). The evaporation was described by Eq. 
\ref{evap} which gives larger rates than those of Cannizzo. In fact, in 
the neutron-star models of Menou et al. (2000) the evaporation is 
unfortunately {\sl much} too strong: the mass used in Eq. (\ref{evap}) 
is 6~M$_{\odot}$ instead of 1.4~M$_{\odot}$ (Dubus, private 
communication). 
 
Menou et al. (2000) found that in black-hole models strong evaporation 
does increase the recurrence times (in agreement with earlier 
calculations by Cannizzo 1998a), but is not sufficient to reproduce the 
longest known recurrence times of LMXBTs when standard values of the 
viscosity parameter $\alpha$ are used in the disc ($\alpha_{\rm hot} 
\sim 0.1$, $\alpha_{\rm cold} \sim 0.01$). The recurrence time could be 
increased to $\sim 50$ years if one assumed $\alpha_{\rm cold} \sim 
0.005$. Because of very slow rises-to-outburst, however, the light-curve 
shapes were not acceptable, even if the lower value of $\alpha$ disposed 
of reflares present for higher $\alpha$ values.  
 
Menou et al. (2000) concluded that `future models' must include 
the missing (but observed in real systems) ingredient: irradiation. This 
conclusion was not new: King \& Ritter (1998) have pointed out that 
LMXBT light-curve shapes are determined by X-ray irradiation. Recently 
DHL have obtained models of irradiated and truncated accretion discs 
describing complete outburst cycles.

\subsection{Irradiation} 
 
Accretion discs in low mass X-ray binaries are irradiated by the X-rays 
emitted by the accretion flow and at high luminosities their optical 
light is due to X-ray reprocessing (van Paradijs \& McClintock 1995). 
The outer disc's emission is then dominated by irradiation. This does not 
mean, however, that the disc's {\sl structure} is dominated by 
irradiation. The condition for irradiation to dominate the vertical 
disc's structure can be obtained from Eq. (\ref{diff3}) and reads 
\begin{equation}  
{F_{\rm irr}\over \tau_{\rm tot}} \equiv {\sigma T^4_{\rm irr} \over  
\tau_{\rm tot}} \gg F_{\rm vis}  
\label{c1}  
\end{equation}  
(Lyutyi \& Sunayev 1976; Tuchman, Mineshige \& Wheeler 1990; Hubeny 
1991; Hur\'e et al. 1994; DLHC) and not just $F_{\rm irr} \gg F_{\rm 
vis}$ as too often assumed in the literature. If this condition  
is satisfied the disc is (in the simplest case) isothermal. In 
other words: irradiation dominates the disc's structure only if the 
irradiation temperature is larger than the {\sl midplane} temperature 
of the disc. 
 
Shakura \& Sunyaev (1973) devised a simple way of describing a disc's irradiation 
by a midplane, central point source: Eq. (\ref{C1}). They wrote $\mathcal 
C$ as: 
\begin{equation}  
{\mathcal C} \equiv (1 - \varepsilon) {H_{\rm irr} \over r}  
\left({d\ln H_{\rm irr} \over d\ln r} - 1\right)  
\label{C2}  
\end{equation} 
where $\varepsilon$ is the X-ray albedo and $H_{\rm irr}$ is the local 
height at which irradiation energy is deposited, or the height of the 
disc ``as seen by X-rays". One often writes $H$ instead of $H_{\rm irr}$ 
but this could lead to misunderstandings because $H_{\rm irr}$ is {\sl 
not} equal to the local pressure scale-height $H$. Therefore one cannot 
use the value $H_{\rm irr}$ estimated from observations to evaluate 
e.g. the value of $\alpha$ from Eq. (\ref{alfn}). According to Shakura \& 
Sunyaev (1973) the rhs of Eq. \ref{C2} should be multiplied by ($H_{\rm 
irr}/r$) if the accreting body is a black hole and not an object with a `hard' 
surface. 
 
When one calculates a self-consistent irradiated-disc structure using 
Eq. (\ref{C1}) with Eq. (\ref{C2}) one finds that it is {\sl not} affected 
by irradiation (Tuchman et al. 1990; DLHC). The reason is that in such a 
model the outer part of the disc that would be most affected by 
irradiation cannot see the irradiating source because of the disc's  
self-screening. Observations, however, show that discs in LMXBs {\sl 
are} irradiated. Therefore the irradiating source is not point-like  
or/and the disc is not planar. In the first case the outer disc 
would be irradiated e.g. by a scattering corona. In the second the disc 
would be warped. There is evidence for both of these effects but 
incorporating them into the DIM is practically impossible. DHL chose to 
use the simplest description of irradiation as given by Eq. (\ref{C1}) 
with a constant $\mathcal C$, whose value is chosen in such a way as to 
correspond to observed X-ray reprocessed fluxes. More precisely, for  
time-dependent calculations DHL took 
\begin{equation} 
\sigma T^4_{\rm irr} = {\mathcal C} \frac{L_{X}}{4 \pi R^2}  
{\rm ~~with~} L_{X}= \eta\ \times{\rm min\ }(\dot{M}_{\rm in}, 
\dot{M}_{\rm Edd})\ c^2 
\label{ill} 
\end{equation} 
$\dot{M}_{\rm in}$ is the mass accretion rate at the inner disc radius 
and $\eta$ is the 
accretion efficiency ($\dot{M}_{\rm Edd}$ is calculated assuming 
$\eta=0.1$). In models including evaporation $\eta$ varies during  
the outburst cycle. 
 
$\mathcal C$ is a measure of the fraction of X-rays that heats the disc and, 
as such, contains information on the irradiation geometry, X-ray albedo, 
X-ray spectrum, etc. Kim, Wheeler \& Mineshige (1999) used an analogous 
prescription for the indirect irradiation flux. On finds that the 
observed optical magnitudes and stability properties of persistent low 
mass X-ray binaries are compatible with a value of $\mathcal C$$\approx$ 
5$\times 10^{-3}$ (with $\epsilon=0.1$, DLHC). However, there is no 
reason for $\mathcal C$ to be a constant in the disc or in time. For 
instance, Ogilvie \& Dubus (2001) showed how $\mathcal C$ could vary in a 
warped disc while Esin, Lasota \& Hynes (2000) argued that $\mathcal C$ 
disminished during the 1996 outburst of the X-ray transient 
GRO J1655-40. For simplicity, DHL assumed $\mathcal C$=5$\times 
10^{-3}$. It should be remembered that this is the simplest assumption but 
not necessarily the most realistic one.

\subsection{Outburst} 
\label{outsxt} 
 
We will base the description of the LMXBT outburst cycle on a model 
discussed in details in DHL. The parameters used in this model are: the 
`black hole' mass $M_1=7$M$_{\odot}$, the mass-transfer rate $\dot{M}_{\rm 
tr}=10^{16}{\rm ~g}\ {\rm s}^{-1}$, $\alpha_{\rm h}=0.2$, $\alpha_{\rm 
c}=0.02$, $<r_{\rm out}>=10^{11}$~cm. Evaporation of the inner disc is 
described by Eq. (\ref{evap}). Below the inner radius $r_{\rm in}$ all 
matter evaporates into an ADAF. Because of the steep dependence of 
$\dot M_{\rm ev}$ on $r$, evaporation above this radius can be neglected. 
Menou et al. (2000) noticed that the detailed functional dependence of 
$\dot M_{\rm ev}(r)$, or of $r_{\rm in}(\dot M_{\rm in})$, has 
essentially no effect on the results; what matters is the value of the 
inner disc radius during quiescence.  
  
The ADAF radiative efficiency scales as $\dot{M}$ (i.e. the luminosity 
scales as $\dot{M}^2$, Esin et al. 1997) . From Eq.\ref{evap}, this is 
equivalent to $\epsilon \propto r^{-2}_{\rm in}$. Irradiation is taken 
into account using Eq.~\ref{ill} with $\eta=0.1$ when $r_{\rm in}=r_{\rm 
min}$ (in outburst) and $\epsilon=0.1\times(r_{\rm min}/r_{\rm 
in})^{-2}$ when $r_{\rm in} \ge r_{\rm min}$ (in quiescence). 
Irradiation is usually negligible in quiescence so these assumptions 
have little importance for the description of this phase of the outburst 
cycle. 
\begin{figure} 
\centering  
{\includegraphics[scale=1,totalheight=16cm]{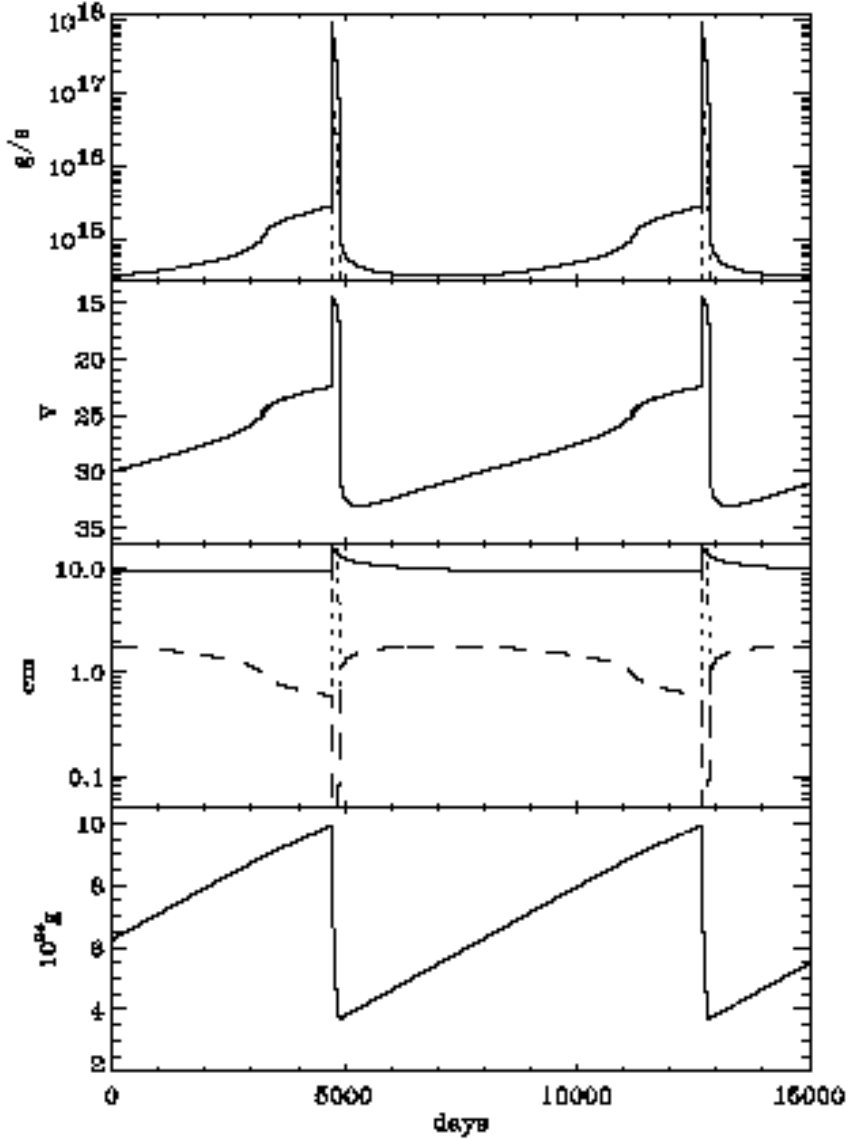}} 
\caption{An outburst cycle with disc irradiation and evaporation  
included. From top to bottom: $\dot{M}_{\rm in}$ (full  
line) and $\dot{M}_{\rm irr}$ (dotted line); V magnitude; $r_{\rm  
out}$ (full line), $r_{\rm trans}$ (dotted line) and $r_{\rm in}$  
(dashed line), $M_{\rm disc}$ (in $10^{24}$ g). }  
\label{tidim}  
\end{figure}  
The outburst cycle for these parameters and assumptions is shown in  
Fig. \ref{tidim}. The recurrence time is 22 years, the peak luminosity 
$L_X=9 \times 10^{37}$ erg s$^{-1}$. The average mass of the disc is 
$7\times 10^{24}$ g and during the outburst $6\times 10^{24}$: due to 
irradiation a much larger fraction than in a standard dwarf-nova disc. 
 
We will now discuss the outburst cycle in more detail.

\subsubsection{Rise to maximum} 
 
A LMXBT outburst starts in the same way as in a dwarf nova since in 
quiescence irradiation is of no importance. As mentioned before, one 
expects these outbursts to be of the `inside-out' type. In a disc 
truncated by evaporation this `inside' can be pretty far from the 
accreting object; in the model presented in Fig. \ref{tidim} just before 
the outburst $R_{\rm in} \approx 6 \times 10^9$ cm, i.e. 2900 $r_{\rm 
S}$. As the inside-out heating front propagates, $\dot{M}_{\rm in}$ rises  
and the outer cold disc is increasingly irradiated (see  
Fig.~\ref{irrise}).  Heating by X-ray irradiation reduces the critical  
density $\Sigma_{\rm min}$ on the hot branch, thus facilitating front  
propagation. Obviously, a larger hot region implies a greater optical  
flux and irradiation always implies larger peak optical luminosities. 
\begin{figure} 
\centering 
{\includegraphics*[scale=1,totalheight=7cm]{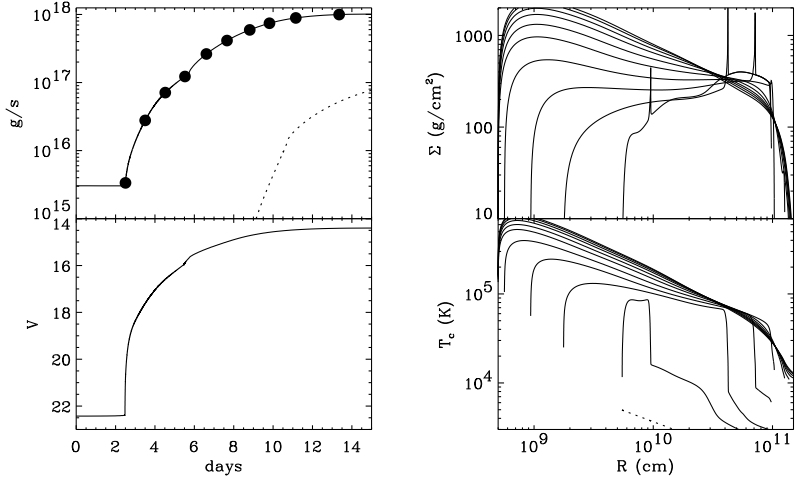}} 
\caption{ 
The rise to outburst according to the model described in Sect. 
\ref{outsxt}. The upper left panel shows $\dot{M}_{\rm in}$ and 
$\dot{M}_{\rm irr}$ (dotted line); the bottom left panel shows the $V$ 
magnitude. Each dot corresponds to one of the $\Sigma$ and $T_{\rm c}$ 
profiles in the right panels. The heating front propagates outwards. The 
disc expands during the outburst due to the  angular momentum transport 
of the material being accreted. At $t\approx 5.5$~days the thin disc 
reaches the minimum inner disc radius of the model. The profiles close 
to the peak are those of a steady-state disc ($\Sigma\propto T_{\rm 
c}\propto R^{-3/4}$). (From DHL). 
} 
\label{irrise} 
\end{figure} 
 
The heating front structure is not modified by irradiation as can be 
seen comparing Figs. \ref{hc} and \ref{irrfront}.  
However, that as explained in detail in DHL, the description of the 
heating front propagation is not complete. The problem is that the outer 
disc regions are `frozen' in the cold state on a timescale on which 
$\dot{M}_{\rm in}$ evolves (see Fig.~\ref{irrise}). If $\dot{M}_{\rm 
in}$ increases on a timescale shorter than the thermal timescale in the 
cold disc, at some radii one can have $T_{\rm irr}>T_{\rm c}$. The code 
used by DHL is not adapted to such a situation, so they assumed that in 
such a case the disc is isothermal at $T_{\rm c}$ and that irradiation 
contributes an additional heating term to the radial thermal equation 
$Q^+_{\rm add}=\sigma (T_{\rm irr}^4-T_{\rm c}^4)$ which to reflect the 
imbalance at the photosphere between the outgoing flux $\sigma T^4_{\rm 
c}$ and the incoming flux $\sigma T^4_{\rm irr}$. This difficulty has 
little practical significance in the context of the DIM but may be a 
sign that some additional physical processes should be included in the 
model. This could be the case irradiation with evaporation of the upper 
layers of the disc (Begelman et al. 1983; Hoshi 1984; Idan \& Shaviv 
1998; de Kool \& Wickramasinghe 1999).  
\begin{figure} 
\centering 
{\includegraphics*[scale=1,totalheight=7cm]{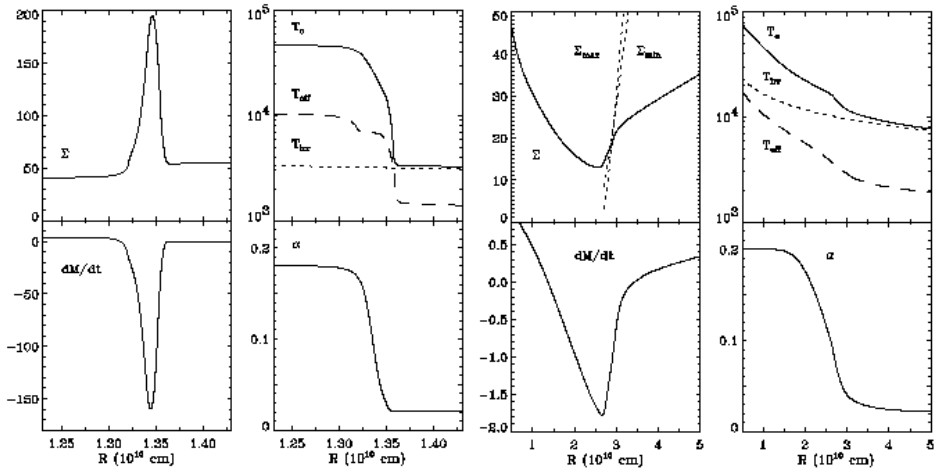}} 
\caption{Left panel: the structure of an `inside-out' heating front  
in an irradiated disc. $\Sigma$  
is in g$\cdot$s$^{-1}$, temperatures are in K, $\dot{M}$ is in units  
of $10^{16}$~g$ $s$^{-1}$.  The cold outer disc is almost  
isothermal with $T_{\rm c}\approx T_{\rm irr}$. 
Right panel: a cooling front in an irradiated disc. The front is  
at the position for which  
$T_{\rm irr}$ (dotted line) is $\approx 10^4$~K. At this point  
$\Sigma_{\rm min}\approx\Sigma_{\rm max}$ since there is no cold  
branch for higher $T_{\rm irr}$ (smaller $R$). As $T_{\rm irr}$  
decreases the two critical $\Sigma$ separate and converge to their  
non-irradiated values. (From DHL)} 
\label{irrfront} 
\end{figure} 
 
The arrival of the heating front at the outer disc rim does not end
the  rise to outburst. What happens then can be seen in Fig. \ref{rad}
which corresponds to a model with parameters slightly different from
those  discussed earlier in this section (the main difference is a
higher  mass-transfer rate which lengthens outburst's
timescales). Fig.  \ref{rad} shows that after the whole disc is
brought to the hot state  a surface density (and accretion rate
`excess') forms in the outer disc.  The accretion rate in the inner
disc corresponds to the critical one but   is much higher near the
outer edge. While irradiation keeps  the disc hot the excess difffuses
inwards until the accretion rate is  roughly constant. During this
last phase of the rise to outburst maximum  $\dot M_{\rm in}$
increases by a factor of 3. In Fig.  \ref{irrise} this phase begins at
the 4th dot on the light-curve.  Irradiation has little influence on
the actual   vertical structure in this region, as discussed in DLHC
and we duly find $T_{\rm c}\propto\Sigma\propto   R^{-3/4}$, as in a
non-irradiated steady disc. Only in the outermost   disc regions does
the vertical structure becomes   irradiation-dominated,
i.e. isothermal.  
 
Now the disc is ready to begin the decay phase of the outburst cycle. 
\begin{figure} 
\centering 
{\includegraphics[totalheight=12truecm,angle=90]{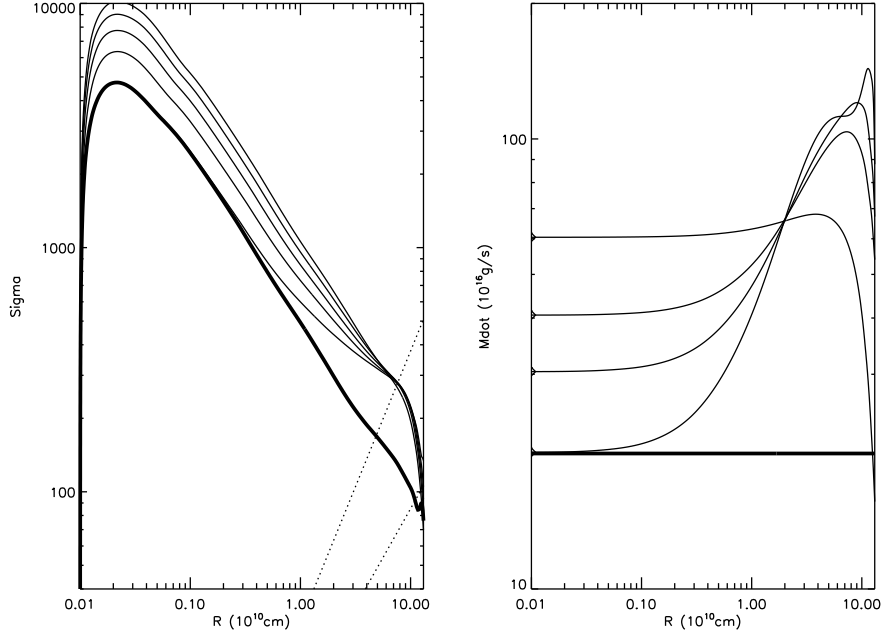}} 
\caption{Surface density and accretion rate during the final stage of 
the rise to outburst maximum in an irradiated accretion disc around a 6 
M$_{\odot}$ `black hole'. The mass-transfer rate is $3\times 10^{16}$  
\gs. The thick continuous line corresponds to the critical mass-transfer 
rate for stability in a hot state. The $\Sigma$ and $\dot M$ profiles 
correspond (from bottom upwards) to 45, 47, 54 and 100 days after the beginning of 
the outburst. (Dubus, private communication) 
} 
\label{rad} 
\end{figure} 
 
\subsubsection{Decay} 
\label{decay} 
 
The shapes of decay lightcurves in LMXTBs have generated a considerable 
amount of theoretical activity. The reason was that their alleged 
`exponential' slopes were not a ``natural" outcome of DIM calculations. 
Various speculations and models involving viscosity-law modifications 
and other attempts to modify the accretion disc's physics were put an 
end to (not immediately, however) by King \& Ritter (1998), who pointed 
out that X-ray irradiation of the disc produces ``naturally" exponential 
decays. As mentioned above, previous attempts to include irradiation in 
the disc's models used Eq. (\ref{C2}) and arrived at the correct 
conclusion that because of self-screening irradiation cannot be 
important in LMXBTs. King \& Ritter took a more realistic approach, 
noticing that since there is extremely strong observational evidence for 
disc irradiation, the inability of the model to reproduce this feature 
just highlights its `limitations'. King \& Ritter used Eq. (\ref{C2}) 
but this did not change the validity of their main conclusion: by 
heating the outer disc regions X-ray irradiation prevents the 
propagation of the cooling front that would switch off the outburst: 
only a viscous decay is then possible. The corresponding lightcurve has 
an exponential slope. Objections were put forward to this last 
conclusion (Cannizzo 1998b) but they were based on models not relevant in 
the context of LMXBT oubursts, as we will see below. 
\begin{figure} 
\centering 
{\includegraphics*[scale=1,totalheight=7cm]{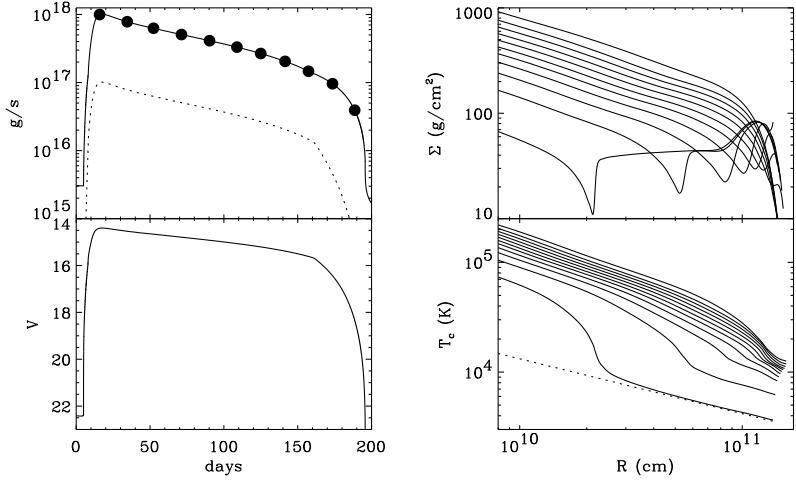}} 
\caption{ 
Decay from outburst peak. The decay is controlled by irradiation until 
evaporation sets in at $t\approx170$~days ($\dot{M}_{\rm 
in}=\dot{M}_{\rm evap}(R_{\rm min})$). This cuts off irradiation and the 
disc cools quickly. The irradiation cutoff happens before the cooling 
front can propagate through most of the disc, hence the 
irradiation-controlled linear decay ($t\approx 80-170$~days) is not very 
visible in the lightcurve. $T_{\rm irr}$ (dotted line) is shown for the 
last temperature profile.(From DHL)}  
\label{irrdecay}  
\end{figure} 
 
Fig. \ref{irrdecay} shows the sequel to what was described in Fig. 
\ref{irrise}. One can see that in general decay from the outburst peak 
can be divided into three parts: \begin{itemize} 
 
\item First, X-ray irradiation of the outer disc inhibits cooling-front 
propagation. Obviously the peak accretion rate is much higher than the 
mass-transfer rate so the disc is drained by viscous accretion of 
matter, as proposed by King \& Ritter (1998).  
 
\item Second, the accretion rate becomes too low for the X-ray 
irradiation to prevent the cooling front from propagating. The 
propagation speed of this front, however, is controlled by irradiation. 
 
\item Third, irradiation plays no role and the cooling front 
switches off the outburst on a local thermal time-scale.  
\end{itemize} 
 
\subsubsection*{`Exponential decay'} 
 
In Fig. \ref{irrdecay} the first phase lasts until roughly day 80-100 
and corresponds to the ``exponential decay" of the light-curve. At the 
ouburst peak the accretion rate is almost exactly constant in the disc 
(Fig. \ref{rad}). 
The subsequent evolution is self-similar: the disc's radial structure 
evolves through a sequence of quasi-stationary ($\dot M(r)=const$) 
states. Therefore $\nu \Sigma \sim \dot{M}_{\rm in}(t)/3\pi$ and the total 
mass of the disk is thus 
\begin{equation} 
M_{\rm d}=\int 2\pi R\Sigma dR \propto \dot{M}_{\rm in} \int {2\over 3} {r 
\over \nu} dr 
\label{md} 
\end{equation} 
 
At the outburst peak the whole disc is ionized and, as mentioned above, 
except for the outermost regions its structure is very well represented 
by a Shakura-Sunyaev solution. In such discs, as well as in irradiation 
dominated discs, the viscosity coefficient satisfies the relation $\nu \propto 
T \propto \dot{M}^{\beta/\left(1+\beta\right)}$. 
In `hot' Shakura-Sunyaev discs $\beta = 3/7$, and  in irradiation 
dominated discs $\beta$ = 1/3. During the first decay phase the outer 
disc radius is almost constant so that using Eq. (\ref{md}) 
the disc-mass evolution can be written as: 
\begin{equation} 
\frac{{\rm d} M_{\rm d}}{{\rm d} t}=-\dot{M}_{\rm in} \propto M_{\rm d}^{1+\beta} 
\label{eq:expo} 
\end{equation} 
showing that $\dot{M}_{\rm in}$ evolves almost exponentially, as long 
as $\dot{M}_{\rm in}^\beta$ can be considered as constant (i.e. over 
about a decade in $\dot{M}_{\rm in}$, as found by King \& Ritter 1998 
for irradiated discs, and by Mineshige, Yamasaki \& Ishizaka 1993 for 
non-irradiated discs with loss of angular momentum). The belief that 
viscous decay must produce power-laws with indices close to -1 is based 
on models in which it is assumed that the disc evolves with constant 
angular momentum. In such models no tidal torques are present and the 
outer disc radius must expand indefinitely (Lyubarski \& Shakura 1987; 
Cannizzo, Lee \& Goodman 1990; Mineshige et al. 1993). Clearly such 
models are not appropriate for discs in close binary systems. 
 
`Exponential' decays in the DIM are only approximately exponential, as 
pointed out by King (1998), who considered an `analytical' model in which  
the disc structure was totally dominated by irradiation. The quasi- 
exponential decay is due to {\sl two} effects: 1.) X-ray irradiation keeps the 
disc ionized, preventing cooling-front g propagation, 2.) tidal torques 
keep the outer disc radius roughly constant. 
 
\subsubsection*{`Linear' decay} 
 
The second phase of the decay begins when a disc ring cannot remain in 
thermal equilibrium. Locally this corresponds to a fall onto the cool 
branch of the \bS-curve. In an irradiated disc this happens when the 
central object does not produce enough X-ray flux to keep the $T_{\rm 
irr}(R_{\rm out})$ above $\sim 10^4$~K. A cooling front appears and 
propagates down the disc at a speed of $V_{\rm front}\approx \alpha_{\rm 
h} c_s$ (see Sect.~\ref{fronts}). 
 
In an irradiated disc, however, the transition between the hot and cold 
regions is set by $T_{\rm irr}$ because the cold branch exists only for 
$T_{\rm irr}\lta 10^4$~K. In an irradiated disc a cooling front can 
propagate inwards only down to a radius at which $T_{\rm irr}\approx 
10^4$~K, i.e. as far as there is a cold branch to fall onto.  Thus the 
decay is still irradiation-controlled. The hot region remains close to 
steady-state but its size shrinks $R_{\rm hot}\sim \dot{M}_{\rm 
in}^{1/2}$ (as can be seen in Eq.~\ref{ill} with $T_{\rm irr}(R_{\rm 
hot})= \mathrm{const}$). The structure of the cooling front is not 
different from the non-irradiated case. Also here the cooling front 
propagation produces both inflow {\sl and} outflow (see Fig 
\ref{irrfront}) of matter so the decline is steeper than what some 
analytic approximations would predict (e.g. King 1998). 
 
\subsubsection*{Thermal decay} 
 
In the model shown in Fig. \ref{irrdecay} irradiation is 
unimportant after $t\gta 170-190$~days because $\epsilon$ becomes 
very small for $\dot{M}_{\rm in}<10^{16}$~g$\cdot$s$^{-1}$.  
The cooling front thereafter propagates freely inwards, on a thermal time 
scale. In this particular case the decrease of irradiation is caused by 
the onset of evaporation which lowers the efficiency. In general there 
is always a moment at which $T_{\rm irr}$ becomes less than $10^4$~K; 
evaporation just shortens the `linear' decay phase. 
 
\subsection{Visual lightcurves} 
\begin{figure} 
\centering 
{\includegraphics[scale=1,totalheight=6cm]{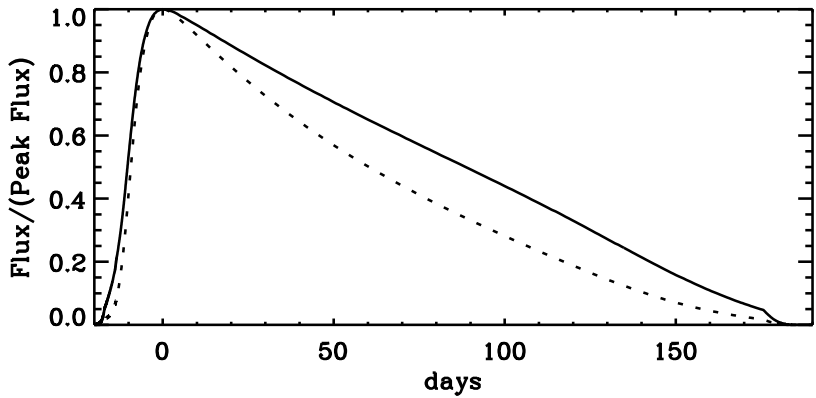}} 
\caption{V-band optical flux (full line) compared to  
$\dot{M}_{\rm irr}=\epsilon \dot{M}_{\rm in}$ (dotted line)  
during an outburst of the model (Fig.~\ref{tidim}).  
$\dot{M}_{\rm irr}$ is taken as an indicator of the X-ray flux. Both  
fluxes are normalized to their peak value. The optical flux is halved in  
$\approx 100$~days while the X-ray flux is halved in $\approx 60$~days. 
(From DHL).}  
\label{optical}  
\end{figure} 
In FRED-type lightcurves the decay in visual light is slower than in 
X-rays. This is also a property of the irradiated DIM 
(Fig.~\ref{optical}). This effect is due to irradiation which, under 
very general conditions, can keep the cold outer disc hotter than usual 
during the propagation of the cooling front. Fig.~\ref{irrdecay} shows 
the disc behind the front is dominated by irradiation and almost 
isothermal. Obviously, the peak optical flux is much higher when the 
disc is irradiated than when it is not.  
 
\subsection{Recurrence times} 
 
\begin{figure}
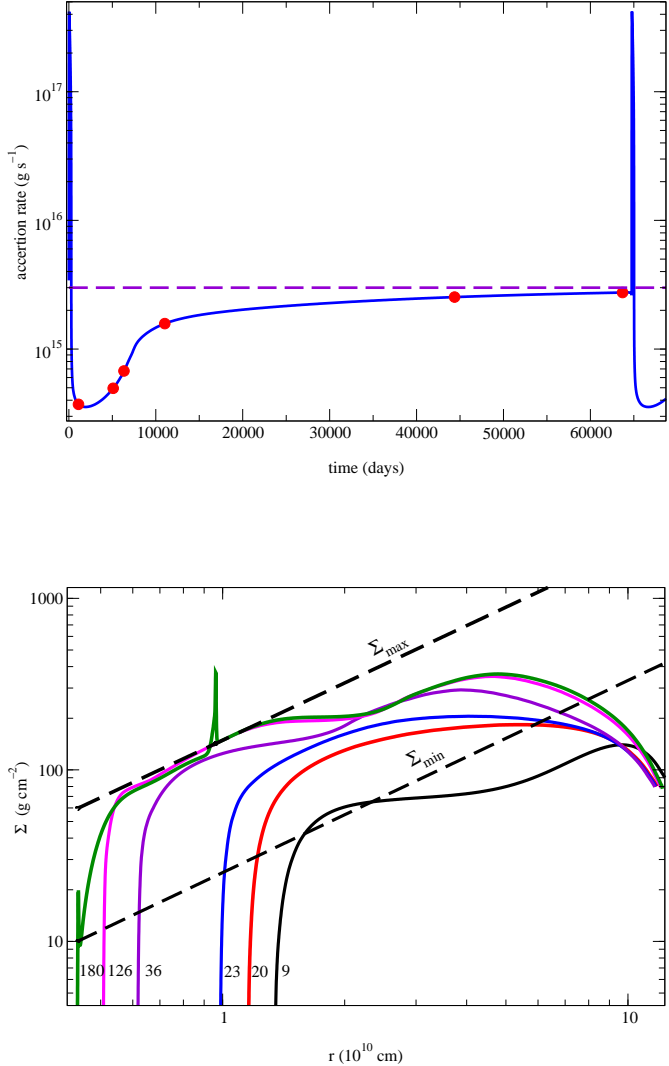
 
\centering 
{\includegraphics[scale=0.75,totalheight=9cm,angle=-90]{fig34a.eps}} 
 {\includegraphics[scale=0.75,totalheight=9cm,angle=-90]{fig34b.eps}} 
\caption{The evolution of the surface density profile in a quiescent,  
truncated accretion disc around a 6 M$_{\odot}$ black hole (lower panel) withtransport 
$\alpha_{\rm hot}=0.1$, $\alpha_{\rm cold}=0.02$, the mean outer disc radius is  
$1.7 \times 10^{11}$ cm. The mass transfer rate is $3\times 10^{15}$ g 
s$^{-1}$, corresponding to the horizontal dashed line on the upper 
panel, which shows the light-curve on which phases of the evolution 
shown in the lower panel are marked by circles. Numbers marking the 
$\Sigma$ profiles are years from the accretion rate minimum. (Based on 
DHL). 
} 
\label{reclc} 
\end{figure} 
Models with truncated and irradiated discs produce recurrence times from 
1 to 180 years. Since observations indicate that LMXBTs have recurrence  
times from $\sim$ 0.5 to $>$ 50 years this is very encouraging. It is  
therefore interesting to see why these recurrence times are so long and what 
are the respective contributions of truncation and irradiation to  
this longevity of quiescence. Figure \ref{reclc} shows an extreme case 
where this can be studied. The recurrence time is 180 years. 
 
The first $\Sigma$ profile in Fig. \ref{reclc} shows the density profile 
about 9 years after the disappearance of the cooling front. (Similar 
profiles can be found in Meyer-Hofmeister \& Meyer \& Liu 2000, but in 
their case only the quiescent phase of the cycle is calculated). It shows 
that the post-outburst surface-density is lower than $\Sigma_{\rm min}$. 
This is, of course, the result of the first phase of the decline from 
outburst during which the cooling front propagation is inhibited and the 
disc is emptied on a viscous time. Therefore, the `usual' assumption 
that at the start of the quiescent phase $\Sigma \sim 2 \Sigma_{\rm 
min}$ (Osaki 1996) is clearly invalid in this case, as is, at least 
numerically, the quiescence-time estimate given by Eq. (\ref{tb}). The 
subsequent evolution proceeds through a sequence of rather flat 
surface-density profiles. The disc is small $r_{\rm out}/r_{\rm in} \sim 
10$ so that the whole disc feels the effect of the boundary condition. 
As the disc fills up with matter the accretion rate increases and the 
inner disc radius decreases. After 36 years of this evolution the inner 
accretion rate is $\sim 1.8\times 10^{15}$ g s$^{-1}$ and  roughly 30\% of 
the mass accreted during the outburst ($5\times 10^{24}$ g) has been 
refilled into the disc. A significant change in the $\Sigma$ profile has 
taken place. A look at Fig. \ref{sigr} elucidates what is going on: the 
outer, cold part of stationary discs has the same profile. Therefore, a 
large part of the disc becomes stationary. This is not surprising because 
the accretion rate is now close to the mass-transfer rate. After another 
90 years the accretion rate arrives very close to the mass-transfer rate 
and now the disc is `full' and quasi-stationary. It will take another 
fifty years before the $\Sigma_{\rm max}$ line is crossed 
at last.  
\begin{figure} 
\centering 
{\includegraphics[totalheight=8cm]{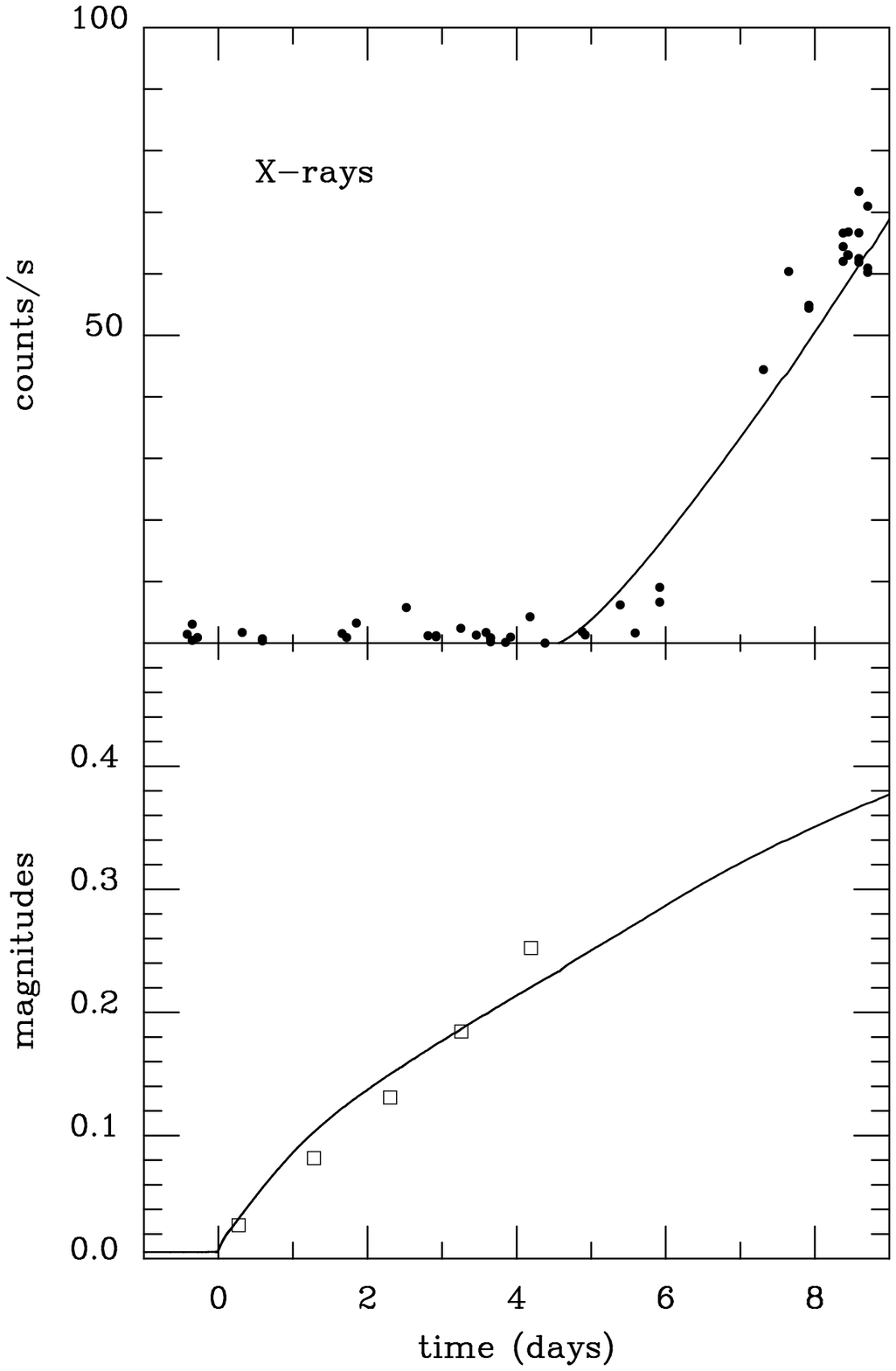}} 
\hspace*{0.5truecm} 
{\includegraphics[totalheight=8cm]{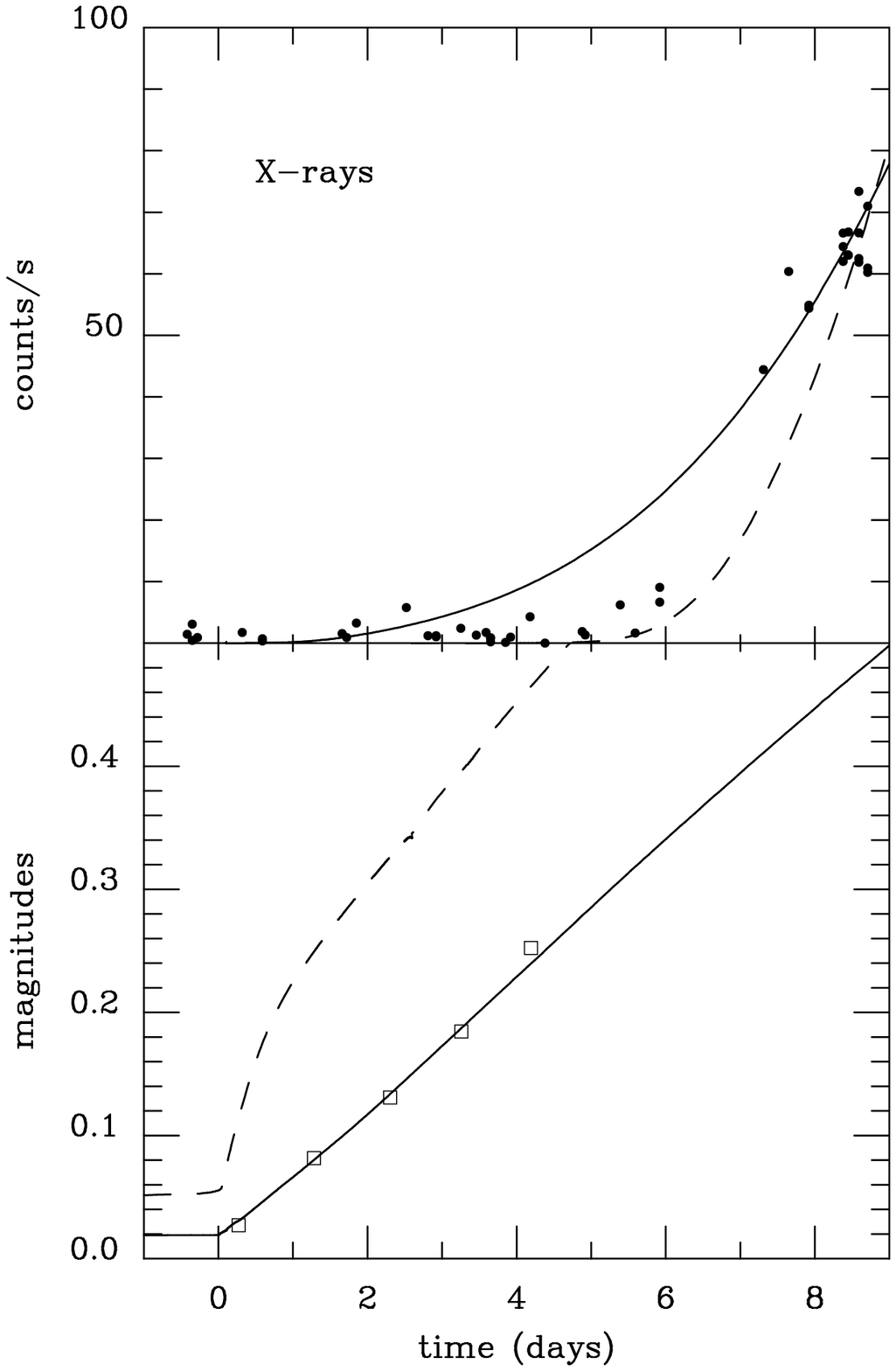}} 
\caption{Observed 2-12 keV X-ray  (1 day averages) and V light curves 
during the rise to the April 1996 outburst of GRO J1655-40 (from Orosz 
et al. 1997). In the left panel continuous lines represent the light 
curves calculated assuming that a thermal-viscous instability operates 
in a disc truncated in quiescence at $10^{10}$ cm. The right panel shows 
results of models in which the quiescent disc extends down to  $4\times  
10^{8}$ cm. The solid curves correspond to an inside-out outburst, the 
dashed curves represent an outside-in outburst triggered by  
mass-transfer enhancement. The V light curves take into account the dilution 
by the light emitted by the 48$L_{\odot}$ secondary}. 
\label{rise55} 
\end{figure} 
 
It is probably impossible to reach such longevity in real 
systems. The reason is that in the last stages of this long quiescence 
even a small fluctuation in the mass transfer from the secondary (which 
will propagate inwards in less than a hundred days) will be able to 
trigger an outburst. If we allow fluctuations of a factor of two, the `real' 
quiescence time would rather be $\sim 40$ years. The refilling 
time-scale for the disc is $\sim 80$ years (see Eq. (\ref{treccu} with 
$\dot M_{\rm tr} \gg \dot M_{\rm in}$. One can therefore conclude that 
the DIM model enriched with the effects of evaporation and irradiation 
can produce the recurrence times observed in LMXBTs).  
\begin{figure} 
\resizebox{\hsize}{!}{\includegraphics[angle=-90]{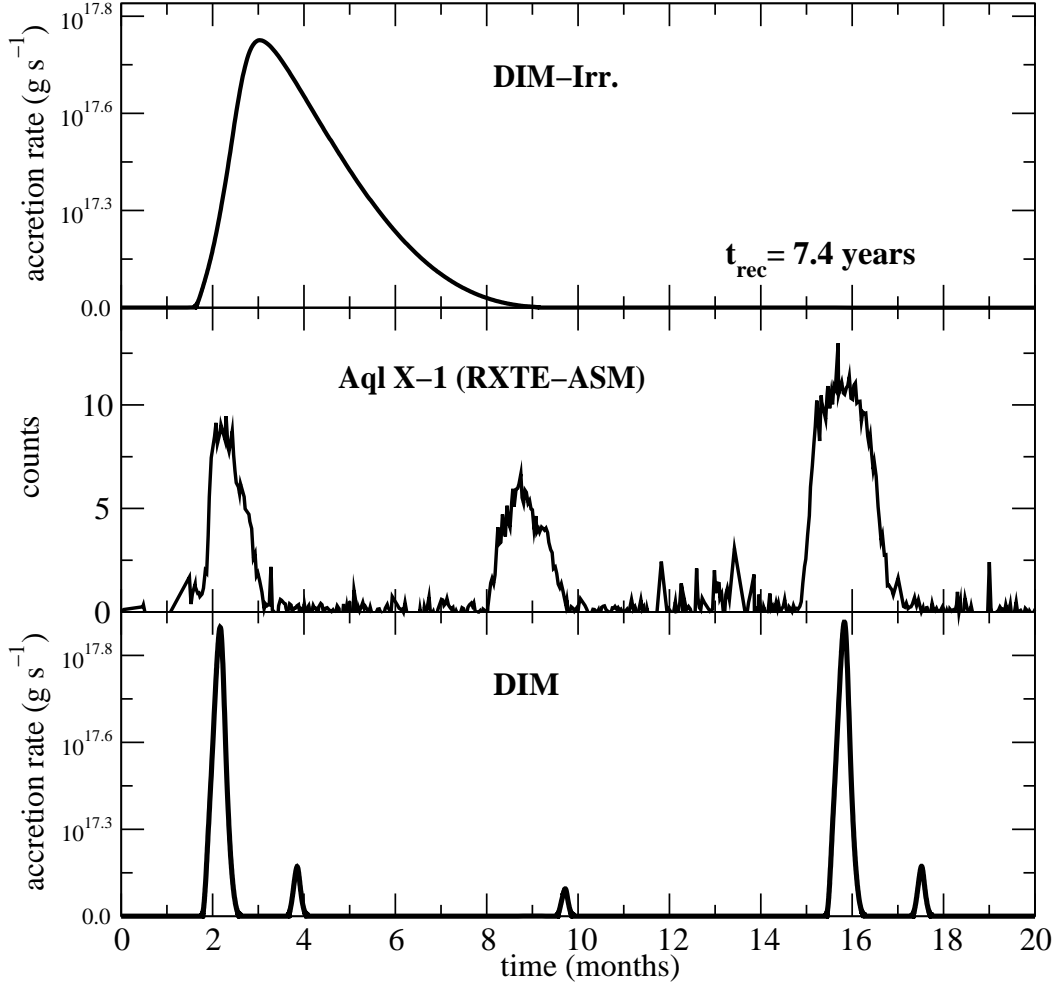}} 
\caption{The observed (RXTE-ASM) light curve of Aql X-1 (middle panel) and two 
model light curves. The upper panel shows what is predicted by a model 
in which the disc is irradiated with $\mathcal C=5\times 10^{-3}$. In the bottom 
panel the disc is not irradiated. The mass of the accreting object 
is 1.4 M$_{\odot}$, the mass-transfer rate $2\times 10^{16}$ \gs, 
$\alpha_{\rm cold}=0.02$, $\alpha_{\rm hot}=0.1$. $<r_{\rm out}>= 
1.81 \times 10^{11}$ cm for the non-irradiated disc and $<r_{\rm out}>= 
1.62 \times 10^{11}$ cm when the disc is irradiated. 
} 
\label{ax1} 
\end{figure} 
 
\subsection{The X-ray delay} 
\label{xdel} 
 
In contrast to the dwarf-nova UV-delay, of which several well documented 
occurrences were observed, there is only one example of an X-ray delay in 
a LMXBT. This observation was almost a miracle considering how rare are 
outbursts in these systems. The April 1996 observation of the rise to 
outburst the black-hole transient source GRO J1655-40 showed a 6 day 
delay between the beginning of the rise of the BVRI light and the {\sl 
RXTE-ASM} X-rays (Orosz et al. 1997). Orosz et al. (1997) also reported 
a delay between I, R, V and B light-curves and considered this to be an 
indication of the outside-in nature of the outburst. The DIM, however, 
predicts no observable delay between the rise of light in these 
wavelengths. The heat front produced by the instability increases the 
effective temperature to $\gta 10000$ K, for which the relevant colour 
indices are close to zero, so no delay is expected. Hameury et al. 
(1997) explain that the impression of a successive delay between the I, 
V, R and B light is given by the different slopes of the respective 
light-curves. These wavelength-dependent slopes are determined by the 
presence of the diluting emission from the luminous, F4-type companion 
star. 
 
Figure \ref{rise55} shows the results of calculations by Hameury et al. 
(1997), who compared multi-wavelength light-curves obtained with DIM with 
and without inner disc truncation. Only the model in which the inner 
disc radius is $\sim 10^{10}$ cm can reproduce both the 6-day X-ray 
delay {\sl and} the slopes of the X-ray and optical light-curves. In 
addition, obtaining an outside-in outburst with a full disc requires a 
specially designed initial surface-density profile. This is a strong, 
theoretical argument in favour of the presence of inner `holes' in 
quiescent discs of LMXBTs. (Arguments based on the X-ray reflection  
spectra are ont bery compelling, see Done \& Nayakshin 2000). 
 
\subsection{Comparison with observations} 
 
The truncated and irradiated DIM should be considered as providing a 
basic framework for the description of LMXBT outbursts rather than a 
successful model. Its successes are limited to a few points: it 
reproduces rise- and decay- times of a subclass of LMXTBs and the 
recurrence times obtained compare well with observed ones. The model in 
its simplest form, however, cannot reproduce other properties or other 
subclasses of LMXTBs. For example, in Fig.~\ref{ax1} we compare the 
observed (RXTE) X-ray lightcurve of Aql X-1 (a neutron-star binary) with 
two theoretical `lightcurves'. One should bear in mind that theoretical 
`X-ray lightcurves' shown here represent just the accretion rate onto 
the central body and not the X-ray luminosity in some specified energy 
domain. These lightcurves can be thought of as bolometric X-ray 
lightcurves only if a 0.1 efficiency is assumed. Therefore in models 
with evaporation the accretion rate does not correspond to the 
bolometric luminosity for $\dot M \lta \dot M_{\rm Edd}$. The upper 
panel of Fig.~\ref{ax1} corresponds to a model with a `standard' value 
of $\mathcal C=5\times 10^{-3}$; the lower one to a `standard' DIM with 
no irradiation or evaporation. The mass of the accreting body is 
1.4~M$_{\odot}$, the mass transfer rate is $2\times 10^{16}$ g\ s$^{- 
1}$, $\alpha_h=0.1$ and $\alpha_c=0.02$. The average outer radius is 
equal to $1.81 \times 10^{11}$ cm in the irradiated-disc model, and 
$1.62\times 10^{11}$ cm in the DIM. In the irradiated-disc case the 
recurrence time is 7.4 years, much too long since the recurrence time of 
Aql X-1 is $\sim 0.5$ year; the standard disc gives 0.8 year, closer to 
the observed value. However, outburst duration is too short in the 
standard DIM and too long in the irradiated-disc case. This suggests 
that by adjusting the value of $\mathcal C$ one could obtain a better 
agreement between model and observation. Of course playing with 
parameters is of no interest if it is not supported by physical 
arguments; and reliable physical arguments could be difficult to find, 
because determining the value(s) of $\mathcal C$ presumably involves not 
only solving the transfer problem for irradiating X-rays but also 
modeling the changing shape of the warped disc (Ogilvie \& Dubus 2001). 
 
Comparisons of the model with other types of LMXBTs such as GRO J1655-40 
may require the addition of other effects to the irradiated-truncated DIM. 
Esin et al. (2000a) have suggested that the plateau phase of the outburst  
in this system could be due to mass-transfer enhancement triggered by 
the (observed) X-ray irradiation of the secondary. As for dwarf novae, a 
successful model will have to include several additional physical 
mechanisms not taken into account in the `standard' version. 
 
Many X-ray light-curves of LMXTBs show secondary maxima (``kinks"). They 
don't have an explanation in the framework of the irradiated-truncated 
DIM. The explanation proposed by King \& Ritter (1998) (a two-phase 
irradiation of the disc) is not confirmed by model calculation (DHL). 
Cannizzo (2000) noticed that the fast rise-time of the secondary maximum 
precludes a viscous origin of this phenomenon and in consequence  
cannot be due to mass-transfer rate fluctuations. Cannizzo proposed that the 
secondary maximum is due to irradiation-induced evaporation and 
re-filling of the inner disc during the decay from outburst. However, Dubus 
(2001) showed that in models in which irradiation is directly related to 
the accretion rate at the inner disc edge the resulting secondary maxima 
should have shapes different from observed ones. 
 
Several LMXBTs show also `secondary outburst' (sometimes also called 
`reflares' but we have reserved this term for the features discussed in 
Sect. \ref{reflares}) similar to those observed in WZ Sge-type systems  
(Kuulkers et al. 1996). No convincing model exist for this phenomena. 
 
\subsection{X-rays in quiescence} 
\label{quiescx} 
 
As mentioned in Sect. \ref{quidn}, the DIM rund into the same 
difficulties with quiescent LMXBTs as it does with quiescent dwarf 
novae. In particular, quiescent X-ray luminosities are too large compared 
to the predictions of the model.

The truncated DIM provides a natural explanation of quiescent X-ray 
fluxes: they are emitted by the inner hot, optically thin ADAF (Narayan 
et al. 1996; Lasota et al. 1996; Narayan et al. 1997a). As discussed in 
detail above, disc truncation is also necessary to make the DIM work 
in the other phases of the LMXBT outburst cycle (Hameury et al. 1997; 
Menou et al. 2000; DHL). Therefore according to the truncated DIM 
quiescent X-rays are produced by accretion. 
 
However, since observed X-ray luminosities are often rather low, one 
should be sure that the X-rays are not emitted by other sources, in 
principle less powerful than accretion. Dwarf novae observations show that, 
quiescent X-rays are emitted by the accretion flow and not by the 
secondaries coronae (see Sect. \ref{quidn}).  
 
In the case of neutron-star LMXBTs, Brown, Bildsten \& Rutledge (1998) 
attribute the quiescent X-rays to thermal emission from the neutron-star 
surface. This emission would be due to repeated deposition during the 
outbursts of nuclear energy deep in the crust. This could be a viable 
alternative to the accretion model (Rutledge et al. 1999). In this case 
X-ray observations would allow the quiescent disc to extend down to the 
neutron star (or the last stable orbit). Menou et al. (1999c) discuss in 
detail the model in which the inner part of the quiescent accretion flow 
around a neutron star forms an ADAF. In this case the advected thermal 
energy must be emitted from the stellar surface; so, for the same 
mass-transfer rate, quiescent neutron-star LMXTBs should be more 
luminous than those containing black holes. This difference in 
luminosities is indeed observed. The problem, however, is that the ADAF 
model predicts quiescent neutron-star LMXTB luminosities much higher 
than observed. In order to reduce the accretion rate onto the neutron 
star (three orders of magnitude are needed) an ADAF model has 
therefore to involve the presence of winds and/or the ejection of the 
accretion flow by the magnetic field of a rapidly rotating neutron star: 
the `propeller effect' (Menou et al. 1999c). According to Chandler \& 
Rutledge (2000) no pulsations were observed in the system Aql X-1 in a 
phase of the outburst cycle where the propeller effect should be at 
work. This could just constrain the geometry of the system, or it could support 
the Rutledge et al. (1999) model. On the other hand Colpi et al. (2000) 
found that this model applied to Cen X-4 would require the neutron star 
in this system to be much heavier than that of e.g. Aql X-1, a difference 
which is not easy to understand.  
 
Recent {\sl Chandra} observations of quiescent Cen X-4 (Rutledge et al. 
2000) show an X-ray thermal plus power-law component spectrum. The 
properties of the thermal component are compatible with the Brown et al. 
(1998) model. The origin of the (variable) power-law component is not 
clear. It could come from an ADAF but Menou \& McClintock (2000) have 
pointed out several difficulties faced by the ADAF model of neutron-star 
LMXTBs. They show that if such a model is applied to Cen X-4, the 
quiescent optical-UV emission can be due neither to the disc nor to the 
ADAF. Lasota et al. (1996) and Narayan et al. (1996) argued that in 
black-hole LMXTBs the optical-UV emission should originate in the ADAF. 
Menou \& McClintock (2000) show that in Cen X-4 this emission could be 
due to the `bright-spot'. If this were also the case for black-hole 
LMXBTs, conclusions based on the Narayan et al. (1996) model would have 
to be revised (there may also be other reasons for revising them, in 
view of Quataert \& Narayan 1999 results). Menou \& McClintock (2000) 
also show that the hard X-ray component observed in Cen X-4 and in 
neutron-star LMXBTs (Barret \& Vedrenne 1994) cannot be only the result 
of Compton up-scattering in an ADAF. They conclude that if an ADAF is 
present around the neutron star in quiescent LMXBTs it does not leave 
any signature of its presence. All these problems are not really 
embarrassing for the irradiated-truncated DIM for LMXBT outbursts: it 
requires only a truncation, an inner `end' of the disc. What exactly 
fills the hole is of lesser importance. 
 
This implies that arguments against the propeller hypothesis based on 
the DIM should be treated with care. Zhang, Yu \& Zhang (1998) and 
Campana et al. (1998) observed in Aql X-1 a three-order-of-magnitude 
decay of the X-ray flux in less than 10 days. They thought this implied 
a turned-on rotation-powered pulsar. An even faster decay has been 
observed by Gilfanov et al. (1998) during an ouburst of the millisecond 
X-ray pulsar SAX J1808.4-3658. In this case, of course, the propoller 
explanantion seems obvious but Gilfanov et al. also consider the 
possibility that the fast decay is due to the propagation a cooling 
wave. Although fast decays (``linear decay") are observed in some 
dwarf-novae they are never as drastic as in Aql X-1 and SAX 
J1808.4-3658. In any case, in several dwarf-novae the propeller effect 
could be at work (e.g. Lasota et al. 1999). Gilfanov et al. (1998) state 
that ``abrupt cut-offs of the light curves are common" in black-hole 
LMXBTs, but consulting for example Chen et al. (1997) rather confirms 
Campana et al. (1998) remark that ``no indication of sudden steepening" 
is seen. As shown in the previous sections, a consistent DIM for LMXBTs 
requires a truncated inner disc which favours a propeller's action ... 
if a magnetic field is present. For the moment there are no predicted 
X-ray light-curves but they could appear (Dubus, Menou \& Esin 2001).

Bildsten \& Rutledge (2000) assert that in the case of black-hole LMXTBs 
the quiescent X-rays may be due to coronal emission from stellar 
companions. They argue that in these systems the ratio of the X-ray flux 
to the stellar bolometric flux is $\lta 10^{-3}$ as in RS CVn stars, 
which are active, close, detached binaries of late-type stars (a G of K 
type giant or subgiant in orbit with a late-type main-sequence or 
subgiant) in which, for orbital periods of $\lta 30$ days, the rotation 
of both components is synchronous with the orbit. Their coronal X-ray 
emission may be as large as $10^{31}$ erg s$^{-1}$ (Dempsey et al. 
1993). However, Lasota (2000a) showed that all the available evidence 
makes it is very unlikely that companions of black-hole LMXTBs are the 
source of the quiescent X-ray luminosity. In fact, as clearly seen in 
Fig. 1 of Lasota (2000a), the luminosities of the three X-ray detected 
black-hole LMXTBs are much higher than those of RS CVn stars and well 
above the observationally determined (and rather well understood 
theoretically) upper limit of the X-ray luminosity of rapidly rotating 
late type stars. In addition there is no reason for the secondaries in 
black-hole LMXTBs to be different from their equivalents in dwarf novae 
(e.g. King 1999). Therefore as in quiescent dwarf novae, quiescent 
X-rays cannot come from the secondary's coronal activity: they must be 
emitted by the accretion flow. This conclusion was firmly confirmed by 
Garcia et al. (2000) whose {\sl Chandra} detection of GRO J0422+32 at 
$\sim 10^{31}$ erg s$^{-1}$ should put an end to the unfortunate 
story of X-ray coronal emission from  LMXTBs. 
 
Lasota (2000a) noticed that the quiescent X-ray luminosity of the three  
then detected black-hole LMXTBs satisfies the relation: 
\begin{equation} 
L_X \approx L_q=7.3 \times 10^{31} P_{\rm day}^{1.77} {\rm erg \ s^{-1}} 
\label{lx}  
\end{equation}  
which finds a natural explanation in the context of the truncated DIM. 
{\sl Chandra} observations by Garcia et al. (2000) confirm the period 
dependence predcited by Eq. (\ref{lx}). These observvations also show 
that the quiescent X-ray luminosity of LMXBTs varies by factors 
from 2 to $\sim 13$.

\section{Conclusions and perspectives} 
 
The thermal-viscous disc instability model of dwarf novae and low-mass 
X-ray binary transients can describe many aspects of the outburst cycle 
of these systems if it is complemented by additional physical mechanisms 
not taken into account in the original version. Mass-transfer rate 
variations (intrinsic or irradiation-induced), disc heating by stream 
impact and tidal torques, disc irradiation by the the white dwarf, disc 
truncation by magnetic fields or by evaporation, 
enhanced-by-tidal-torques angular-momentum removal, are physical 
processs which when taken into account in the appropriate circumstances, 
can in principle reproduce the wealth of dwarf-nova outburst 
light-curves and spectral variations. Disc (self)irradiation and inner 
truncation due to evaporation must be taken into account for the model 
to reproduce some simple properties of LMXBTs light-curves. Clearly, 
additional structures such as accretion-disc coronae and warps should be 
incorporated into the model if more complex or `non-standard' lightcurve 
shapes are to be reproduced. Since disc X-ray irradiation plays a 
dominant role in LMXBT behaviour, variations of X-ray spectra should be 
included in the model. 
 
The Achilles' heel of the DIM is  quiescence. This phase of  
the outburst cycle is not accurately described by the model and some of 
its basic assumptions might have to be drastically revised. This is the most 
fundamental problem which must be solved before the model can be considered a 
credible description of disc outbursts in close binaries. 
 
From the very beginning, the DIM's main problem has been the origin of 
the angular-momentum transport and energy dissipation mechanisms. 
Although it is often called the ``viscosity problem", its solution might 
imply that other than viscous mechanisms are at play. 
 
Inner disc evaporation is an important ingredient of the `generalized' 
DIM, especially when it is applied to LMXBTs. Although several 
models have been proposed, the exact nature of inner disc truncation 
still awaits eplanation. 
 
In addition one can list the following more detailed problems that remain  
to be solved and questions to be answered (or answered unequivocally): 
 
\begin{itemize} 
 
\item What produces superoutbursts? The most probable answer to this 
question would come from a `hybrid' model combining tidal torques and 
mass-transfer enhancements (Smak 1996). 

\item Why do recurrence times of SS Cyg and U Gem differe by a factor 2.5?
 
\item Why is the recurrence time of WZ Sge so long? 
 
\item The origin and evolution of warps in accretion discs during 
outbursts. This problem is of course more general because steady discs 
should also be warped. 
 
\item How is matter fed into the disc? There is some observational 
evidence that in dwarf novae during decline from outburst (see e.g. 
Steeghs et al. 2000) and in quiescence (absence of hot spot in HT Cas - 
Wood, Horne \& Vennes 1992) part of the stream overflows the outer disc 
rim. Then it should hit the disc somewhere close to the circulation 
radius (Lubow 1989). Similar ideas were proposed for LMXBs by Frank, 
King \& Lasota (1987). 
 
\item What are the reasons for secondary outbursts in WZ Sge-type dwarf 
novae and LMXBTs? Despite similarities between the two types of systems 
(Lasota 1996a,b; Kuulkers et al. 1996) the reasons could be different. 
 
\item Why do some LMXBTs after many years of quiescence enter into an 
active phase during which several outbursts occur within a couple of years 
(e.g. GRO J1655-40)? 
 
\item Does the thermal instability produce only outbursts? Can it 
produce more gentle luminosity modulations as has been observed by e.g. 
Dubus et al. (1997)? The answer seems to `yes' (Dubus \& Lasota 2001). 
 
\item What is the origin of the various spectral components of the 
quiescent emission from LMXBTs? The answer to this question is important 
not only for testing the DIM but for deciding whether quiescent LMXBTs 
which are supposed to host black-holes because of high masses of the 
accreting bodies, show evidence for event horizons as claimed by Narayn, 
Garcia \& McClintock (1997; see also Lasota \& Hameury 1998; Menou et 
al. 1999c; Garcia et al. 2000). 
 
\item How are accretion disc X-rays irradiated? What is the geometry? 
What are the respective roles of the corona and warp.  
 
\item In many (all ?) black hole X-ray binaries powerful jets are observed 
in what is called the ``low/hard X-ray state" (see Fender 2000 for a review). 
Do these outflows influence the outbursts and how? Of course to answer 
these questions one should presumably understand {\sl what} produces  
jets. 
 
\end{itemize}

\section*{Acknowledgments} 
 
I am most grateful to Jean-Marie Hameury, Guillaume Dubus and Kristen 
Menou, without whom this article could not have been written. The reasons 
for this are evident: a quick look at this article shows how much I owe 
them. GD, JMH and KM also read various preliminary versions of this 
article and made very useful comments and suggestions. I express my 
gratitude to Joe Smak who taught me so many things about the disc 
instability model. Thanks are also due to Valentin Buat-M\'enard, Phil 
Charles, Anya Esin, Jean-Marc Hur\'e, Rob Hynes, Irit Idan, Erik 
Kuulkers, Jeff McClintock, Ramesh Narayan, Giora Shaviv, Brian Warner 
and Insu Yi with whom I collaborated on disc-instability models of dwarf 
novae and X-ray transients. Discussions over the years with Marek 
Abramowicz, Steve Balbus, Didier Barret, Axel Brandenburg, John 
Cannizzo, Wolfgang Duschl, Juhan Frank, Charles Gammie, Carole Haswell, 
John Hawley, Keith Horne, Andrew King, Wojciech Krzemi\'nski, Emmie 
Meyer-Hofmeister, Friedrich Meyer, Shin Mineshige, Yoji Osaki, Bohdan 
Paczy\'nski, John Papaloizou, Hans Ritter, Oded Regev, Rob Robinson, 
S{\l}awek Ruci\'nski, Henk Spruit, Caroline Terquem, Craig Wheeler, 
Janet Wood and Jean-Paul Zahn were of great help. This work was 
supported in part by the {\sl Programme National de Physique 
St\'ellaire} of the CNRS. 
  
\newpage

\appendix 
\section{APPENDIX} 
 
Dubus (1998) obtained the following fits of ${\Sigma_{\rm max, min}}$  
(these fits correspond to the case called ``optically thick" in HMDLH): 
\begin{eqnarray}  
\lefteqn{\Sigma_{\rm max} = 13.4 ~ \alpha^{-0.83} \left( {M_1 \over \rm  
M_\odot} \right)^{-0.38} \left( {r \over 10^{10} \ \rm cm} \right)^{1.14}  
~\rm g~cm^{-2}} \\ 
\label{sigmax}  
\lefteqn{\Sigma_{\rm min} = 8.3 ~ \alpha^{-0.77} \left( {M_1 \over \rm  
M_\odot} \right)^{-0.37} \left( {r \over 10^{10} \ \rm cm} \right)^{1.11}  
~\rm g~cm^{-2}} 
\label{sigmin}  
\end{eqnarray} 
of the corresponding accretion rates: 
\begin{eqnarray}  
\lefteqn{\dot{M}_{\rm A} = 4.0 ~ 10^{15} ~ \alpha^{-0.004} \left( {M_1 \over   
\rm M_\odot} \right)^{-0.88} \left( {r \over 10^{10} \ \rm cm} \right)^{2.65}  
~\rm g~s^{-1}} \\ 
\label{mdmax}  
\lefteqn{\dot{M}_{\rm B} = 9.5 ~ 10^{15} ~ \alpha^{0.01} \left( {M_1 \over   
\rm M_\odot} \right)^{-0.89} \left( {r \over 10^{10} \ \rm cm} \right)^{2.68}  
~\rm g~s^{-1}}, 
\label{mdmin}  
\end{eqnarray}  
of the critical effective temperatures: 
\begin{eqnarray}  
\lefteqn{{T}_{\rm eff~A} = 5800 ~ \alpha^{-0.001} \left( {M_1 \over   
\rm M_\odot} \right)^{0.03} \left( {r \over 10^{10} \ \rm cm} \right)^{-0.09}  
~\rm K} \\ 
\label{tecc}  
\lefteqn{{T}_{\rm eff~B} = 7200 ~ \alpha^{-0.002} \left( {M_1 \over   
\rm M_\odot} \right)^{0.03} \left( {r \over 10^{10} \ \rm cm} \right)^{-0.08}  
~\rm K}  
\label{te} 
\end{eqnarray}  
and critical midplane temperatures:  
\begin{eqnarray}  
\lefteqn{{T}_{\rm c~A} = 9000 ~ \alpha^{-0.13} \left( {M_1 \over   
\rm M_\odot} \right)^{0.00} \left( {r \over 10^{10} \ \rm cm} \right)^{-0.01}  
~\rm K} \\  
\lefteqn{{T}_{\rm c~B} = 21700 ~ \alpha^{-0.21} \left( {M_1 \over   
\rm M_\odot} \right)^{-0.02} \left( {r \over 10^{10} \ \rm cm} \right)^{0.05}  
~\rm K}  
\label{tc} 
\end{eqnarray}  
 
For $\alpha < 0.1$ a second critical point appear on the lower branch of 
the \bS-curve (Sect. \ref{stansc}). The critical surface densities and accretion rates 
can be represented as: 
\begin{eqnarray}  
\lefteqn{\Sigma^l_{\rm max} = 6.5 ~ \alpha^{-0.80} \left( {M_1 \over \rm  
M_\odot} \right)^{-0.38} \left( {r \over 10^{10} \ \rm cm} \right)^{1.13}  
~\rm g~cm^{-2}} \\  
\lefteqn{\Sigma^l_{\rm min} = 5.1 ~ \alpha^{-0.88} \left( {M_1 \over \rm  
M_\odot} \right)^{-0.39} \left( {r \over 10^{10} \ \rm cm} \right)^{1.19}  
~\rm g~cm^{-2}} \\  
\lefteqn{\dot{M}^l_{\rm A} = 0.5 ~ 10^{15} ~ \alpha^{-0.09} \left( {M_1 \over   
\rm M_\odot} \right)^{-0.90} \left( {r \over 10^{10} \ \rm cm} \right)^{2.64}  
~\rm g~s^{-1}} \\  
\lefteqn{\dot{M}^l_{\rm B} = 0.6 ~ 10^{15} ~ \alpha^{0.001} \left( {M_1 \over   
\rm M_\odot} \right)^{-0.89} \left( {r \over 10^{10} \ \rm cm} \right)^{2.70}  
~\rm g~s^{-1}},  
\end{eqnarray}  
the critical effective temperatures as: 
\begin{eqnarray}  
\lefteqn{{T}^l_{\rm eff~A} = 3500 ~ \alpha^{0.02} \left( {M_1 \over   
\rm M_\odot} \right)^{0.03} \left( {r \over 10^{10} \ \rm cm} \right)^{-0.09}  
~\rm K}\\  
\lefteqn{{T}^l_{\rm eff~B} = 3600 ~ \alpha^{-0.00} \left( {M_1 \over   
\rm M_\odot} \right)^{0.03} \left( {r \over 10^{10} \ \rm cm} \right)^{-0.08}  
~\rm K}  
\end{eqnarray}  
and the critacl midlplane temperature can be fitted by: 
\begin{eqnarray}  
\lefteqn{{T}^l_{\rm c~A} = 3300 ~ \alpha^{-0.06} \left( {M_1 \over   
\rm M_\odot} \right)^{0.01} \left( {r \over 10^{10} \ \rm cm} \right)^{-0.03}  
~\rm K}\\  
\lefteqn{{T}^l_{\rm c~B} = 3500 ~ \alpha^{-0.18} \left( {M_1 \over   
\rm M_\odot} \right)^{-0.02} \left( {r \over 10^{10} \ \rm cm} \right)^{0.06}  
~\rm K}.   
\end{eqnarray}  
 
When the disc is irradiated (see Sects.\ref{irrsc} \& \ref{global}) one obtains the following fits 
(DHL) for the critical surface densities ($\xi=(T_{\rm irr}/10^4{\rm \ K})^2$): 
\begin{eqnarray} 
\lefteqn{ 
\Sigma_{\rm max}=(10.8-10.3\xi){\rm ~} \alpha_{\rm cold}^{-0.84}  
\left( {M_1 \over   
\rm M_\odot} \right)^{-0.37+0.1\xi} }\\ 
\lefteqn{\hskip 6.5truecm \times 
\left( {r \over 10^{10} \; \rm cm} \right)^{1.11-0.27\xi}{\rm ~~}\rm{ g}\  
{\rm cm}^{-2}}\nonumber\\ 
\lefteqn{\Sigma_{\rm min}=(8.3-7.1\xi){\rm ~} \alpha_{\rm hot}^{-0.77}  
\left( {M_1 \over   
\rm M_\odot} \right)^{-0.37} \left( {r \over 10^{10} \ \rm cm}  
\right)^{1.12-0.23\xi}{\rm ~~}\rm{ g}\ {\rm cm}^{-2}} 
\label{simaxmin} 
\end{eqnarray} 
and for the critical midplane temperatures: 
\begin{eqnarray} 
\lefteqn{T_{\rm c}(\Sigma_{\rm max})=10700{\rm ~} \alpha_{\rm cold}^{-0.1} 
\left( {r \over 10^{10} \; \rm cm} \right)^{-0.05\xi}{\rm ~~K}}\\ 
\lefteqn{T_{\rm c}(\Sigma_{\rm min})=(20900-11300\xi){\rm ~} 
\alpha_{\rm hot}^{-0.22}  
\left( {M_1 \over   
\rm M_\odot} \right)^{-0.01}}\nonumber\\ 
\lefteqn{\hskip 7truecm \times \left( {r \over 10^{10} \rm cm} 
\right)^{0.05-0.12\xi} {\rm ~~K}}. 
\label{tcmaxmin} 
\end{eqnarray}

\end{document}